\documentclass[a4paper,twoside,openanay,11pt]{book}

\usepackage{natbib}
\usepackage{graphicx}
\graphicspath{{Figures/}}

\usepackage{geometry}

\usepackage{todonotes}
\usepackage{etoolbox}
\usepackage{multirow}
\usepackage{diagbox}
\usepackage{hyperref}
\usepackage[utf8]{inputenc}
\usepackage[T1]{fontenc}

\usepackage{caption}
\usepackage{subcaption}
\usepackage{tcolorbox}
\usepackage{bigfoot}
\usepackage{epigraph}
\usepackage[shortlabels]{enumitem}
\usepackage{hhline}
\usepackage{listings}
\usepackage{color}
\usepackage{amsmath}
\usepackage{array}
\usepackage{eurosym}
\usepackage{gensymb}
\usepackage{amsthm}
\usepackage[toc,page]{appendix}
\usepackage{pdfpages}

\theoremstyle{definition}
\newtheorem{definition}{Definition}[section]
\newtheorem{example}{Example}[section]
\newtheorem{usecaseinner}{Use Case}
\newenvironment{usecase}[1]{%
	\renewcommand\theusecaseinner{#1}%
	\usecaseinner
}{\endusecaseinner}

\definecolor{dkgreen}{rgb}{0,0.6,0}
\definecolor{gray}{rgb}{0.5,0.5,0.5}
\definecolor{mauve}{rgb}{0.58,0,0.82}

\newtoggle{hidecomments} 

\iftoggle{hidecomments}{ 
	\newcommand{\adrien}[1]{}
	\newcommand{\tom}[1]{}
	\newcommand{\anis}[1]{}
	\newcommand{\bruno}[1]{}
	\newcommand{\gilles}[1]{}
	\newcommand{\reviewer}[1]{}
}{
	\newcommand{\adrien}[1]{\todo[color=blue!40, inline]{\footnotesize{Adrien: #1}}}
	\newcommand{\tom}[1]{\todo[color=red!40, inline]{\footnotesize{Tom: #1}}}
	\newcommand{\anis}[1]{\todo[color=green!40, inline]{\footnotesize{Anis: #1}}}
	\newcommand{\bruno}[1]{\todo[color=yellow!40, inline]{\footnotesize{Bruno: #1}}}
	\newcommand{\gilles}[1]{\todo[color=gray!40, inline]{\footnotesize{Gilles: #1}}}
	\newcommand{\reviewer}[1]{\todo[color=white!40, inline]{\footnotesize{Reviewer: #1}}}
}

\newcommand{\Csharp}{%
	{\settoheight{\dimen0}{C}C\kern-.05em \resizebox{!}{\dimen0}{\raisebox{\depth}{\#}}}}

\usepackage[acronym]{glossaries}

\newacronym{aec}{AEC}{Architecture, Engineering and Construction}
\newacronym{ad}{AD}{Algorithmic Design}
\newacronym{cad}{CAD}{Computer-Aided Design}
\newacronym{caad}{CAAD}{Computer-Aided Architectural Design}
\newacronym{wimp}{WIMP}{Windows, Icons, Menus and Pointers}
\newacronym{hmd}{HMD}{Head-Mounted Display}
\newacronym{ar}{AR}{Augmented Reality}
\newacronym{vr}{VR}{Virtual Reality}
\newacronym{dse}{DSE}{Design Space Exploration}
\newacronym{ga}{GA}{Genetic Algorithm}
\newacronym{gan}{GAN}{Generative Adversarial Network}
\newacronym{mr}{MR}{Mixed Reality}
\newacronym{xr}{XR}{eXtended Reality}
\newacronym{imu}{IMU}{Intertial Measurement Unit}
\newacronym{bim}{BIM}{Building Information Modelling}
\newacronym{gd}{GD}{Generative Design}
\newacronym{rv}{RV}{Reality Virtuality}
\newacronym{6dof}{6-DoF}{Six Degrees of Freedom}
\newacronym{brep}{B-rep}{Boundary Representation}
\newacronym{csg}{CSG}{Constructive Solid Geometry}
\newacronym{nurbs}{NURBS}{Non-uniform rational B-spline}
\newacronym{gdl}{GDL}{Geometric Description Language}
\newacronym{vpl}{VPL}{Visual Programming Language}
\newacronym{hci}{HCI}{Human-Computer Interaction}
\newacronym{tcp}{TCP}{Transmission Control Protocol}
\newacronym{tui}{TUI}{Tangible User Interface}
\newacronym{mqtt}{MQTT}{Message Queuing Telemetry Transport}
\newacronym{swot}{SWOT}{Strengths, Weaknesses, Opportunities, Threats}
\newacronym{sdk}{SDK}{Software Development Kit}
\newacronym{dag}{DAG}{Directed Acyclic Graph}
\newacronym{gpm}{GPM}{Generalized Parametric Model}
\newacronym{api}{API}{Application Programming Interface}
\newacronym{srgs}{SRGS}{Speech Recognition Grammar Specification}

\usepackage{titlesec}
\usepackage{titletoc}
\titleformat{\chapter}[display]
{\bfseries\huge}
{\filleft\Large\MakeUppercase{\chaptertitlename} \  \rlap{ \resizebox{!}{0.8cm}{\thechapter} \rule{5cm}{1.2cm}}}
{0ex}
{
	\vspace{2ex}%
	\filleft}
[\vspace{2ex}%
\titlerule
]

\titleformat{\section}
{\bfseries\Large}
{\thesection.}
{0.5em}
{}
[\vspace{0.5ex}
\titlerule]

\title{Integrating Immersive Technologies for Algorithmic Design in Architecture}

\author{Adrien Coppens}

\newpagestyle{front}[\small]{
	\headrule
	\sethead[\usepage][][\textsl{\chaptertitle}]
	{}{}{\usepage}}

\newpagestyle{main}[\small]{
	\headrule
	\sethead[\usepage][][\textsl{Chapter \thechapter \ -- \chaptertitle}]
	{\thesection \ -- \sectiontitle}{}{\usepage}}

\newpagestyle{back}[\small]{
	\headrule
	\sethead[\usepage][][\textsl{\chaptertitle}]
	{}{}{\usepage}}

\newpagestyle{ccl}[\small]{
	\headrule
	\sethead[\usepage][][\textsl{Chapter \thechapter \ -- \chaptertitle}]
	{}{}{\usepage}} 

\makeglossaries
\setglossarypreamble[acronym]{Various acronyms will be used throughout this dissertation to abbreviate frequent terms, some of which will even find usages across all sections. The expansion will be given at least on the first occurrence of each acronym in the text, but the following list of acronyms can be used as a reference if needed.}

\begin{document}

\frontmatter

\cleardoublepage
\thispagestyle{empty}
\thispagestyle{empty}

\includegraphics[width=3cm]{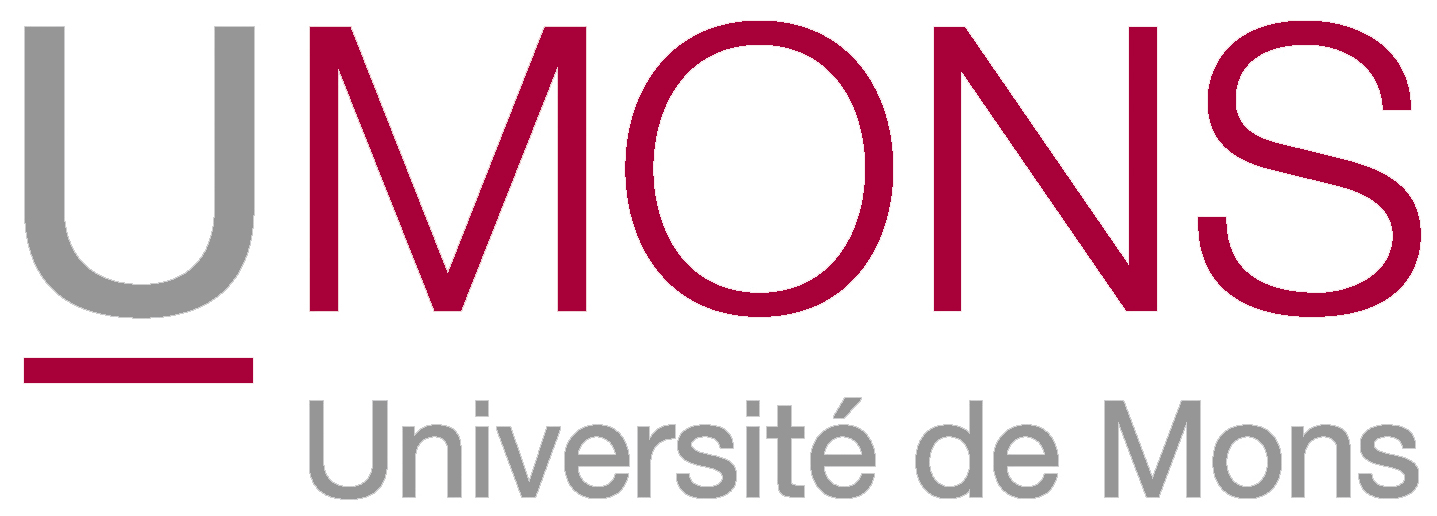}  
\hfill
\begin{tabular}[b]{c}
{\small Université de Mons}\\
{\small Faculté des Sciences}\\
{\small Département d'Informatique}
\end{tabular}
\hfill
\includegraphics[width=2.8cm]{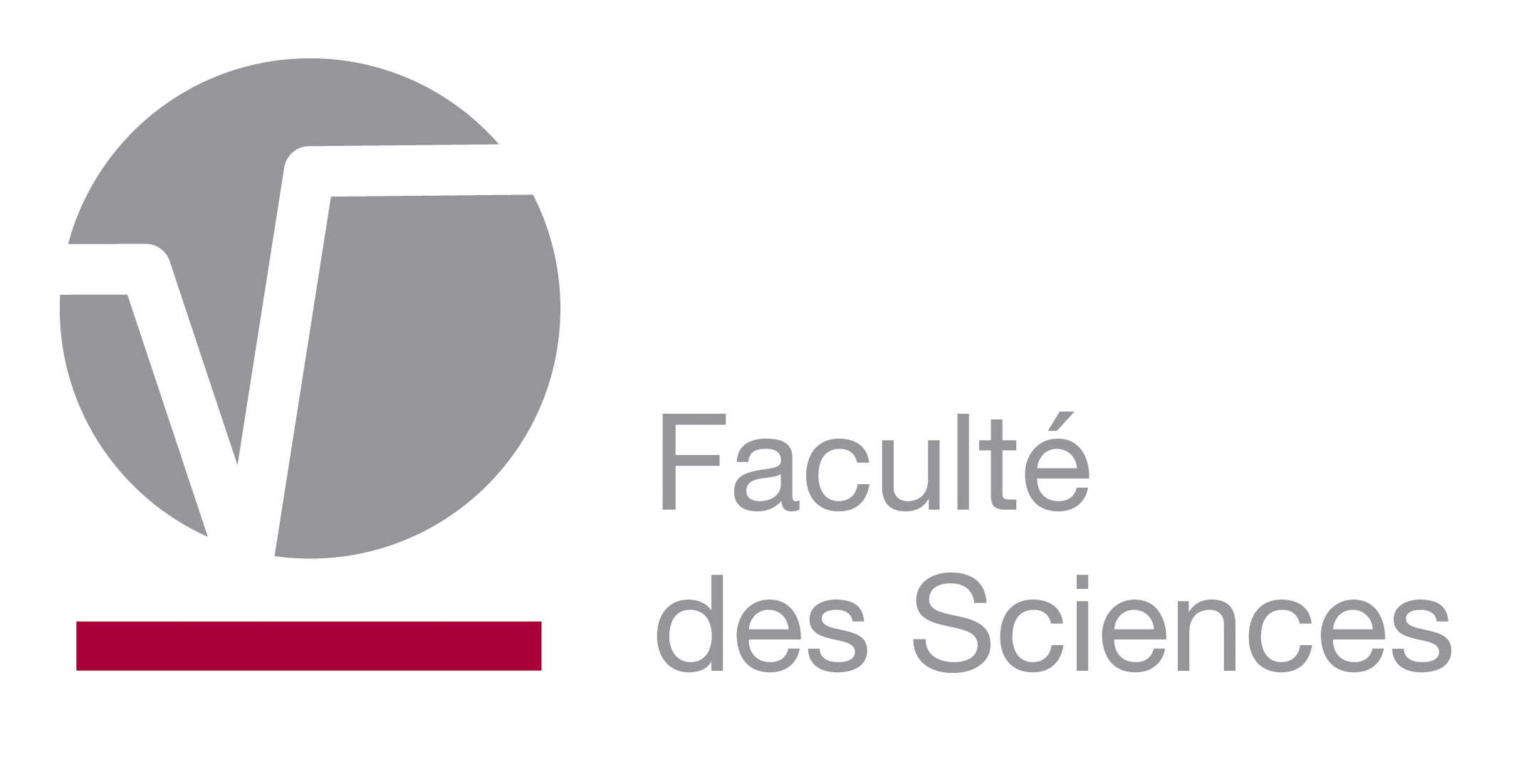}

\begin{center}

\vfill

{\bf \LARGE Integrating Immersive Technologies\\for Algorithmic Design in Architecture} \\ 

\bigskip

{\Large \bf Adrien Coppens}\\

\bigskip

A dissertation submitted in fulfilment of the requirements of\\
the degree of {\em Docteur en Sciences}

\vfill

\begin{center}
\begin{small}

\begin{tabular}{l l}
{\bf Advisors}\\
Dr.\ \textsc{Tom Mens} &  Université de Mons, Belgium\\
Dr.\ \textsc{Mohamed-Anis Gallas} &  Université de Mons, Belgium\\
\\
{\bf Jury}\\
Dr.\ \textsc{Bruno Quoitin} & Université de Mons, Belgium \\
Dr.\ \textsc{Gilles Halin} & Ecole d'Architecture de Nantes, Université de Lorraine, France\\
Dr.\ \textsc{Annie Guerriero} & Luxembourg Institute of Science and Technology, Luxembourg
\end{tabular}
\end{small}

\vfill

February 2022 
\end{center}

\end{center}
\thispagestyle{empty}
\thispagestyle{empty}
\newpage
\textcolor{white}{.}

\thispagestyle{empty}

\pagestyle{front}
\thispagestyle{empty}

\chapter*{Acknowledgements}

As an important chapter of my life is about to end, I would like to express my gratitude towards those who directly or indirectly have been part of this four-year journey.

First and foremost, I would like to thank Tom, my advisor and head of the lab I have been part of this whole time.
Thank you for giving me the opportunity to work with all these technologies I wanted to explore, even though they are outside of your main research interests,  but also for your incredible availability (how many dozens of days do you have in a week?) as well as the swiftness, quantity and quality of your invaluable feedback and guidance along the way; and thank you for letting me borrow your most excellent (immersive?) research kit during the CAMPaM workshop; so, for all these reasons and many more, bedankt!\footnote{Sentence intentionally left long and complex}

Anis, my second adviser, thank you for providing a context in which I could experiment with immersive technologies. Thank you for helping me benefit from your network of contacts in a field I knew nothing about, but also for turning your students into ``guinea pigs'' to test our prototypes and for accompanying me to these conferences and seminars.

Thank you as well to all jury and ``comité d'accompagnement'' members, for accepting to review my work and provide valuable feedback on it. Your comments and suggestions helped make this dissertation better.
More specifically, thank you Annie for inviting me at LIST during these 3 months by combating the sanitary situation. Thank you Gilles for the discussions we had at various events, and for showing me that hybrid specimens in between computer scientists and architects do exist. ACK to you Bruno, for providing the external view on my work and for giving me an excuse to talk about technical details.

``Merci'' (in Luxembourgish) to Nico, Elie, Sylvain, Calin and the people at LIST, for welcoming me in your office in order to collaborate on that prototype.

Thank you to all the colleagues I met during this time, and to the multiple co-workers I had the pleasure to share an office with. I would likely forget some of you so I would rather keep it generic but you all helped make our work environment such a nice and pleasant place.

Thank you to Robin, for choosing to work on my student project proposal, and for carrying it out nicely so that it contributed to the present dissertation.
Thank you to all the ``guinea pigs'' that tried various versions of our prototypes and provided feedback on them, and to all survey respondents for their valuable time.

Thank you to Ivan Sutherland, for inspiring and pioneering both head-mounted displays and computer-aided design. As a nod to your research achievements and how their combination matches the subject of this dissertation, each chapter starts with an epigraph with quotes from you that felt appropriate.

On a more personal level, thank you to all my tennis and padel partners, hitting the ball with you allowed me to exert myself and helped clear my mind from all those 0's and 1's. Thank you to my friends, for providing all those relaxing and enjoyable times.
Thank you to my family for your unwavering support during these four years but also since I was born.

Finally, thank you to all the ``et al.'' that I may have forgotten.

\chapter*{Abstract}

Architectural design practice has radically evolved over the course of its history, due to technological improvements that gave rise to advanced automated tools for many design tasks. Traditional paper drawings and scale models are now accompanied by 2D and 3D Computer-Aided Architectural Design (CAAD) software.

While such tools improved in many ways, including performance and accuracy improvements, the modalities of user interaction have mostly remained the same, with 2D interfaces displayed on 2D screens. The maturation of Augmented Reality (AR) and Virtual Reality (VR) technology has led to some level of integration of these immersive technologies into architectural practice, but mostly limited to visualisation purposes, e.g. to show a finished project to a potential client.

We posit that there is potential to employ such technologies earlier in the architectural design process and therefore explore that possibility with a focus on Algorithmic Design (AD), a CAAD paradigm that relies on (often visual) algorithms to generate geometries. The main goal of this dissertation is to demonstrate that AR and VR can be adopted for AD activities. 

To verify that claim, we follow an iterative prototype-based methodology to develop research prototype software tools and evaluate them. The three developed prototypes provide evidence that integrating immersive technologies into the AD toolset provides opportunities for architects to improve their workflow and to better present their creations to clients. Based on our contributions and the feedback we gathered from architectural students and other researchers that evaluated the developed prototypes, we additionally provide insights as to future perspectives in the field.
	
\tableofcontents

\mainmatter

\addcontentsline{toc}{chapter}{List of Acronyms}
\printglossary[type=\acronymtype, title=List of Acronyms]

\chapter{Introduction}
\epigraph{
``It’s not an idea until you write it down.'' }{Ivan Sutherland}

During my computer science studies, I had the opportunity to work with gestural interaction trackers and immersive technologies on several projects. I have to say I was hooked to these innovative interaction devices as soon as I got a hold of them. When the opportunity to pursue a PhD on these technologies arose, it was obvious to me I should go for it.

After a few meetings with both Prof. Tom Mens and Dr. Mohamed-Anis Gallas, we converged to an interesting area that combined my interests for the aforementioned technologies with an actual research problem: integrating immersive technologies into architectural design and Parametric Modelling in particular or, as I prefer to call it, \gls{ad}.

Due to its interdisciplinary nature, the current thesis was co-directed by Prof. Tom Mens from the Faculty of Sciences and Dr. Mohamed-Anis Gallas from the Faculty of Architecture and Urban Planning.

The present document presents my work on the subject, that was carried out during the last four years, and aims to help the reader understand why and how I proceeded as well as what perspective my work offers.

\pagebreak

\section{Context}
\begin{figure}[h]
	\includegraphics[width=.95\textwidth]{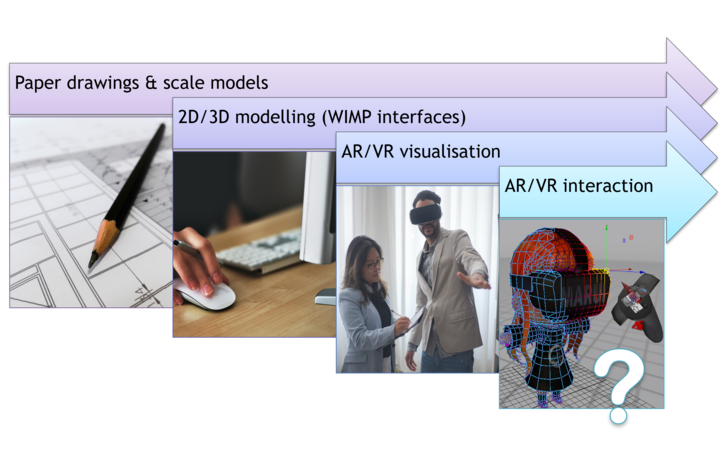}
	\centering
	\caption[]{\label{fig:context} A brief overview of the evolution of architectural tooling. Created based on pictures from online sources\footnotemark.}
\end{figure}

The \gls{aec} industry has radically evolved due to technological improvements that gave rise to advanced automated tools for many design tasks. Traditional paper drawings and scale models are now accompanied by 2D and 3D \textbf{\gls{cad}} software.

These tools have consistently improved in terms of performance and accuracy, and new features are constantly added to them, such as real-time (lighting, structural) simulations and photo-realistic renderings.

\footnotetext{This figure was created using three CC0 pictures from \href{https://www.pexels.com/}{www.pexels.com} and one (the last one) from \href{https://uploadvr.com/marui-plugins-bring-vr-support-to-3d-tools-maya-and-blender/}{uploadvr.com/marui-plugins-bring-vr-support-to-3d-tools-maya-and-blender}}

In fact, the inclusion of immersive technologies in architectural practice is currently mostly restricted to enhancing the visualisation of final models \citep{blach1998flexible}, built using traditional (non-immersive) desktop software. At that point, the design process is essentially over, and changes to the model based on the immersive feedback are unlikely.

We believe that there is potential to employ immersive technologies before the process is finished, in order to take advantage of the visualisation characteristics during design activities. These include the sense of scale, the immersion, the spatial perception as well as the interaction opportunities provided to the user \citep{delgado2020research, jayaram2001assessment, louis2020high}.

\section{Algorithmic Design in architecture}
\label{sec:intro-ad}

Immersive technologies provide opportunities to visualise a geometry in a three-dimensional (virtual) environment. While this can be beneficial for architecture in general, this is particularly attractive to design processes that would benefit from rapid iteration cycles, where the designer would like to quickly transition from the editing software to an updated view of the rendered model, and vice versa.

So as to limit the scope of our work to a reasonable area, we chose to focus on \textbf{\glsfirst{ad}} \citep{woodbury2010elements, MonederoParametricdesignreview2000}. We made that choice because that design paradigm inherently favours parametrised adjustment of designed models and therefore is a suitable candidate for the rapid iterations we suggest.

Furthermore, we identified a clear lack of support for immersive technologies with the most popular software tools supporting \gls{ad}, leading to more opportunities for us to explore the field. We will consequently mostly discuss the integration of immersive technologies with \gls{ad} in particular.
But what is \gls{ad} exactly?

\gls{ad} is an architectural design paradigm that involves generating geometries using algorithms which are often driven by parameters that can be changed, allowing designers to explore different solutions by tweaking the values of these parameters. Algorithms can be represented in textual or visual forms, sometimes interchangeably, and may correspond to different programming paradigms (see section \ref{sec:computational-design} for more information on the paradigms covered by \gls{ad} languages and tools). 

\begin{figure}[t]
	\includegraphics[width=.95\textwidth]{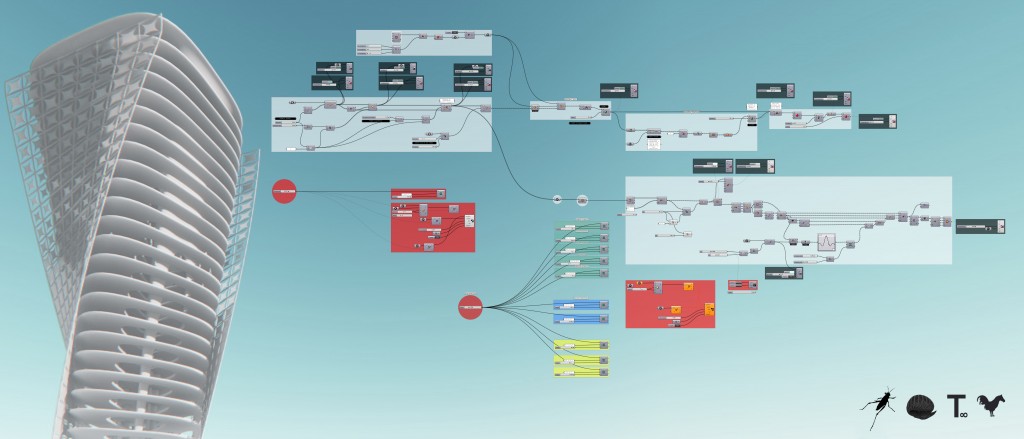}
	\centering
	\caption[]{\label{fig:gh-example} Example of an architectural project that relies on \gls{ad} (through Grasshopper and additional plug-ins) to generate a tower. Reproduced from an example project by Mario Alberto Espinoza\footnotemark.}
\end{figure}

This enables the designer to explore different solutions by modifying the values of parameters from these algorithms, and facilitates the design process of complex structures, such as the control tower from the Managua Airport, represented in Figure \ref{fig:gh-example}.

\pagebreak
\footnotetext{\url{http://www.iaacblog.com/programs/managua-tower/}} 

In architectural practice, the most common representation form is flow-based programming \citep{morrison1994flow} through a visual interface. Tools such as Grasshopper\footnote{\url{http://grasshopper3d.com}}, GenerativeComponents\footnote{\url{http://bentley.com/products/product-line/modeling-and-visualization-software/generativecomponents}} and Dynamo Studio\footnote{\url{http://autodesk.com/products/dynamo-studio}}, stand out as the most popular software solutions \citep{cichocka2017optimization}. 

\begin{figure}[h]
	\includegraphics[width=.95\textwidth]{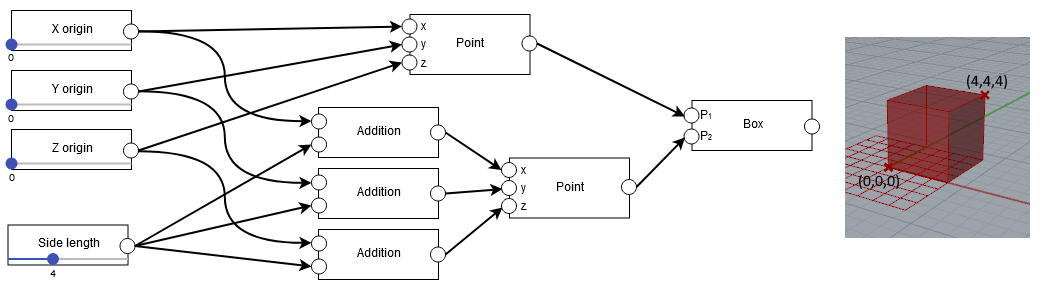}
	\centering
	\caption{\label{fig:simple-cube} A basic example of an \gls{ad} visual algorithm with its associated geometry.}
\end{figure}

Figure \ref{fig:simple-cube} shows an abstract representation of such a visual algorithm that generates a simple cube, with the corresponding rendered geometry on the right.

In flow-based programming, the final output is constructed by connecting processes that have an internal behaviour and return an output value. The idea is that in order to produce a result, one does not need to know the details about the inner workings of each process. The processes can be thought of as ``black boxes'', whose output values can simply be reused by other processes.
Here, the output value of each process is often a geometry that is passed from one component to the next in order to construct the final output.

We chose Grasshopper as our target \gls{ad} system for the real prototypes we aim to develop, because it is one of the most popular \gls{ad} systems used in both research and industrial contexts \citep{cichocka2017optimization, stals2018influence}. Additionally, its software development toolkit supports \Csharp, which is convenient considering my familiarity with Unity, that also supports \Csharp . Choosing Grasshopper also means we have an easy access to a license and informal evaluators since the usage of that software is taught at the University of Mons, as part of one of Dr. Gallas' courses.

\subsection{Terminology}

On a terminological note, \gls{ad} is often referred to as \emph{computational design} or \emph{parametric modelling}. Both of these terms appear too generic to us since they could apply to non-algorithmic design as well: the former simply informs that a computer was used, while the latter signals that the design is driven by parameters; parametric modelling (or parametric design) is in fact regularly confused with \gls{bim} (see section \ref{sec:bim}).
A related term whose definition and relation with \gls{ad} is sometimes unclear is \gls{gd}. Some consider it to be a superset of \gls{ad}, while others define it as a subtype of \gls{ad} \citep{caetano2020computational}. Most publications mentioning \gls{gd} are related to (multi-criteria) optimisation \citep{VillaggiSurveyBasedSimulationUser2018}, machine learning techniques \citep{NagyProjectDiscoverApplication2017} and Domain Space Exploration \citep{calixto2015literature}, all of which necessarily involve algorithms that generate designs. Hence, we would be tempted to agree with the latter group and consider \gls{gd} as a subset to \gls{ad}.

Since the most popular software tools for \gls{ad} offer visual representations of the algorithms (similar to what is shown in Figure \ref{fig:simple-cube}), more specific terms could be used to emphasise on that aspect. Based on computer science terminology, we could indeed consider these visual algorithms as graph structures (and therefore come up with a term such as ``graph-based design'') or even highlight the fact that we rely on flow-based programming (which would lead to a term such as ``flow-based design'' or ``dataflow-based design'' to disambiguate with other types of flows such as control flow). However, these terms are not standard in the architectural field and appear to be too technical to be broadly used. Another issue that distances us from relying on terms that highlight the visual nature of these representations is that \gls{ad} software generally also support textual programming. We will therefore simply refer to the previously defined modelling paradigm as \gls{ad} from now on.
This choice of term is consistent with a recent paper that results from a literature review \citep{caetano2020computational}.

Additionally, we note that the ``model'' term in the context of \gls{ad} can be ambiguous: it could be used to designate a visual \gls{ad} program or the generated (or even rendered) geometry. For that reason, in the remaining of this document we will try to avoid mentioning ``model'' and instead use ``algorithm'' and ``geometry'' (or geometrical representation) to circumvent this ambiguity whenever possible.

\section{Thesis statement} 

Since we consider the integration of immersive technologies to be particularly lacking in Algorithmic Design editors in architecture, our aim is to create and evaluate prototypical tools that show the potential of using these technologies as part of that design paradigm's toolkit.

The main question that this thesis dissertation tries to answer is: ``How can immersive technologies be adopted for Algorithmic Design activities in architecture?''. This leads to multiple sub-questions: Which immersive technologies are the most appropriate? How are these technologies perceived by practitioners? How can these technologies be integrated with existing tools, both technically and in terms of user interaction, so that they are adapted to \gls{ad} practice ?

\begin{center}
	\begin{tcolorbox}[sharp corners, colback=blue!10, colframe=gray!80!blue, coltitle=white,fonttitle=\sffamily\large, title=Thesis statement, width=.97\textwidth]
		Integrating \gls{ar} and \gls{vr} technologies into the Algorithmic Design toolset provides opportunities for architects to improve their workflow and to better present their creations to clients.

	\end{tcolorbox}
\end{center}

\section{Research methodology}
\label{sec:methodology}

To provide evidence for the thesis statement, we demonstrate that both \textbf{\glsfirst{ar}} and \textbf{\glsfirst{vr}} can be integrated into \gls{ad} practice, through the implementation of several prototypes. They show that such technologies are useful additions to the toolset at the architects' disposal for \gls{ad} activities. We substantiate that usefulness through surveys and interviews.

\begin{figure}[h]
	\includegraphics[width=.5\textwidth]{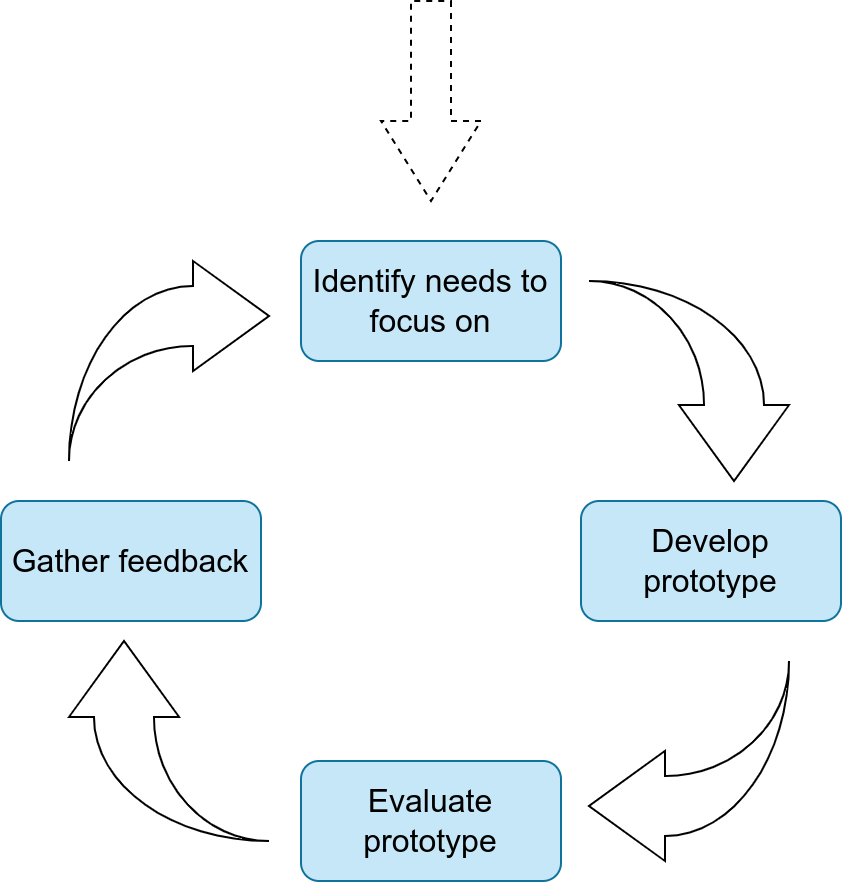}
	\centering
	\caption{\label{fig:iterative-process} The iterative research process we followed to support the thesis statement.}
\end{figure}

To structure our course of action, we generally follow an iterative and prototype-based process. Figure \ref{fig:iterative-process} depicts that process, which closely resembles the action research spiral \citep{kemmis2014action} (plan, act, observe, reflect, re-plan, etc.).

First, we identify a gap in existing tooling, that we attempt to fill with a prototype. This allows us to organise evaluations of that prototype with students and academics, through workshops and visits in universities. Based on the feedback we thereby gather, we identify further needs and therefore start another iterative cycle with a new prototype.

During these four years, we repeat that cycle three times, with three different prototypes being developed. 
The first prototype enables geometry streaming and parameter adjustment in \gls{vr}, through a bridge between Grasshopper and a virtual environment.
The second one lets users have greater control over their design, with actual algorithm editing capabilities. That control necessitates novel interaction mechanisms and we therefore explore a few options via different techniques and devices.
The third and last prototype goes back to parameter adjustment but operates through more accomplished visualisation and interaction systems.

\section{Thesis structure}
\label{sec:structure}

The thesis is split into \ref*{chap:conclusion} chapters. In between the current introduction (chapter~1) and the conclusions (chapter~\ref{chap:conclusion}), the reader will find a background literature review followed by two chapters (\ref{chap:immersive-adjustment} and \ref{chap:immersive-design}) describing and discussing the developed prototypes.

Specifically, chapter \ref{chap:sota} presents and discusses the state of the art and the state of the practice in several domains that are heavily tied with our work. This comprises an introduction to the terminology in use for immersive technologies, a general overview on computer-aided (architectural) design, and a section on how to interact within three-dimensional environments. Aggregated together, these knowledge pieces form the basis on top of which we caved out an iterative research process.

Chapter \ref{chap:immersive-adjustment} and \ref{chap:immersive-design} describe the prototypes developed during each iteration cycle, grouped by interaction level i.e., how much control they offer to the user over their design. Chapter \ref{chap:immersive-adjustment} focuses on the two prototypes that cover parameter adjustment, while chapter \ref{chap:immersive-design} introduces a third prototype that enables control over the visual algorithm itself.
In both chapters, we present the building blocks the prototypes are made of, discuss how they were received by evaluators, and explain the rationale behind the decisions we made to move on from each prototype to a new one.

Finally, chapter \ref{chap:conclusion} summarises our contributions towards the thesis statement, reflects on the developed prototypes, their use cases and usage scenarios, and opens perspectives on how our research could be continued. We additionally provide recommendations and projections of where the field should head towards.

It should however be noted that this thesis structure does not exactly reflect the methodology or chronology of our research developments. It is rather organised based on the relation between the features and use cases that our iterations aim to cover.

\chapter{State of the art} 
\label{chap:sota}
\epigraph{
``I just need to figure out how things work.'' }{Ivan Sutherland}


Before diving into our work, section \ref{sec:sota-immersive} introduces the display and interaction technologies commonly used for creating immersive experiences, as well as the corresponding terminology.

Section \ref{sec:visual-modelling} presents three-dimensional visual modelling techniques that have been proposed for various application domains, so that we may draw inspiration from such techniques and apply them in the Algorithmic Design context.

Section \ref{sec:sota-cad} goes through the history of Computer-Aided Design and its architectural counterpart, before focusing on architectural design through programming and Algorithmic Design itself.

Then, Section \ref{sec:sota-3dui} presents an overview of three-dimensional interaction techniques that are adapted to immersive manipulations, so that we may choose the techniques that are most appropriate to the Algorithmic Design context. 

Finally, Section \ref{sec:new-ways-archi} addresses new ways of interacting with architectural designs, including optimisation techniques, exploratory approaches, and the use of immersive technologies, to help us better picture the landscape of existing integrations of these technologies and techniques in the architectural design field.

\pagebreak

\section{Immersive technologies}
\label{sec:sota-immersive}
Terms such as Augmented Reality and Virtual Reality are now familiar to a lot of people, thanks to well-known applications such as Pokémon Go and popular devices such as HTC Vive\footnote{\url{www.vive.com}} and Oculus Rift\footnote{\url{www.oculus.com/rift}}. 
In the last few years, these immersive technologies have been subject to an increasing attention in research, business and society in general \citep{Suhstateimmersivetechnology2018}.
This section clarifies the terminology and how immersive experiences relate to one another, then presents the underlying technologies and provides examples of use cases.

\subsection{Terminology}
\label{sec:immersive-terminology}
\begin{figure}[h]
	\includegraphics[width=\textwidth]{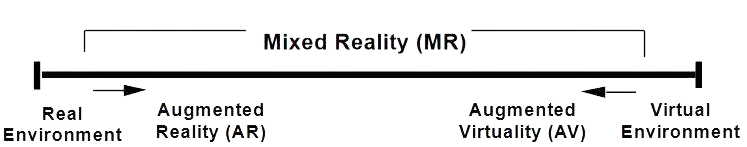}
	\centering
	\caption[The original \gls{rv} continuum]{\label{fig:rv-continuum} The original \gls{rv} continuum \citep{Milgramtaxonomymixedreality1994}}
\end{figure}

A very frequently used taxonomy is the \gls{rv} continuum \citep{Milgramtaxonomymixedreality1994} represented on Figure \ref{fig:rv-continuum}.
Experiences that mix real and virtual elements can be placed on that continuum, whose reach goes from the real world to an entirely virtual one.

\begin{figure}
	\centering
	\begin{subfigure}[t]{.45\textwidth}
		\centering
		\includegraphics[width=.7\linewidth]{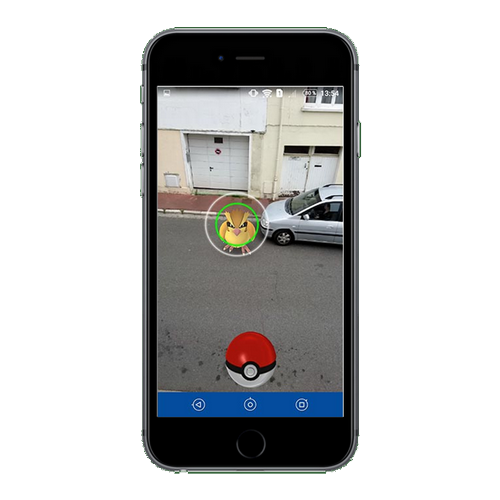}
		\caption{Superimposing a Pokémon on top of the camera feed (\gls{ar} feature).}
		\label{fig:ar-example}
	\end{subfigure}\hfill%
	\begin{subfigure}[t]{.5\textwidth}
		\centering
		\includegraphics[width=.9\linewidth]{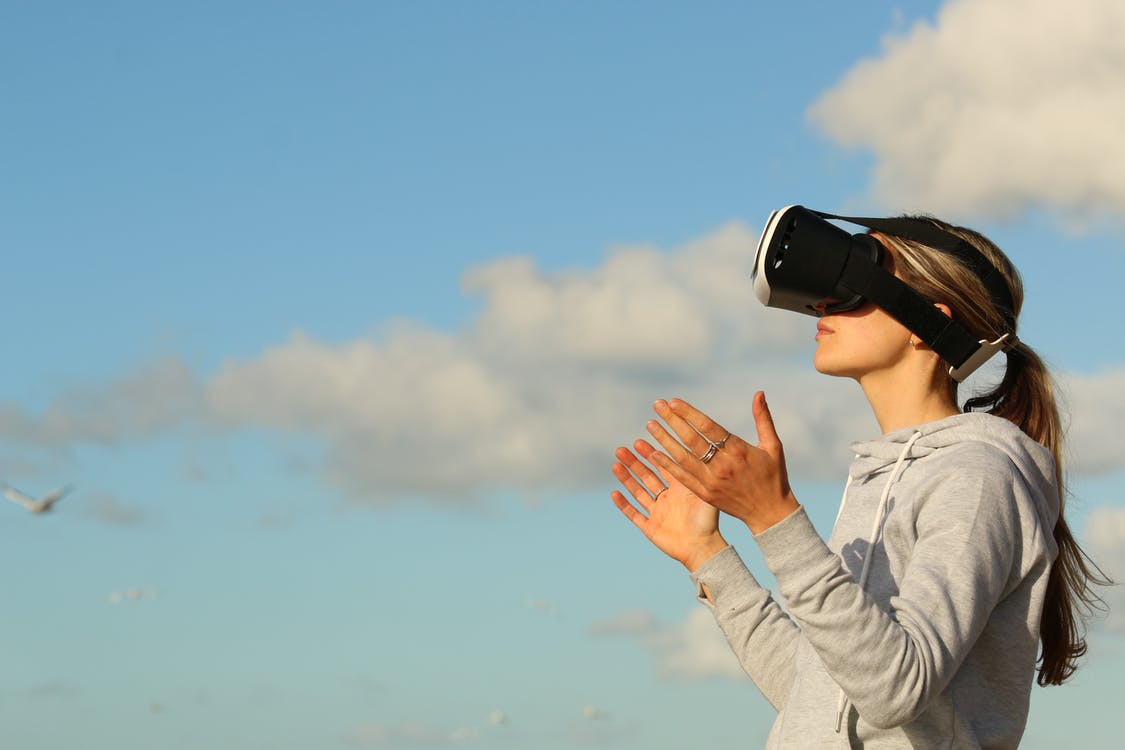}
		\caption{A person wearing a \gls{vr} headset, occluding that person's field of view.}
		\label{fig:vr-example}
	\end{subfigure}
	\caption[]{Example images\footnotemark on the two most commonly mentioned immersive technologies: \gls{ar} and \gls{vr}.}
	\label{fig:ar-vr-examples}
\end{figure}

This includes \gls{ar}, that describes the superimposition of virtual elements onto the real world. It can basically be seen as ``adding virtual things on top of the real world’s perception''. A typical example of an \gls{ar} device is Google Glass\footnote{\url{https://google.com/glass/start/}}. The game Pokémon Go\footnote{\url{https://pokemongolive.com/}} helped popularise \gls{ar} with its integration of Pokémon creatures on top of the live camera feed as if these creatures were there in the real world (as shown on Figure \ref{fig:ar-example}).

On the other end of that continuum, we find \gls{vr}, that fully immerses users into a three-dimensional virtual world. While the paternity of the concept is unclear, \gls{vr} as a term is generally attributed to \citeauthor{lanier1988vintage} who worked actively \citep{lanier1988vintage, conn1989virtual} in the domain in the late 1980’s. The usual equipment used for \gls{vr} experiences involves a visualisation system, typically either a \mbox{\gls{hmd}} or multiple wall-sized displays, as well as some interaction device, such as a controller, a glove or a tracking system.

On the \gls{rv} continuum, two other terms are mentioned: Augmented Virtuality (AV) and \gls{mr}.
AV can be seen as the ``opposite'' of \gls{ar}, since it describes experiences that augment the virtual world with real world elements. We will not talk much about it, not only because it is much less commonly encountered, but also because its frontier with \gls{ar} and \gls{vr} can be blurry, especially with the ever-improving technologies that will make virtual elements harder to distinguish from real ones.

\begin{figure}[h]
	\includegraphics[width=.8\textwidth]{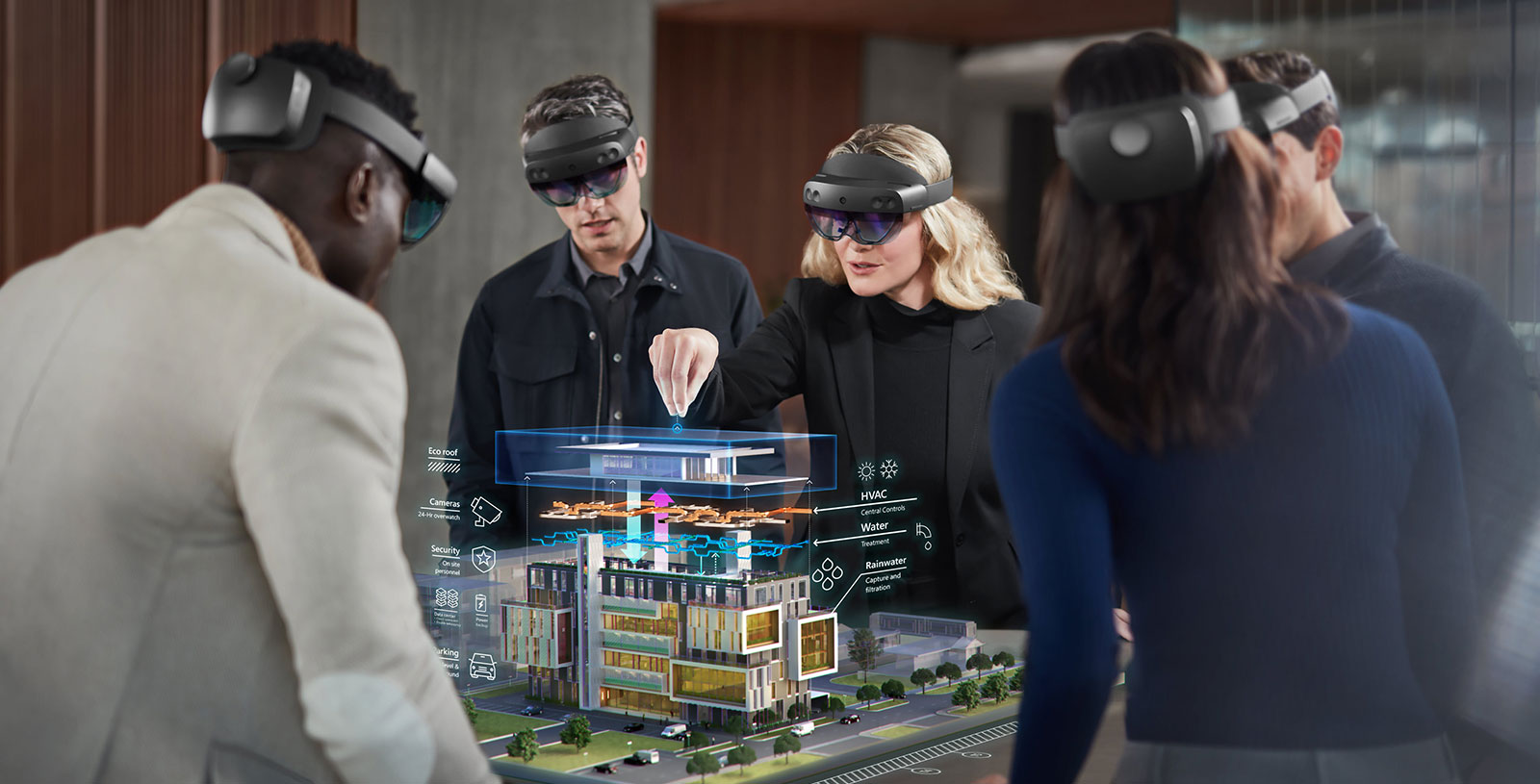}
	\centering
	\caption[]{Commercial visual of an \gls{ar} experience with proper spatial mapping of virtual elements, that would be classified as \gls{mr} by some\footnotemark.}
	\label{fig:mr-example-wrong}
\end{figure}

The original definition of \gls{mr} encompasses everything that falls in between the extrema of the aforementioned \gls{rv} continuum: a \gls{mr} environment is ``one in which real world and virtual world objects are presented together within a single display'' \citep{Milgramtaxonomymixedreality1994}. 
However, the term \gls{mr} is often used nowadays to describe \gls{ar} devices and experiences that show an advanced degree of spatial understanding, generally thanks to an environment scanning system. Figure \ref{fig:mr-example-wrong} depicts such an experience, with a virtual building properly anchored to a real table. This would be described by some as a \gls{mr} experience. While that statement is not fundamentally false (since \gls{ar} is a subset of \gls{mr} according to the original definition), it leads to confusion and may prompt unfamiliar individuals to reject the classification of that experience as \gls{ar}.

\footnotetext{The figure was created using images from \href{https://www.phonandroid.com/pokemon-go-guide-astuces-progresser-facilement.html}{www.phonandroid.com/pokemon-go-guide-astuces-progresser-facilement.html} and \href{https://www.pexels.com}{www.pexels.com}, respectively.} 

We believe the communication and the media coverage around the Microsoft Hololens (an \gls{ar} headset capable of scanning its surroundings) played a big role in that deviation from the original definition, since it is mostly referred to as a \gls{mr} device. 
The misuse of that term, with regards to its original meaning, became so common that a new term was coined to describe exactly what \gls{mr} initially was: \gls{xr}.

\footnotetext{Picture from Microsoft's Mixed Reality documentation portal, found on\\\url{https://docs.microsoft.com/en-us/windows/mixed-reality/design/color-light-and-materials}.}

However, we admit that the ability to properly understand the surroundings as part of an \gls{ar} experience enables a whole new range of applications. We therefore suggest to introduce a new term as a subset of \gls{ar}, to recognise that particularity without borrowing \citeauthor{Milgramtaxonomymixedreality1994}'s original \gls{mr} term. Three options we suggested in \citep{CoppensMergingrealvirtual2017} were ``Spatial-Aware \gls{ar}'', ``Surroundings-Aware \gls{ar}'' and ``Spatial-mapped \gls{ar}''. 

Many other taxonomies have been proposed; there is even another one suggested in the paper that also describes the \gls{rv} continuum \citep{Milgramtaxonomymixedreality1994}: a three-dimensional hyperspace that places \gls{mr} systems according to 3 axes. These 3 axes are: Extent of World Knowledge (EWK: how much we know about the - real or virtual - world in which the experience happens), Reproduction Fidelity (RF: how realistic the augmented content is) and Extent of Presence Metaphor (EPM: how immersed the user is). Other examples of classifications are based on location and temporality \citep{fuchs2006traite}, on intended purpose \citep{dubois2000combinons}, on the type of entity that is being augmented \citep{hugues2011new,mackay1996augmenting}, or on who is in control of the experience (either the system or the user) \citep{renevier:tel-00008264}.

\subsection{Technological building blocks for immersive experiences}
In order to create such experiences, there are a few technological needs, mainly in terms of display and tracking devices. This section will cover the most common techniques currently in use.

\subsubsection{Displays for VR}

Since their purpose is to immerse the user into an entirely synthetic world, VR-enabling displays have to occlude a large part of the user's field of view. For that reason, the typical representation of a \gls{vr} user as of today is a person wearing a \gls{hmd}, as pictured on Figure \ref{fig:vr-example}. These \glspl{hmd} are generally binocular, meaning that each eye can be presented with a (slightly) different image. This results in the ability to produce stereoscopic imaging, that creates a three-dimensional effect, fooling the wearer's perception into feeling immersed in a three-dimensional virtual space.

\begin{figure}[h]
	\includegraphics[width=.6\textwidth]{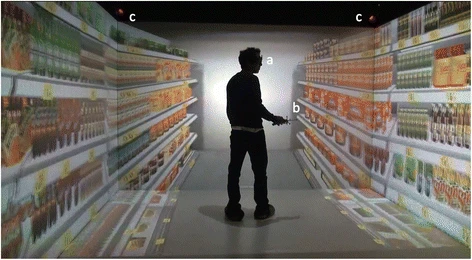}
	\centering
	\caption{A CAVE-like setup used to create a virtual grocery store. Reproduced from \citep{borrego2016feasibility}.}
	\label{fig:cave}
\end{figure}

Although more expensive, wall displays can also be used to create \gls{vr} experiences. These projection-based systems are usually referred to as CAVE-like setups, since the first occurrence of such a system was the Cave Automatic Virtual Environment (CAVE) \citep{Cruz-NeiraCAVEAudioVisual1992}, which dates back to the early 1990's. A more recent example of a CAVE-like arrangement is shown in Figure \ref{fig:cave}, where we can see the projections on the walls of a cubic room as well as on the floor.
These types of setups provide a better sense of presence \citep{juan2009comparison} but are typically a lot more costly. Furthermore, considering we work in the context of architectural design, it should be noted that the spatial understanding and distance perception is not necessarily better than those experienced with \glspl{hmd} \citep{ghinea2018perception}.

\subsubsection{Displays for AR}

From the user's point of view, the simplest form of \gls{ar} displays are monitor-based ones, where the augmentation happens on a distant screen such as a television or a mobile device (a more precise term is then available: hand-held \gls{ar}). The display is treated as a window to the augmented world, hence the alternative name ``Window-on-the-World'' from \citep{milgram1995augmented}. Examples of experiences relying on such displays are analysis tools for sport broadcasting (with virtual elements such as names and arrows being superimposed to the camera feed), or the previously mentioned \gls{ar} feature in Pokémon Go (Figure \ref{fig:ar-example}).

Another non-intrusive alternative is Spatial \gls{ar}, that is sometimes referred to as projective \gls{ar} since it is about augmenting reality by projecting images directly onto real objects. It was first introduced in \citep{raskar1998office} and has the advantage that it naturally provides multi-user experiences. Combined with the non-intrusive aspect, this technology becomes a good candidate for cultural contexts e.g., for museum exhibitions and monumental projections.

While it is possible to create an \gls{ar} experience by using a \gls{vr} headset mounted with a camera (to project the camera's input to the displays), it typically creates latency, since there is a delay between the recording of an image and the moment it is being displayed to the wearer. These displays are classified as video see-through but the display delay can create cybersickness (a term used to cover various symptoms generally including nausea that resemble motion sickness). In fact, cybersickness is believed to be heavily tied \cite[p.~49]{LaViola3duserinterfaces2017} with mismatches between visual and vestibular (body balance and movement) information, and solutions that limit such mismatches are usually preferred to mitigate the issue.

\begin{figure}[h]
	\includegraphics[width=.8\textwidth]{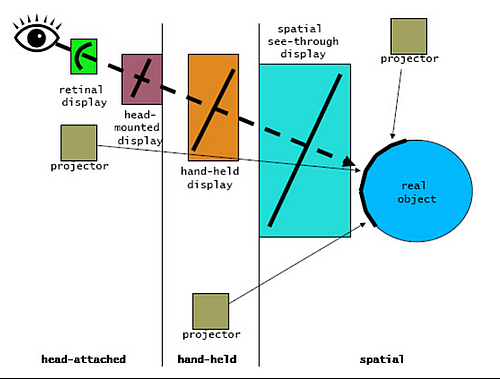}
	\centering
	\caption{Main options to produce \gls{ar} experiences. Reproduced from \citep{bimber2006modern}.}
	\label{fig:ar-displays}
\end{figure}

Another option is to use optical see-through displays, that present the user with a direct view of the real world, in the sense that the real world is not occluded, obviating the need to project the world on the display. In that case, the augmentation happens on a transparent surface placed in front of the observer, and virtual elements are integrated into the observer's view of the real world. There are multiple ways to place the augmentation surface relative to the observer. Figure \ref{fig:ar-displays} helps in picturing them, while also serving as a visual summary of the terms introduced in the present section.

An additional term is introduced in Figure \ref{fig:ar-displays}: retinal display. Sometimes named Retinal Projection Display (RPD) or Virtual Retinal Display, such a device directly projects \gls{ar} content on the user's retina, using low-power laser beams. The technology has not matured yet but several prototypes are available \citep{schowengerdt2003binocular, takahashi2008stereoscopic}. Theoretically, when the technology develops into a small form-factor, wide field of view, high fidelity display, it should supersede most other \gls{ar} technologies and could even lead to devices capable of switching back and forth between \gls{ar} and \gls{vr} modes.

It should also be noted that we only talked here about visual \gls{ar} as it is the most common type of \gls{ar}. Other kinds of augmentations exist, targeting other senses: smell (olfactory \gls{ar}), touch (haptic \gls{ar}), taste (gustatory \gls{ar}) and hearing (audio \gls{ar}). These technologies fall out of scope of the current dissertation since we did not rely on them for the experiments we conducted.

\subsubsection{Tracking needs}

Both \gls{ar} and \gls{vr} experiences typically require some form of motion and positional tracking. Common techniques to enable suck tracking are hereby presented.

\begin{figure}[h]
	\includegraphics[width=.4\textwidth]{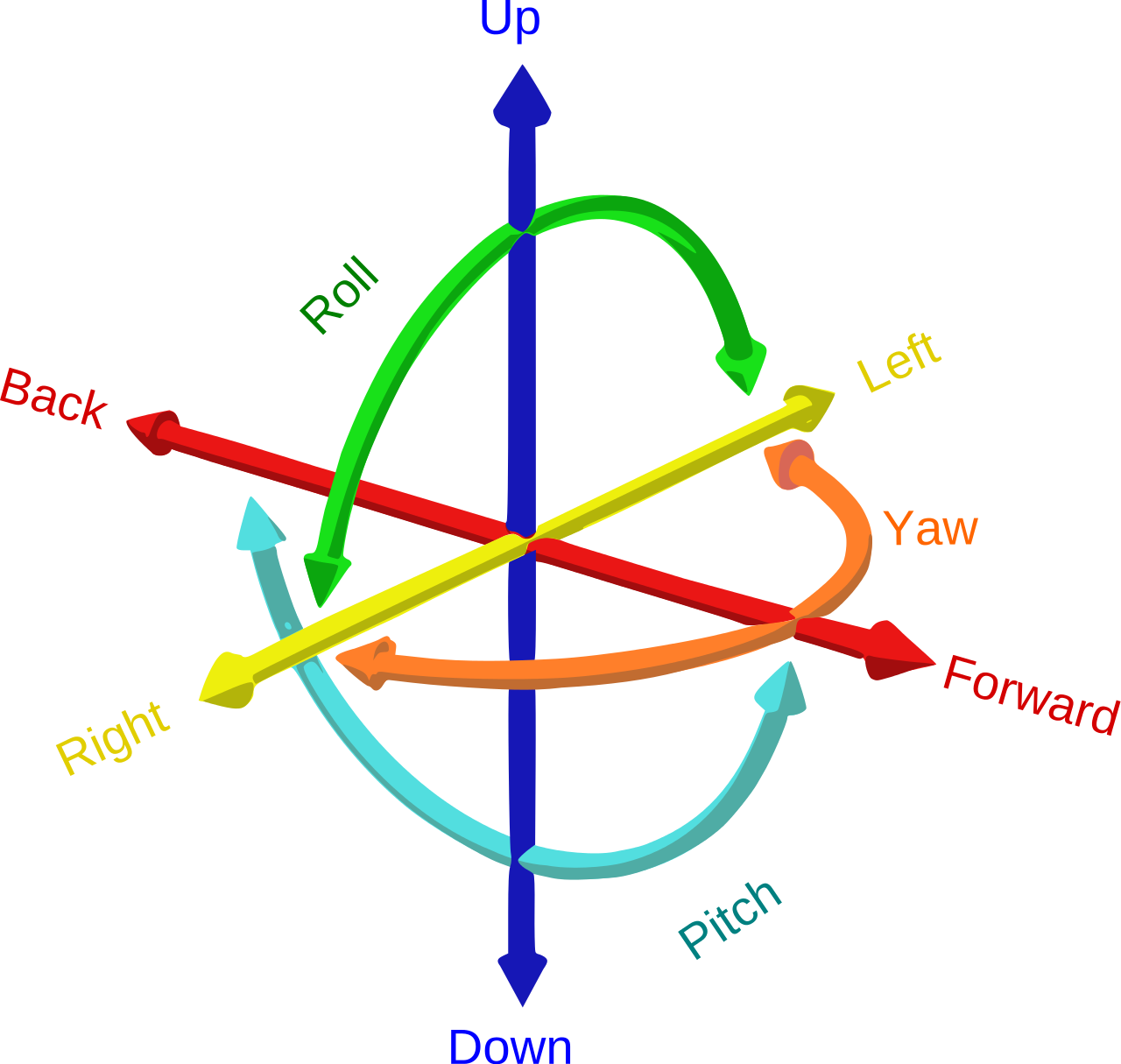}
	\centering
	\caption[]{The six degrees of freedom we have access to in 3D space. Reproduced from Wikimedia Commons\footnotemark.}
	\label{fig:6dof}
\end{figure}

The bare minimum to enable HMD-based experiences is rotational tracking for the wearer's head. High quality headsets (such as the HTC Vive or the Oculus Rift) also include positional tracking capabilities. This allows the wearer's head to be fully-tracked in 3D space i.e., we know its location along the three (x,y,z) positional axes as well as its orientation along the three usual (yaw, pitch, roll) rotational axes.
Since this gives six separate and independent ways to the head to modify its situation, a device with these tracking capabilities is usually labelled as a \gls{6dof} tracking system.
Figure \ref{fig:6dof} clarifies the concept, with each color representing one degree of freedom.

The ideal tracking system should be accurate, precise (no jitter), fast (low latency and high refresh rate), robust (immune to environmental factors), and provide great mobility (lightweight, wireless or even autonomous, small form factor) in a wide area. This is obviously ambitious, but the next section describes major tracking technologies that attempt to reach that goal.

\subsubsection{Tracking technologies}
\label{sec:sota-tracking}

\footnotetext{\url{https://commons.wikimedia.org/wiki/File:6DOF.svg}} 

\begin{figure}[h]
	\includegraphics[width=.4\textwidth]{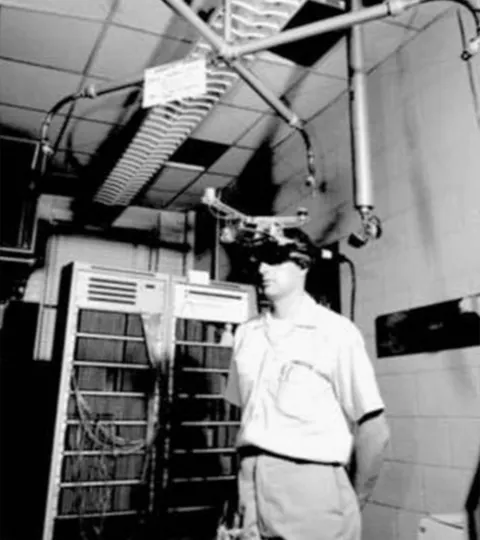}
	\centering
	\caption{The sword of Damocles. A mechanical tracking system designed for an early \gls{hmd}. Reproduced from \citep{sutherland1968head}.}
	\label{fig:damocles}
\end{figure}

The simplest tracking method is likely mechanical tracking, where the tracked object is directly attached to the tracking system so that an object's position and rotation can be determined from sensors placed on the joints of the tracking system. This technique can lead to bulky and cumbersome systems with mechanical arms made up of articulated pieces, such as the Sword of Damocles, pictured in Figure \ref{fig:damocles}.

Another tracking option that is (one of) the most popular on the market is inertial tracking, driven by accelerometers, gyroscopes and, sometimes, magnetometers.

An accelerometer measures an object's acceleration along one axis, taking gravity into account, whereas gyroscopes measure rotation around a single axis. Combining three sensors of each type therefore theoretically suffices to obtain a \gls{6dof} tracker that provides relative position and rotation data, reporting how much the position and rotation has changed since the tracking process started. Magnetometers can additionally provide a reference heading to stabilise other sensors' measurements, since they measure magnetic fields and can indicate their direction (e.g., the Earth's magnetic field's direction for a compass). The combination of these types of sensors in a tracking system is often referred to as \gls{imu} since they are based on the principle of inertia ($F = ma = m\frac{dv}{dt}$), used to derive a relative position and rotation.

The inertial tracking solution is widely used since the necessary sensors are relatively cheap and small, and are often already integrated in most smartphones.
In practice though, an \gls{imu} has to be coupled to another tracking technology to provide positional information, since the drifting of accelerometers very quickly leads to unusable data. This is due to the relative nature of the measurements that necessarily accumulate errors, coupled with the fact that an accelerometer measures an acceleration that is then used to calculate a movement, meaning that the accumulated error on the acceleration is reported quadratically on the relative position output. The inability of smartphone \glspl{imu} to provide reliable positional tracking on their own is the reason smartphone-based \gls{vr} headsets can only produce limited experience where real world movement cannot be taken into account (the virtual point of view therefore is fixed or controlled externally) and it can also cause additional cybersickness compared to higher-end systems, since a user's head generally slightly moves when rotating.

Another popular technique is optical tracking, where some kind of sensor (e.g., a camera) tracks known patterns, features, or markers in the surrounding environment. The sensor can be external to the tracked object (outside-in tracking) or attached to it (inside-out tracking). 

\begin{figure}[h]
	\includegraphics[width=.6\textwidth]{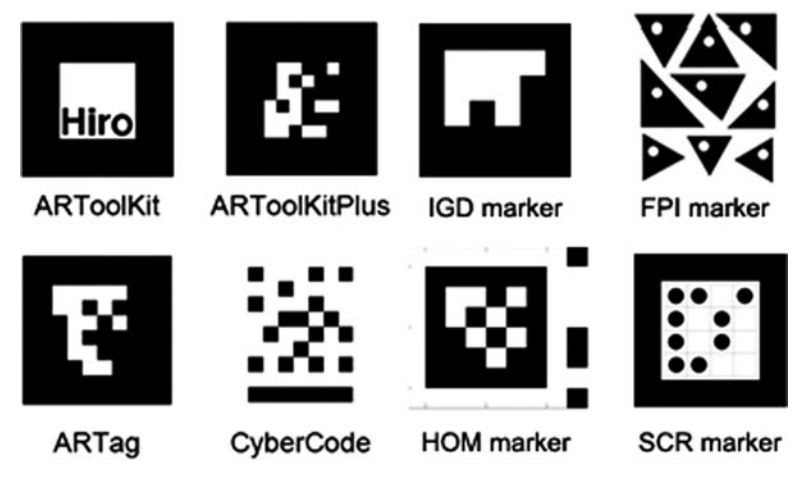}
	\centering
	\caption{Some well-known \gls{ar} marker types, sharing similarities with QR codes. Reproduced from \citep{kan2011qr}.}
	\label{fig:paper-markers}
\end{figure}

Marker-based systems rely on markers placed on the tracked object. These markers can be active (light-emitting sources) or passive (easily identifiable items that do not emit light themselves) and are typically not visible to the human eye (e.g., using infrared lights), except when relying on paper markers, such as the ones presented on Figure \ref{fig:paper-markers}.
The latter option provides a cost-efficient way to produce \gls{ar} experiences with printed papers and a simple camera. A very common marker-based library amongst \gls{ar} developers is ARToolKit, that was initially developed as part of an \gls{ar}-based conferencing system \citep{kato1999marker}.

Pattern-based solutions (such as Microsoft's Kinect sensor\footnote{\url{https://developer.microsoft.com/windows/kinect}}) project a known pattern on the environment and observe the distortion of that pattern to derive spatial data, while feature-based solutions identify (and track) features to construct a 3D representation of the surroundings that ultimately allows the system to build a 3D plan.

That last option is strongly related to the simultaneous localisation and mapping (SLAM) problem \citep{durrant2006simultaneous}, that describes the mapping of an unknown environment by a mobile robot. Since it has no previous knowledge of the surroundings, the robot must solve two related and interdependent problems at once: localising itself and mapping the environment. The problem is typically solved using feature descriptors and feature-tracking algorithms to derive the position of landmarks in the environment, combined with a variety of mapping methods. Recently, neural networks have been extensively used to solve SLAM instances \citep{chaplot2020learning, zhang2017neural}.
Relying on SLAM-based techniques can allow for good tracking quality while suppressing the need for markers, but is harder to develop than most tracking solutions and is typically sensitive to environmental changes (e.g., moving objects). Some \gls{vr} headsets such as the Oculus Quest\footnote{\url{www.oculus.com/quest}} rely on SLAM techniques to provide tracking capabilities.
A more complete description of the SLAM problem and the associated state of the art is out of scope of the current thesis, but it remains an important component of spatial-aware \gls{ar} \glspl{hmd}.

\begin{figure}[h]
	\includegraphics[width=.6\textwidth]{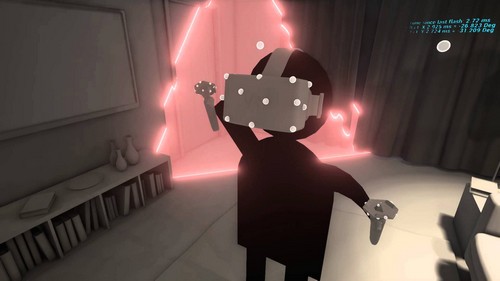}
	\centering
	\caption[]{Valve's Lighthouse tracking system, using optical (infrared) tracking and active markers. Reproduced from a YouTube video\footnotemark that explains how the system works.}
	\label{fig:lighthouse}
\end{figure}

\footnotetext{\url{https://www.youtube.com/watch?v=J54dotTt7k0}} 

Every tracking technique has its drawbacks and combining the strengths of different technologies produces the best results \citep{welch2002motion}. For that reason, lots of trackers rely on hybrid solutions. A relevant example would be Lighthouse, Valve's tracking system for the HTC Vive, that relies on an \gls{imu} for rapid updates while the more accurate position is obtained from an optical tracking system (using infrared and active markers) to limit the \gls{imu}'s drifting issue to the few \gls{imu} updates between the optical system's updates. Figure \ref{fig:lighthouse} pictures a red laser beam about to hit the \gls{hmd} to provide a one-dimensional update (the system uses two base stations that each emit two beams per update).

Note that the technologies described in this section can be used for head tracking but also for interaction devices e.g., controllers or gloves, and even to provide hand or body tracking.

\subsection{Application domains} 

In general, \gls{ar} and \gls{vr} are most useful when they can provide a virtual equivalent to applications where a real implementation would be too difficult (or even impossible), too costly, or too dangerous to realise. This section covers such scenarios along with other \gls{ar} or VR-based application domains, ignoring architectural design for now since that specific field will be discussed in further details in Section \ref{sec:sota-cad-immersion}.

\subsubsection{Entertainment}
The entertainment sector is a major player in the popularisation of immersive technologies, with a plethora of movies and games being created.
In addition to the previously mentioned smartphone game Pokémon Go, other popular titles have been extended to \gls{ar}, with games such as ARQuake \citep{thomas2002first} that lets students shoot virtual monsters in their university campus. Some of these \gls{ar} games are based on real sports, like tennis \citep{henrysson2005face}, where players facing each other control virtual racquets through their phones. Many \gls{vr} games have been developed in the last few years, some of them by researchers themselves, e.g., a geocaching game \citep{brade2017being} to compare presence and usability in \gls{vr} with the real world equivalent, but most studies rely on commercial games with \gls{vr} support, such as Team Fortress 2 \citep{martel2015diving} or Half-Life 2 \citep{tan2015exploring}.

Outside of video games, \gls{ar} has been used to enhance book experiences \citep{billinghurst2001magicbook} or card and board games \citep{lam2006art, lee2005tarboard, molla2010augmented}.
Researchers have also investigated \gls{vr} movies and the consequences the medium has for movie makers \citep{serrano2017movie}, or how it impacts viewer engagement \citep{gruenewald2020feeling}.

\begin{figure}[h]
	\includegraphics[width=.95\textwidth]{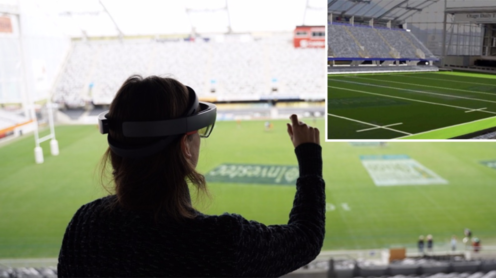}
	\centering
	\caption{\label{fig:ar-sport} An \gls{ar} feature to superimpose virtual line marks on a rugby field, through a Hololens headset. Reproduced from \citep{zollmann2019arspectator}.}
\end{figure}

The sport industry also makes use of \gls{ar}, especially for broadcasting, where overlays are often added on top of camera images to show additional information to the viewer. Such augmentations are common to point out specific athletes, what they did or should have done, or to draw virtual lines in order to indicate distances. Figure \ref{fig:ar-sport} pictures a similar experience but for a spectator that is actually inside the stadium.

That being said, both \gls{ar} and \gls{vr} raise interest in a wide range other domains, some of which are covered in the remaining of this section.

\subsubsection{Healthcare}
Healthcare is a very large field on its own, with many opportunities for immersive technologies to shine. Thanks to the sense of presence induced by \gls{vr} and the actual presence inherent to see-through \gls{ar}, many psychiatric treatments relying on these technologies have been proposed. A plethora of anxiety and phobias are covered by such treatments, since they can expose the patient to their fears in a controlled environment, meaning the practitioner can choose the degree of exposure and even abort the experience at any time. As for VR-based treatments, a term has even been coined to encompass such therapies: Virtual Reality Exposure Therapy (VRET). VRETs have been proposed for flying phobia \citep{botella2004treatment}, fear of heights \citep{krijn2004treatment}, animal phobias \citep{carlin1997virtual} (also in \gls{ar} \citep{botella2005mixing, juan2005using}) and Post-Traumatic Stress Disorders \citep{rizzo2009virtual}.

Other healthcare applications include the assessment and rehabilitation of disabilities, following brain injuries \citep{rose2005virtual}, strokes (with both \gls{vr} \citep{jack2001virtual} and \gls{ar} \citep{mousavi2013spatial}), or amputations (with phantom limbs \gls{ar} treatment \citep{carrino2014augmented})

These technologies can additionally find usages during interventions, with \gls{vr} being used for pain distraction (e.g., during heavy interventions), while \gls{ar} can help surgeons with augmented overlays \citep{fuchs1998augmented, sato1998image}.

\subsubsection{Education and training}
\label{sec:education-training}
Both \gls{ar} and \gls{vr} have been used in educative contexts, e.g., for surgical education \citep{basdogan2007vr}, as part of anatomy courses (with \gls{vr} \citep{nicholson2006can} and \gls{ar} \citep{blum2012mirracle}), or to teach astronomy \citep{fleck2013augmented}. These examples back up our claim that these technologies show their potential when a physical counterpart to the virtual experience would be harder or impossible to implement.

Thanks to the ability of immersive environments to replicate the real world, potentially in a realistic way and including sensible physics simulations, immersive technologies provide opportunities for training applications. Examples include firefighters \citep{xu2014virtual} and astronauts \citep{aoki2007virtual} training in \gls{vr}, as well as individuals learning assembly tasks \citep{reiners1999augmented} or military operations \citep{brown2006augmented}. These types of experiences have proven to be successful in allowing their users to transfer virtually acquired skills to the real-world counterpart of the target activity. As an example, researchers have demonstrated \citep{michalski2019getting} that real table tennis skills can be improved through \gls{vr} training.

\subsubsection{Culture and tourism}
Many touristic locations offer binoculars to visitors, so that they can better observe the surroundings in exchange for a bit of money. Sometimes, these devices are augmented with information on or pointers to specific points of interest, thereby creating an \gls{ar} experience \citep{fritz2005enhancing}.

More advanced usages of immersive technologies in a similar context are also common, with cultural heritage experiences allowing users to visualise monuments \citep{gaitatzes2001reviving} or inhabitants \citep{noh2009review, vlahakis2001archeoguide} that have since disappeared. Similar experiences have also been created for places that still exist but cannot (easily) be visited, while other are simply made for advertising purposes \citep{kim2019hedonic, loureiro202020}.

\subsubsection{Industrial maintenance and complex tasks}

As discussed in section \ref{sec:education-training}, both \gls{ar} and \gls{vr} are used to train workers, including for maintenance tasks \citep{gavish2015evaluating}, but \gls{ar} can also act as a virtual assistant when actually performing these tasks. Specific pieces that the worker has to manipulate can be superimposed with information and related 3D models can even be displayed \citep{schwald2003augmented} to help with the task at hand. That principle has been followed to support various applications such as welding \citep{echtler2004intelligent} and pump maintenance operations \citep{garza2013augmented}.

The augmented information does not necessarily have to be set in stone, as collaborative solutions also have been developed, with remote engineers able to place indicators when needed \citep{bottecchia2010tac, benbelkacem2011augmented}. AR-based solutions could therefore, in some instances, replace lengthy manuals with dynamic, world-anchored and collaborative digital equivalents.


\section{3D visual modelling}
\label{sec:visual-modelling}


Architectural design is not the only field employing visual modelling, as similar approaches have been applied to many domains, and three-dimensional versions of some of the corresponding tools were developed.

\begin{figure}[h]
	\includegraphics[width=.95\textwidth]{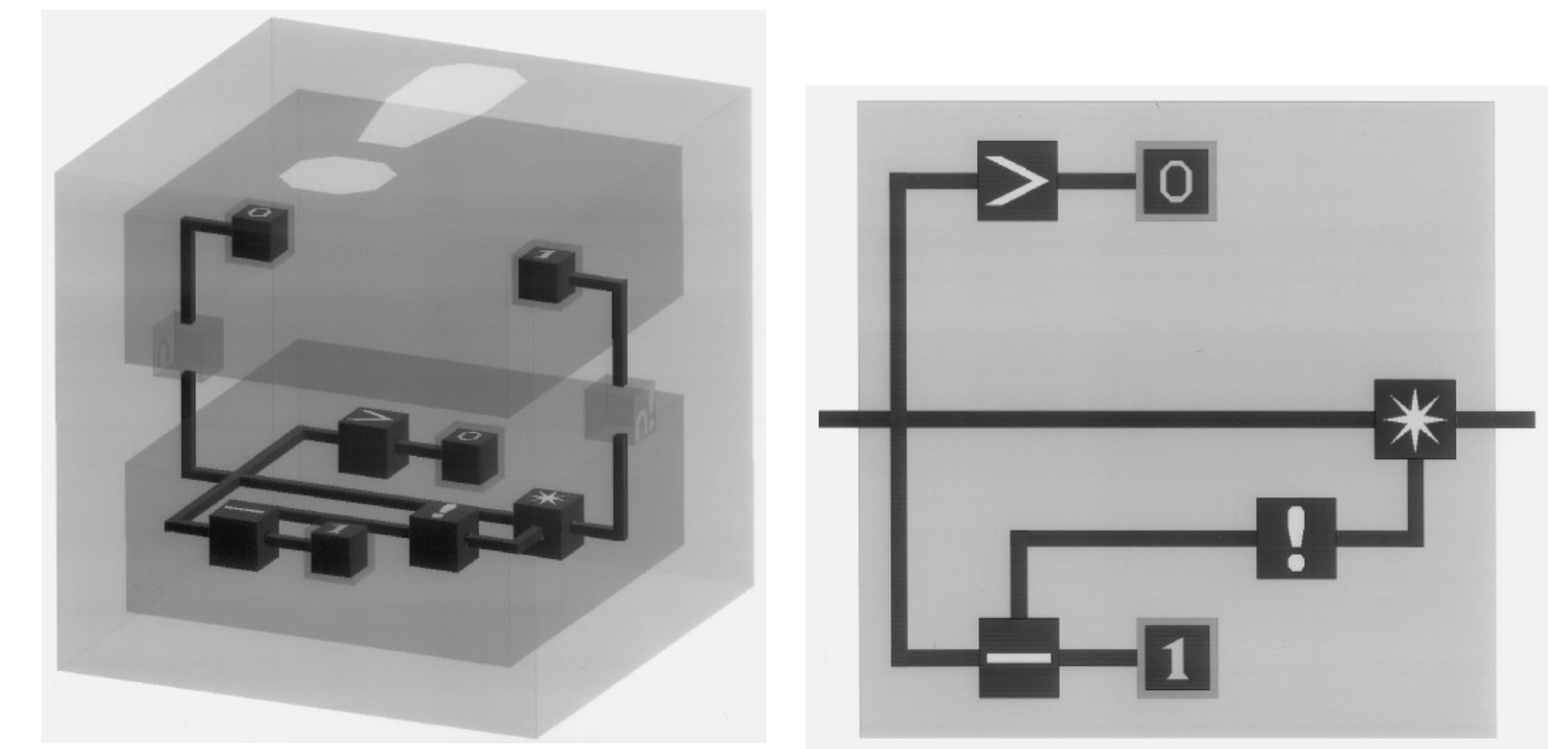}
	\centering
	\caption{\label{fig:cube} The CUBE programming language: a recursive definition named ``$!$'', that calculates the factorial of a given number, with a planar projection of the lower plane on the right. Reproduced from \citep{najork1996programming}.}
\end{figure}

In fact, numerous 3D programming languages for virtual 3D environments have been designed, most of which rely on three-dimensional dataflow diagrams to define programs. Examples from the early 90’s include CUBE/CUBE-II \citep{najork1991cube, najork1996programming}, a functional language based on a dataflow metaphor, and Lingua Graphica \citep{stiles1992lingua}, that translates from/to C++ code. Figure \ref{fig:cube} presents a CUBE definition that returns the factorial of a given number. The upper plane is only executed when the input value $n$ is equal to 0, and outputs 1 in that case. The lower plane is only executed when the input value $n$ is greater than 0 and outputs the product of the input value with the output of the ``$!$'' function for the value $n-1$. This indeed returns $n!$ since $n! = n*(n-1)!$ for $n>0$. We can additionally note that if a negative value were to be provided as input, nothing would be returned since none of the planes' conditions would be satisfied.

The previously mentioned languages target general programming, but the main motivation to create 3D languages and editors is often to match the dimensionality of the program with its output, so as to integrate both of them in a single virtual environment.
As an example, SAM (Solid Agents in Motion) \citep{geiger1998sam} is another early 3D visual language that enables ``parallel systems specification and animation'', targeting animated 3D presentations.

\subsection{Immersive authoring of visual models}
\label{sec:immersive-modelling}

While the previous examples were indeed designed with virtual environments in mind since the authors all mention such environments in their respective papers, they were never adapted to immersive displays (e.g., \gls{vr}), despite the alledged benefit of immersing the programmer in the same environment as the program and its output, an approach sometimes referred to as embodied spatial programming.

On the other hand, \citeauthor{steed1996dataflow} implemented an immersive system that allows users to define object behaviours whilst being immersed \citep{steed1996dataflow}, once again through dataflow graphs. The system could be used to design animations or interactive applications, that conveniently also took place within the virtual environment. With similar goals in mind, \citep{lee2004immersive} presented an \gls{ar} system to define the behaviour of scene objects for \gls{ar} applications.
A more recent example of a more accomplished \gls{vr} authoring system is FlowMatic \citep{zhang2020flowmatic}, that not only lets users create, destroy, or define basic animations for 3D models but also allow them to define an object's behaviour depending on discrete events such as timers or collisions.

\begin{figure}[h]
	\includegraphics[width=.9\textwidth]{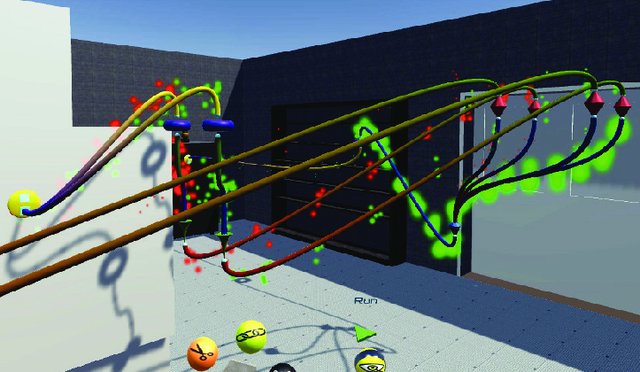}
	\centering
	\caption{\label{fig:ivy} The Ivy programming tool that helps with designing sensor-driven systems. The depicted environment places the user in an industrial fabrication workshop, and the displayed program cuts the workshop's power if excessive vibrations are detected on a machine. Reproduced from \citep{EnsIvyExploringspatially2017}.}
\end{figure}

Another application domain is the Internet of Things (IoT), with prototypes such as Ivy \citep{EnsIvyExploringspatially2017}, a VR-based programming tool that allows its users to define the behaviour of IoT systems that depend on sensor data, through yet another visual dataflow-based representation. Figure \ref{fig:ivy} shows the tool's interface, with coloured particles that represent data flowing through the links.
Aiming towards a similar goal but using \gls{ar}, the Reality Editor \citep{heun2013reality} allows (re)programming of smart objects and their relations to others, so as to define a system's behaviour. A more recent prototype is CAPturAR \citep{wang2020capturar}, that serves a similar purpose but offers an activity-recording feature using a body-tracking system, to help in defining scenarios.
 
\section{Computer-Aided Design}
\label{sec:sota-cad}

\subsection{Evolution of Computer-Aided Design} 
\label{sec:sota-cad-design}

\begin{figure}[h]
	\includegraphics[width=.5\textwidth]{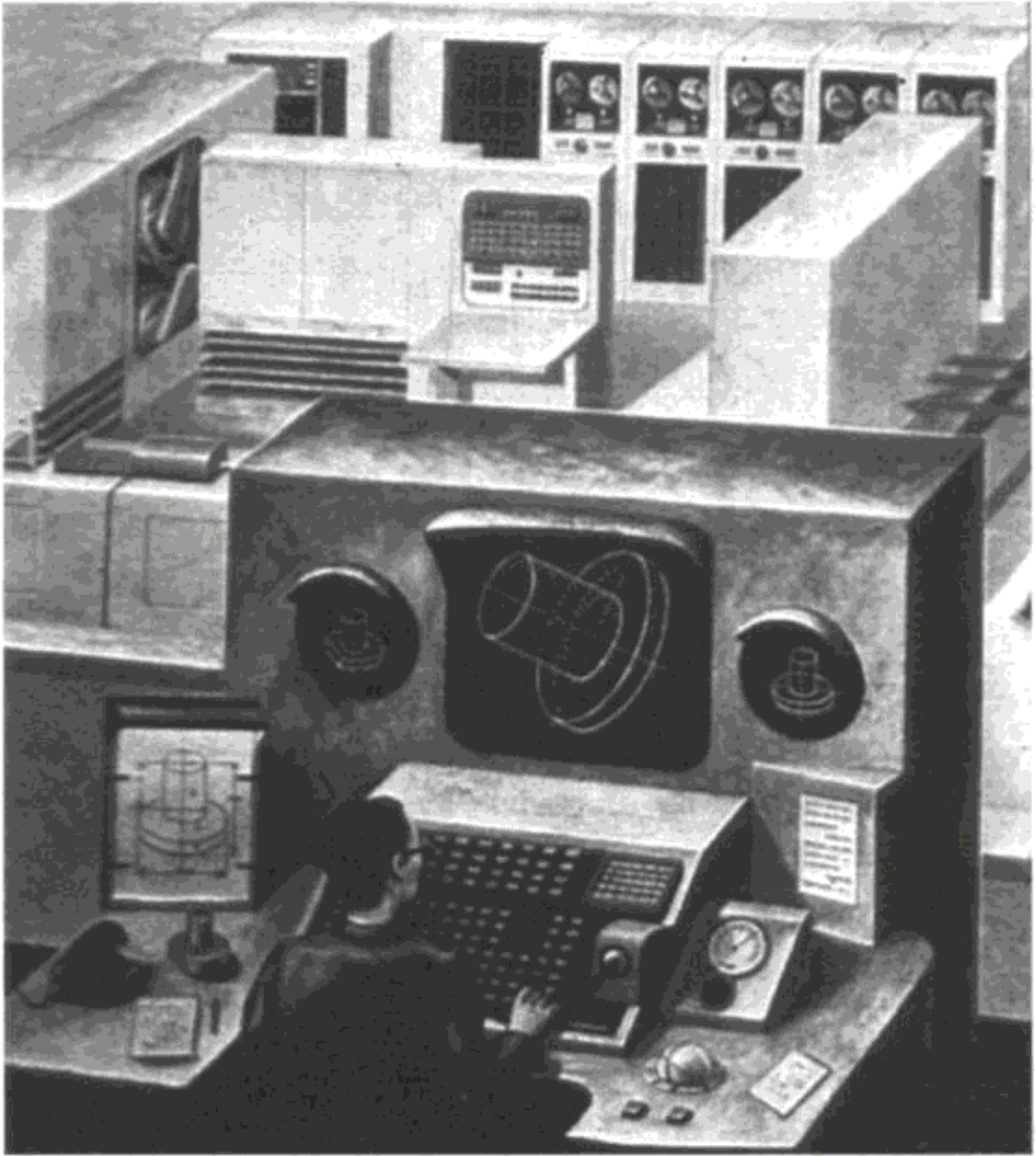}
	\centering
	\caption[Artistic impression of a design workstation, from the Fortune magazine in November 1956 (reproduced from \citep{mitchell1990afterword}) ]{\label{fig:artist-concept-fortune} Artistic impression of a design workstation, from the Fortune magazine in November 1956. Reproduced from \citep{mitchell1990afterword}.}
\end{figure}

The starting point for \gls{cad} likely dates back to the 1950's. During the first half of that decade, computers started to be produced for commercial purposes and people naturally began to imagine and reflect on how they could be used for design activities. Those reflections lead to conceptual developments in the following years; a good example being the artistic impression of a design workstation, depicted in Figure \ref{fig:artist-concept-fortune} and published in the Fortune magazine in 1956.

Academics also picked up interest in the potential of using computers for design activities, with a notable milestone in 1959 when a meeting took place at MIT \citep{coons1963outline} between the ``Computer Applications Group'' and the ``Mechanical Engineering Department''. That meeting concluded that computers had a significant role to play for (engineering) design and notably lead, a year later, to a report \citep{ross1960computer} whose author is often credited with coining the Computer-Aided Design term; a term that was used in the report's title.

Ever since these early concepts, \gls{cad} tools were indeed developed and have matured over the years: they reached new domains, expanded their functionality and their adoption progressed dramatically.
In \citep{horvath2015ubiquitous}, the authors identified 5 periods of evolution for \gls{cad}; the remaining of this section is inspired by that paper's structure, although it should be noted that such a periodisation (dividing history into periods) always involves some degree of subjectivity and arbitrariness.

\subsubsection{Early research developments (1960's)}
\begin{figure}[h]
	\includegraphics[width=.6\textwidth]{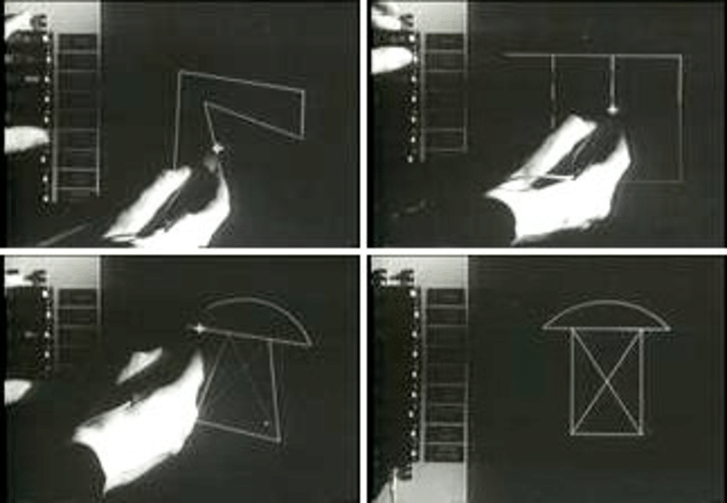}
	\centering
	\caption[Sutherland drawing simple shapes and applying constraints using Sketchpad (reproduced from \citep{fuste2018hypercubes}) ]{\label{fig:sketchpad} Sutherland drawing simple shapes and applying constraints using Sketchpad. Reproduced from \citep{fuste2018hypercubes}.}
\end{figure}

A seminal work in the history of \gls{cad} is Sutherland's PhD dissertation \citep{sutherland1964sketchpad} in \citeyear{sutherland1964sketchpad}. He developed Sketchpad, a computer program that is generally considered as the first \gls{cad} system. It pioneered the use of a graphical user interface, controlled by a light pen (an old version of a stylus pen that works with CRT displays) and allowed users to create lines and curves, as seen in Figure \ref{fig:sketchpad}. In addition to drawing these simple shapes, the user could also create and apply constraints to the drawing and define certain drawings as master objects, that could then be instantiated (any change to the master drawing automatically updates the clones). 

The system was first presented at the 1963 AFIPS conference \citep{johnson1963sketchpad} and a video demonstrating its functionality at the time can be found online\footnote{\url{https://www.youtube.com/watch?v=6orsmFndx_o}}. The conference covered information science in general but the 1963 edition had a track dedicated to \gls{cad}, in which other \gls{cad} systems were presented, together with more conceptual contributions (e.g., \citep{coons1963outline} outlining the requirements for \gls{cad} systems). 

The first conference fully dedicated to \gls{cad} was created shortly after, in 1964 \citep{horvath2015ubiquitous}. Contemporaneously to those research-driven explorations, some large industrial firms also noticed the potential of the technology and developed in-house applications for their own needs (e.g., the DAC-1 system at General Motors).

Another important milestone for \gls{cad} history during that period was the development of Bézier curves (with their mathematical representation) that are still in use nowadays. That family of curves was invented independently by two mathematicians working for French car manufacturing companies: Bézier, employed by Renault, and de Casteljau, working for Citroën \citep{shah1995parametric}. They intended to create a mathematical form for a curve that would be easy and intuitive to modify (thanks to control points that define the curvature), so as to allow for experimentations by a user on a graphical system. Because de Casteljau was not allowed to publish his work that was kept secret by his employer \citep{farin2002history}, the curves are named after Bézier, who published extensively on the matter and even wrote a PhD dissertation \citep{bezier1977essai} based on his work at Renault.

Those developments mostly focused on two-dimensional sketching, even though there were some exceptions that also integrated some 3D capabilities (notably in Sketchpad itself).

\subsubsection{Industrial adoption (1970's)}
\label{sec:cad-70}
During the seventies, commercial 3D modellers grew into a more tangible reality, with systems running on workstations. \gls{cad} therefore became subject to a broader industrial adoption, although mostly limited to large companies. 

More design applications were also considered in that period, including architecture but also Computer-Generated Imagery (CGI) for the entertainment (film) industry, with programs such as MAGI's SynthaVision, that allowed users to make 3D animations. A promotional video of the software's capabilities from 1974 can be found online\footnote{\url{https://www.youtube.com/watch?v=jwOwRH4JpXc}} but the system was notably used, a few years later, for the original (1982) Tron movie, including most of its action sequences.

Overall, the \gls{cad} field matured in that decade, to a point where an overview of the state-of-the-art \citep{eaglesham1979cad} and the impact of the technology in the industry was published in a 1978 edition of the Computer-Aided Design journal. 

\begin{figure}[h]
	\includegraphics[width=.5\textwidth]{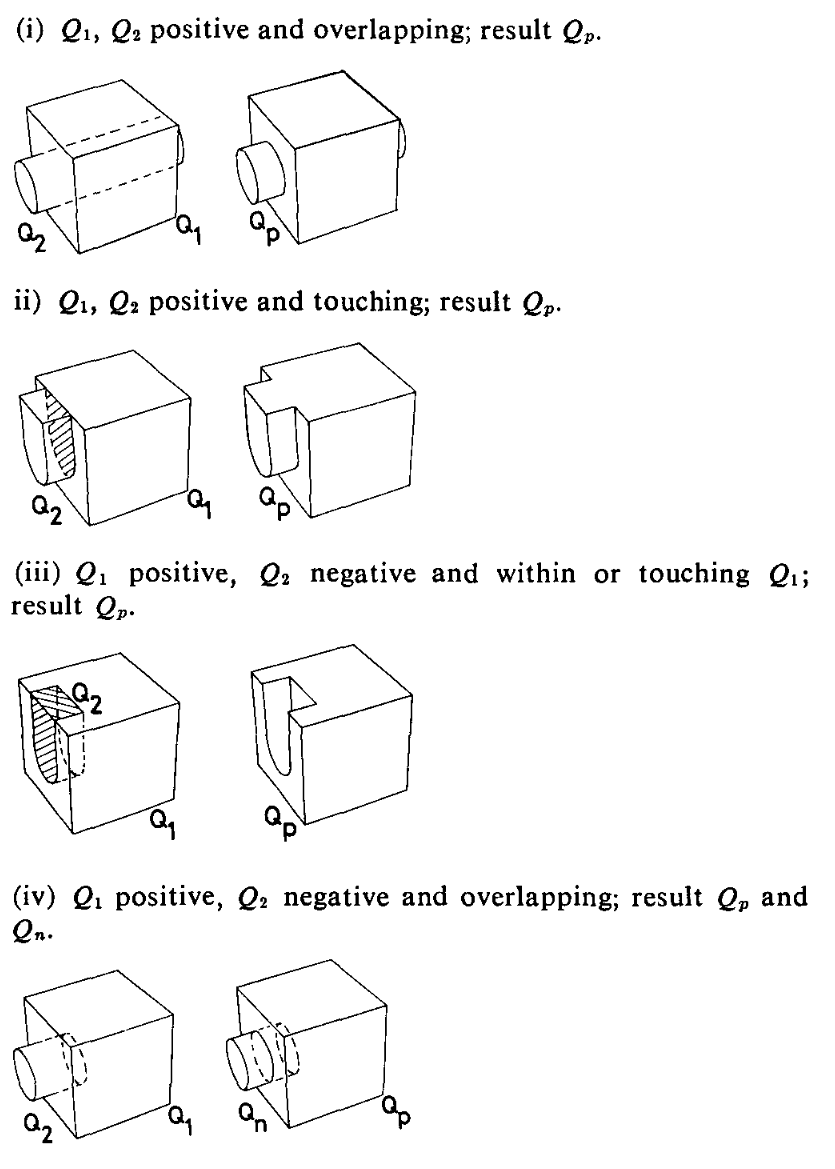}
	\centering
	\caption[Braid's addition (and substraction) of objects (reproduced from \citep{braid1975synthesis}) ]{\label{fig:brep} Braid's addition (and substraction $Q_2$ is negative) of objects. Reproduced from \citep{braid1975synthesis}.}
\end{figure}

In terms of important research developments during that era, one of the most prominent was the concept of \gls{brep} \citep{braid1975synthesis}, a technique that allows for solid objects to be described (and therefore stored on a computer) using primitive solids that are moved, scaled or rotated, and then combined with or subtracted from one another, as shown with simple examples in Figure \ref{fig:brep}.

Similar ideas were developed in the same period, with Voelcker's paper \citep{voelcker1977geometric} often cited as the original reference for \gls{csg}. Both \gls{brep} and \gls{csg} represent complex objects by combining primitive ones (with what is often referred to as boolean operations) but, whereas \gls{brep} internally stores faces, edges and vertices, \gls{csg} directly stores the primitive solids.

Both systems are used today, sometimes jointly to take advantage of their respective benefits \citep{hoffmann1989geometric}, such as \gls{csg}'s insurance that an object is always closed (and defines an interior volume) and \gls{brep}'s efficiency for graphical rendering.

\subsubsection{Mass adoption and networking (1980-1995)}
Based on earlier developments, the first version of AutoCAD was released in 1982. It was one of the first \gls{cad} systems to run as an application on microcomputers (as opposed to mainframes and minicomputers that had to be shared and accessed through terminals) and became very popular. The availability of such software combined with the advent of (more affordable) personal computers lead to a much larger adoption of \gls{cad} tools during the eighties and the first half of the nineties. 

Another key factor was the development of computer networks, which enabled better cooperation and data exchange. Many researchers were therefore developing specifications for exchange formats at the time \citep{wilson1987short}, with IGES (Initial Graphics Exchange Specification) \citep{smith1983iges, smith1986initial} being the most widely used. Those research advances led to a more concerted effort to develop a single standard specification, named STEP (Standard for the Exchange of Product Model Data) \citep{pratt2001introduction, pratt2005iso}. That specification eventually became an ISO standard, thereby reaching a wider industrial adoption.

In addition to these technical computer-to-computer communication advances, the early nineties also pushed \gls{cad} forward thanks to the invention of an alternative technique to draw curves using control points: the \gls{nurbs}. The use of that technique in \gls{cad} is hard to attribute to one particular individual since it results from the work of several separate researchers iterating over a concept that then lead to a later industrial adoption, but a PhD thesis in particular \citep{versprille1975computer} can be considered as seminal work on the subject.

While Bézier curves start and end at control points (respectively the first and the last ones) defined by the user, \gls{nurbs} curves do not reach their first/last control points (they bend from/towards them). Although both approaches are rather intuitive to use, this can make Bézier curves slightly easier to handle, even though \gls{nurbs} curves are more efficient for complex curves (in terms of computing power to calculate them). They can also exactly represent circles, whereas Bézier curves can only approximate them \citep{nurbs-vs-bezier}.

Even though Sutherland's Sketchpad could already apply certain constraints to drawings, the eighties saw \textbf{constraint-based design} arise as a paradigm. 
As is often the case with innovations, constraint-based design was further developed in a research context, with theoretical advances on how to enable variations of a (constrained) geometry \citep{hillyard1978analysis, light1982modification}.

Another paradigm also appeared in that period: \textbf{feature-based modelling}. It was first proposed by Pratt \citep{pratt1984solid} and was subsequently integrated into research prototypes \citep{cunningham1988designing,shah1988expert}. The idea of feature-based modelling is that the design is composed of features (such as a hole) that carry a semantic meaning instead of simply being geometric shapes (e.g., a circle). These features can also impose constraints or define properties.

Both constraint-based design and feature-based modelling were adopted by the industry and became part of most commercial \gls{cad} systems and suites by the mid 1990s \citep{shah1998designing}.

\subsubsection{Maturation and collaboration (1995-2005)}
Even though 3D modelling tools existed before, they truly started to shine and be used in the industry starting from the mid nineties \citep{baba1998towards}.
The sophistication of these tools, with more features and improved user interfaces, combined with a larger availability of personal computers and better compatibility (of both the hardware and the software) led to a rise in usage. In \citep{asanowicz1999evolution}, the author considers that we may only talk about mature \gls{cad} starting from that period, which coincides with the first mass adoption of \gls{cad} tools \citep{horvath2015ubiquitous}.

\begin{figure}[h]
	\includegraphics[width=.8\textwidth]{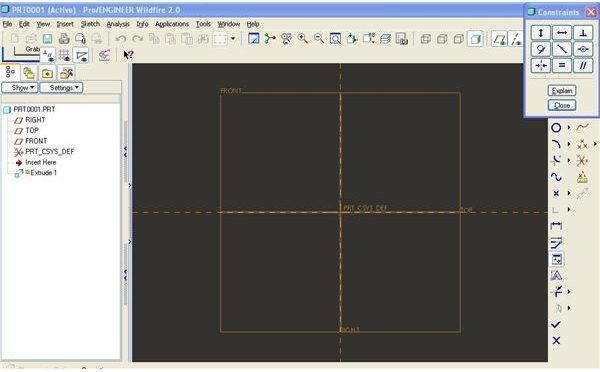}
	\centering
	\caption{\label{fig:proengineer}Sketcher constraints in Pro Engineer Wildfire 2.0 (2004). Reproduced from brighthubengineering.com\protect\footnotemark.}
\end{figure}

Constraint-based design was integrated into commercial \gls{cad} tools such as Pro/Engineer, and enabled automatic solving of geometric constraints. Even though the tool started as yet another \gls{brep} system, it received an update with sketch-based constraint definition during the nineties. This allowed designers to easily add constraints and annotate them to specify dimensional values. The sketches were then instantiated using a constraint solver \citep{hoffmann2005constraint}. Figure \ref{fig:proengineer} shows the Sketcher interface in Pro Engineer Wildfire 2.0.

Together with feature-based modelling, constraint-based design led the way to a new paradigm: \textbf{history-based modelling} that is sometimes also referred to as \textbf{parametric modelling}. History-based tools remember the designer's actions on the features and allow for later modifications to the parameters of the modification, before replaying the steps that previously followed that action, to rebuild and update the geometrical representation accordingly.

Another paradigm was placed in opposition to history-based design: \textbf{direct modelling}. Since creating parametric relations requires specific training, direct modelling was presented as an easier-to-grasp alternative. The approach is also called history-free because the designer directly works on the geometry with no knowledge of previous actions and no relations to define. It is therefore easier and faster to produce results in early stages of design but is not necessarily the better option in the long-term, especially for ``families of designs'' (that differ only by a set of parameter values) or highly configurable models \citep{tornincasa2010future}.

The research on collaborative design that was carried out during the second half of the nineties and beyond was also pushed and helped by the tremendous growth of the internet. \gls{cad} systems therefore started to include better support for online collaboration.

\footnotetext{\url{https://www.brighthubengineering.com/cad-autocad-reviews-tips/22421-why-3d-cad-modeling-software-is-parametric/}} 

\subsubsection{CAD sub-specialisation and input/output enhancements (2005-now)}

While each of the previous periods saw evolutions and breakthroughs that were crucial to the evolution of \gls{cad} systems, the beginning of the 21st century was mostly about consolidation and enhancements, in terms of Product Data Management (allowing better integration between different systems and therefore facilitating cooperation \citep{liu2001review}), but also with regards to portability (with the advent of tablets and smartphones). 

In the meantime, a plethora of specialised \gls{cad} tools were created, therefore extending the reach of computer-aided design to new domains. Specialised \gls{cad} tools were already available for fields such as mechanical engineering, architecture or electronics design, but \gls{cad} started to spread to other areas like life sciences. Examples include dentistry \citep{davidowitz2011use} and medicine in general \citep{bibb2014medical} but also specialisations that need to work at a miniature scale (often molecular or even atomic-level design), such as pharmacology (drug design \citep{macalino2015role}) or chemical biology (e.g., protein design \citep{mandell2009computer}).

\begin{figure}
	\includegraphics[width=.6\textwidth]{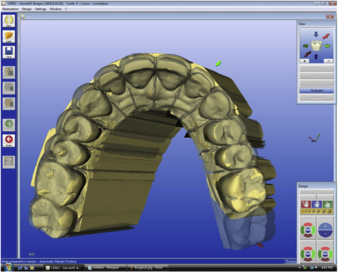}
	\centering
	\caption{\label{fig:dental-cad} A version of the CEREC software, used to design dental restorations and implants. By watching carefully, one can see the overlay of 2 models on the picture: the blue one is the scan result and the yellow one represents the model being worked on. Reproduced from \citep{davidowitz2011use}.}
\end{figure}

Another aspect that greatly evolved in the recent years is the integration with external data and tooling. As for the input side, 3D scanning and photogrammetry are increasingly employed, either to create a basis to work on (rough shape of an existing similar structure or even something the designer wants to replicate or reverse engineer) \citep{kucs2009implementation}, or simply in order to provide context to the design (e.g., surroundings) \citep{wolf2014elements}. Figure \ref{fig:dental-cad} shows an example of a specialised \gls{cad} tool for dentistry, that also illustrates the use of scanned data to help the designer.

The popularisation of 3D printing technology led to an increase in their use for Computer-Aided Manufacturing (CAM), the next logical step in the production process for many CAD-enabled domains. For example, manufacturers have used the technology to make dental implants \citep{dawood20153d} or even bone tissues \citep{bose2013bone}. \gls{cad} tools have also evolved with regards to visualisation features, with photo-realistic renderings becoming more commonplace, for instance to better picture violations of requirements in autonomous vehicle safety assessment \citep{o2017computer} or to simulate coating appearance depending on lightning \citep{jhamb2020review}.

\subsection{Computer-Aided Architectural Design}
\label{sec:sota-cad-caad}

The evolution of \textbf{\gls{caad}} is mostly consistent with the history of \gls{cad} in general \citep{mitchell1990afterword, Moubilerealtimeparametricstructural2018}, as expected since both fields can benefit from the same technological advances. 

The concept of \gls{caad} as a research area started to emerge during the 1970's, with overviews of the state of the art and the practice in that period, but is only truly recognised as a separate field in the 1980's \citep{koutamanis2005biased}. 

\subsubsection{Generations of CAAD}

A common way to structure the evolution of \gls{caad} (as well as \gls{cad} in general) is to divide its history in generations of tools \citep{kale2005diffusion} but many different divisions have been proposed and the suggested timelines often contradict one another. 
The three generations presented in this section are therefore a subjective summary based on our literature review. 

The first generation would be computer-aided drafting, where the computer essentially serves as an alternative and potentially more precise tool to draw lines and shapes. 
Drawings are then stored in an electronic format and can therefore be easily modified and shared. Three-dimensional wireframe versions of such tools were also developed but the underlying design process did not fundamentally change.

A second generation of \gls{caad} technology has, in a sense, raised the level of abstraction, by allowing designers to work with more complex objects and transformations.
Regardless of which technology is used to store these objects, i.e., \gls{brep} or \gls{csg} (see section \ref{sec:cad-70}), or even a combination of both, the geometrical representation is created using parametrised primitive (3D) shapes directly, that are then combined to construct a composite object.

A third generation comprises Computational Design methods, that will be further described in section \ref{sec:computational-design}. The general idea is that these methods allow architects to act on rules, constraints, or instructions that will generate or act on the designed geometry. The level of abstraction of the artefacts manipulated by the architect for that generation of tools is therefore even further raised, since the designer no longer has to act on the geometrical representation directly to produce changes.

It should be noted that each new generation of \gls{caad} technology never entirely replaces the previous one. This is particularly true for the last two generations described, that are both prominent in today's architectural practice and their respective benefits are often combined in the same (suite of) tools.

\subsubsection{Building Information Modelling}
\label{sec:bim}

\begin{figure}[h]
	\includegraphics[width=.7\textwidth]{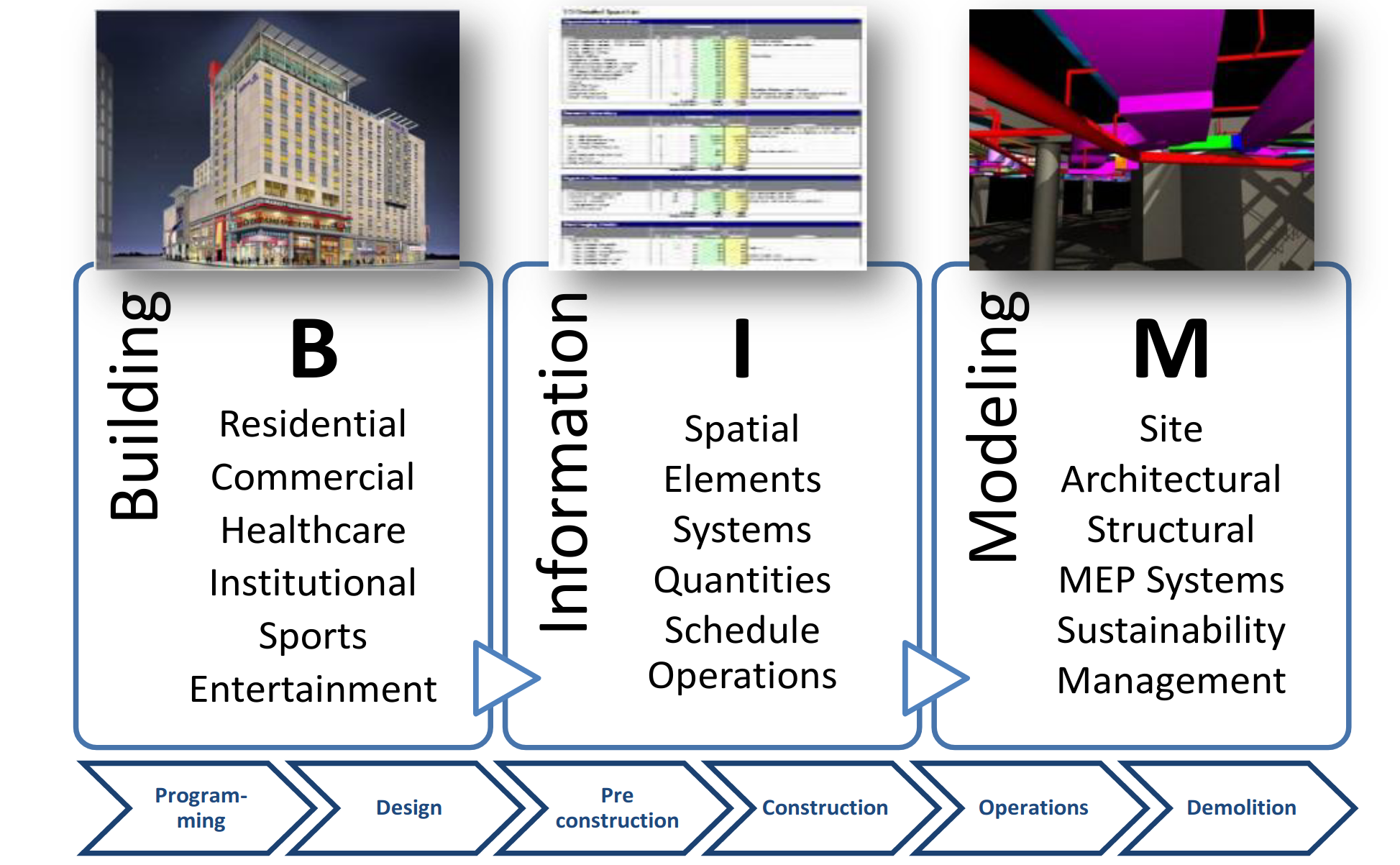}
	\centering
	\caption{\label{fig:bim} The \gls{bim} concept. Reproduced from \citep{kymmell2008building}.}
\end{figure}

An important milestone in \gls{caad} history is the release of Autodesk Revit in 2000. The software helped popularise \gls{bim} in the mid-2000s \citep{Moubilerealtimeparametricstructural2018, azhar2012building}, although the conceptual basis dates back to the 1980's \citep{aish2017evolution} with the Building Description System \citep{eastman1975use}. 
\gls{bim} describes both the software and the process that encourages a better integration of stakeholders on a project in the \gls{aec} sector \citep{azhar2012building}. These stakeholders include owners, designers, constructors, engineers, contractors, suppliers and facility managers. The core concept is that a project consists of 3D models that are interlinked, shared with all stakeholders, and connected to the information from these stakeholders and the different phases of the project's development. That concept is depicted in Figure \ref{fig:bim}. 

Using the technology tends to lower the risk of inconsistencies and errors, but also helps provide semantic meaning to building elements. In a way, the concept of \gls{bim} relates to the Enterprise Resource Planning (ERP) technology in use in many businesses, where the company would be the construction project, and the departments and divisions would correspond to the project's stakeholders. They all share a common database where everyone is able to access and modify his part.

\subsection{Programming architectural models}
\label{sec:computational-design}

\begin{figure}[h]
	\includegraphics[width=.7\textwidth]{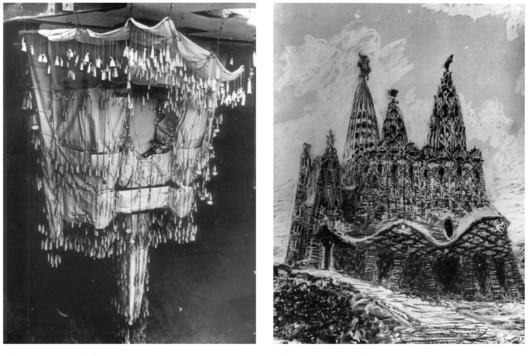}
	\centering
	\caption{\label{fig:gaudi} Gaudí's hanging model and resulting rendered form. Reproduced from \citep{maher2003parametric}.}
\end{figure}

Parametric definition of geometries and mathematical expressions to describe volumes clearly predate the computer era.
The work from Antoni Gaudí at the turn of the 20th century, in particular the Colònia Güell, is also often cited as an early occurrence of algorithmic design thinking. This is likely due to the innovative way the architect realised the form-finding step of this architectural masterpiece: through chains hung from a ceiling of other chains and burdened with small weights, depicted in Figure \ref{fig:gaudi}, so as to let the laws of physics curve the chains to distribute the load evenly. Once flipped upside-down, the resulting shapes define a structure of arches superimposed on each other. If a chain anchor's point or load changes, the entire structure is changed by the natural optimisation process. This resembles the way \gls{ad} parameters impact the whole geometry and may explain why Gaudí's work is sometimes associated with \gls{ad}, although it would likely be more accurate to talk about ``analog computing'' when mentioning that sort of natural optimisation.

Even way before these constructions were designed, during antiquity, civilisations were using mathematical patterns and even rule-based (i.e., algorithmic even though no computer was involved) methods for architecture and art, with Islamic patterns \citep{agirbas2020algorithmic} coming to mind as a primary application of these concepts.

The \gls{caad} tools we will talk about in this section are more heavily tied to the concept of programming, with the explicit existence of an algorithm that can be edited.
This comes as a profound paradigm change that may not be natural for conventionally-educated architects, since they have to learn to ``think algorithmically'' i.e., to decompose a design idea into a set of instructions that are simple enough for a computer to execute. These instructions can be processed with variable parameter values or even ignored depending on specific conditions declared by the designer-programmer through what is called flow-control mechanisms. This is one of the main benefits of using such algorithm-based design tools: a plethora of possible (geometric) solutions can be generated by these algorithms, allowing the designer to explore a variety of different outcomes. Depending on the particular project on which it is used, designing with algorithms can reduce human errors \citep{burry2011scripting} and costs \citep{woodbury2010elements}.

\subsubsection{Different representations: textual vs visual}

Programming is typically carried out through (advanced) text editors, at least for professional developers, since it seems to be the appropriate choice for productivity and scalability \citep{myers1990taxonomies}.
That being said, designing with algorithms does not come naturally for traditionally-trained architects. While an increasing number of university curricula focus on, or at least include, courses about \gls{ad}, architects are not usually programmers. An architect starting to learn \gls{ad} through a textual language therefore first has to get familiar with programming concepts such as variables, functions and scopes, in addition to the language's syntax. 


\begin{figure}
	\includegraphics[width=.8\textwidth]{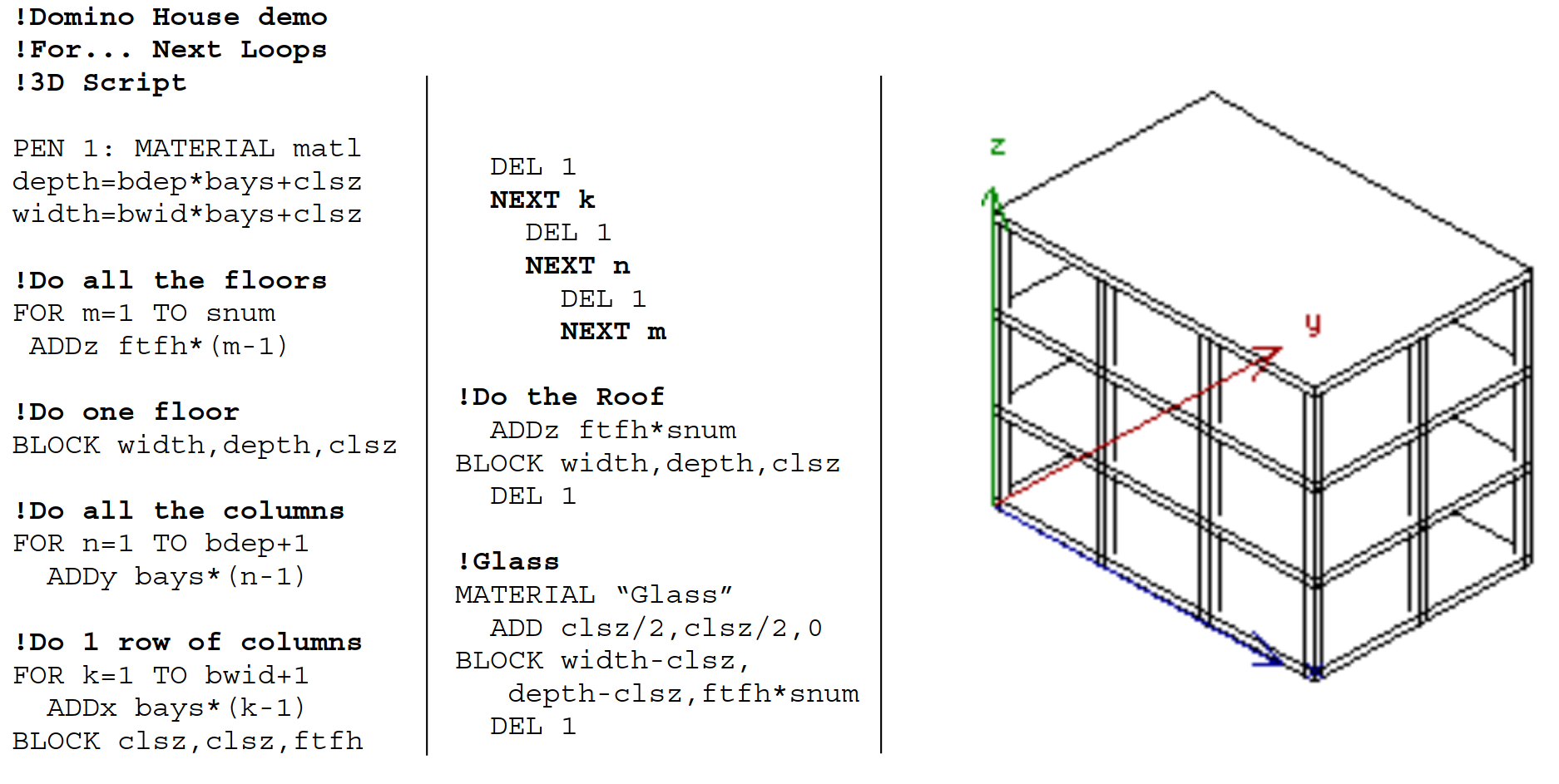}
	\centering
	\caption{\label{fig:gdl} A \gls{gdl} script to generate a domino house. Reproduced from \citep{nicholson1998gdl}.}
\end{figure}

\begin{figure}
	\includegraphics[width=.8\textwidth]{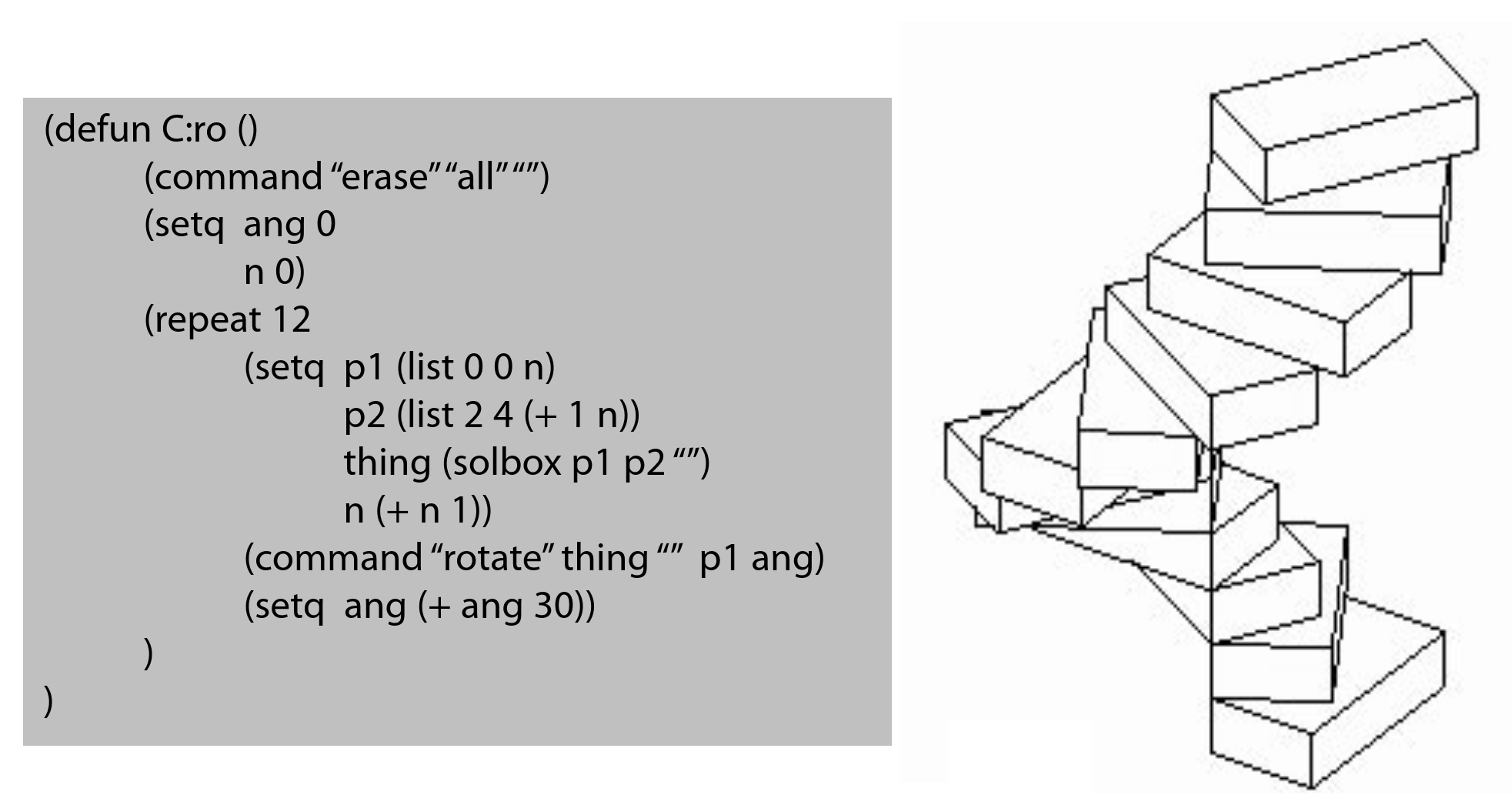}
	\centering
	\caption{\label{fig:autolisp} An AutoLISP script to generate a spiral staircase. Reproduced from \citep{coates1995generative}.}
\end{figure}

Figure \ref{fig:gdl} depicts an example in \gls{gdl}, a script language for ArchiCAD, while Figure \ref{fig:autolisp} shows a staircase model in AutoLISP (for AutoCAD). Both languages were released in the 1980's.

\begin{figure}[h]
	\includegraphics[width=.7\textwidth]{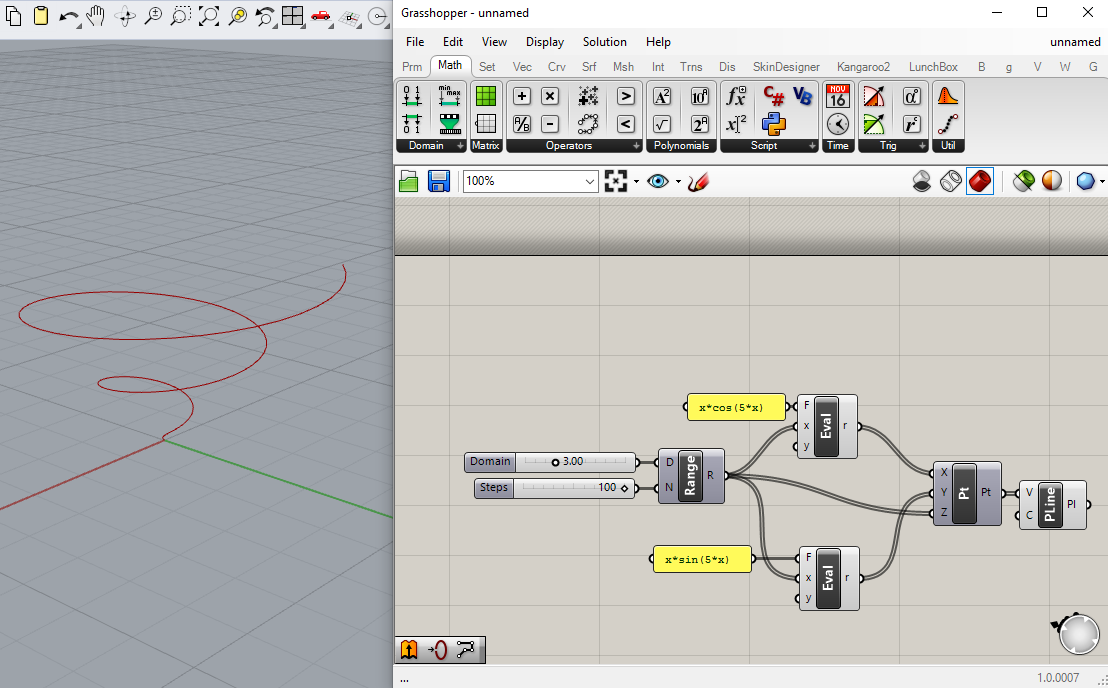}
	\centering
	\caption{\label{fig:gh} Example of a Grasshopper definition for a conical spiral. The spiral (output by the PLine component) is created from a list of "vertices" (input parameter V) that are effectively points (Pt), whose coordinates are defined by evaluating (Eval components) two formulas on a list of values, themselves generated from a Range component. Reproduced from \citep{leitao2012programming}.}
\end{figure}

Because of the difficulty to learn textual programming for non-programmers, \textbf{\glspl{vpl}} have been developed for architectural design and are more popular than the textual alternatives in that context \citep{sammer2019visual}. 

The most popular ones as of today include GenerativeComponents, Dynamo Studio and Grasshopper. An example created with the latter is depicted in Figure \ref{fig:gh}.

All these tools aim to make \gls{ad} more accessible by providing an intuitive representation of the algorithm that better satisfies the visual nature of architects. This makes the learning curve more gentle, although it comes at the cost of a reduced scalability. In fact, contrary to the common adage that ``a picture is worth a thousand words'', \gls{vpl} models tend to become hard to understand and manipulate \citep{leitao2012programming} when they grow in size and complexity. This is partly due to the lack of advanced abstraction mechanisms that tends to induce redundancy in the visual code (copying-and-pasting parts of an algorithm). 

Furthermore, many of these visual languages do not natively support recursivity\footnote{Such functionality is however usually available within \gls{ad} solutions, using plug-ins or through a textual programming language supported by the solution.}. The choice is understandable since they usually target non-programmers that would need to get familiar with the concept first and could easily end up with infinite loops in the meantime, but may further increase the scalability issue in certain instances.

In an attempt to benefit from the best of both worlds, some editors (in fact, most of the popular \gls{ad} software tools) enable hybrid programming approaches, that rely on a combination of visual and textual representations. In most cases, the visual language is used to define the outline of the algorithm (i.e., the overall logic on how to construct the target geometrical representation), using some nodes or blocks that themselves may contain code in textual form.

\subsubsection{Programming paradigms}
\begin{figure}[h]
	\includegraphics[width=.8\textwidth]{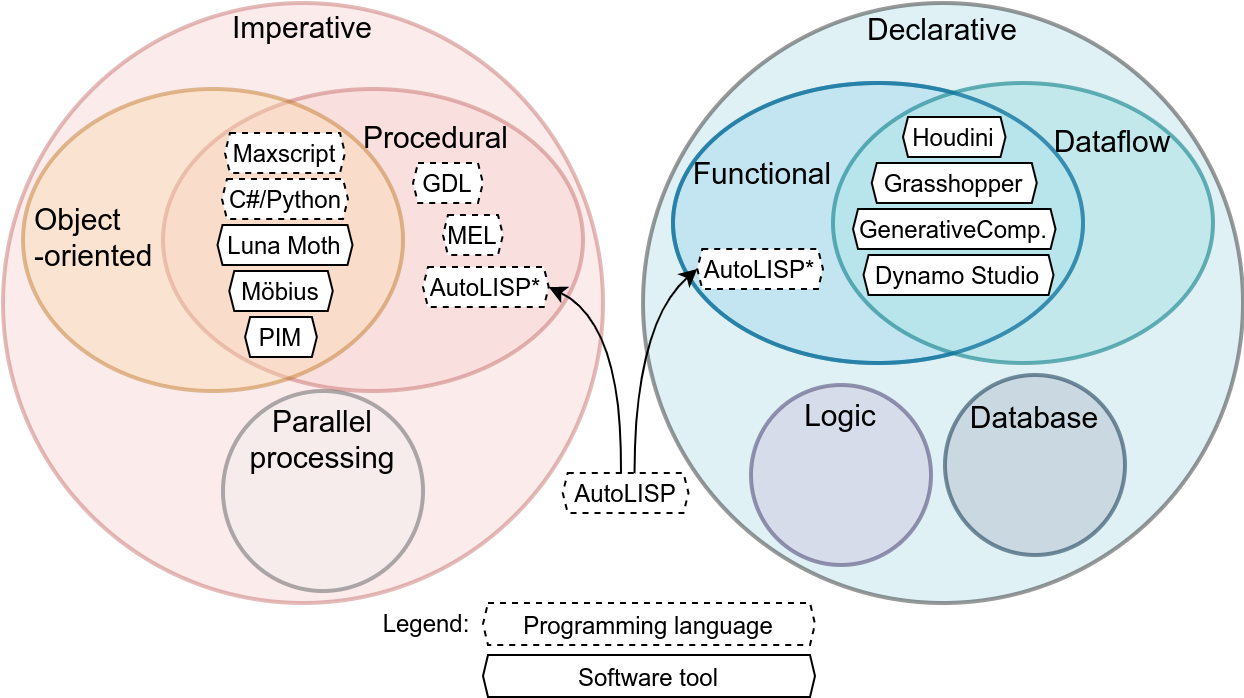}
	\centering
	\caption[]{\label{fig:caad-programming-languages} Algorithmic Design tools and programming languages (adapted from \citep{davis2013modelled} and based on \citep{appleby1991programming})\footnotemark.}
\end{figure}

Another way to classify the plethora of \gls{ad} tools that have been created is by grouping them by the paradigms of their underlying programming languages. Such a taxonomy is presented in \citep{appleby1991programming} and further depicted in \citep{dave2013gesture}, where it is annotated with examples of programming languages and \gls{ad} tools made for or at least used by architects. Figure \ref{fig:caad-programming-languages} presents a reworked and updated diagram of that taxonomy, with a few examples of currently popular or historically-relevant \gls{ad} tools and programming languages.

On the left-hand side, the red circle represents the set of \textbf{imperative languages} i.e., languages that focus on how a program has to operate, through ordered statements that alter the state of a program. Within that set, procedural programming is a subtype of imperative programming that relies on functions (or procedures) to split a program into separate pieces that can be called from different parts of the overall program. Examples of \gls{ad} languages that fall into this category are the previously mentioned \gls{gdl} for ArchiCAD as well as MEL (Maya Embedded Language), that both use control flow statements (e.g., IF/ELSE) to conditionally decide which code blocks to execute and loops (e.g., FOR) to execute the same code block a specified number of times (typically until a certain condition is met).

\footnotetext{The image is an intentionally oversimplified view of the relations between programming paradigms. Many hybrid languages have been created and integrate features from paradigms that do not overlap in the figure, including AutoLISP that is in fact mentioned twice, in both the procedural and the functional sets.} 

Another subset of imperative languages is the set of object-oriented programming languages, that rely on the notion of objects that contain data and behaviour, typically through object-specific procedures that explain why object-oriented languages are very often also procedural languages. These objects help package specific functionalities with the relevant data states, and isolate parts of the code that are unrelated. An example of an object-oriented \gls{ad} language is Maxscript for 3DS Max, but popular general-purpose programming languages (broadly used outside of \gls{ad}) such as \Csharp \, and Python can also be used with many \gls{caad} tools. In fact, these languages are supported by major solutions such as Rhino, AutoCAD, Maya, Houdini, Dynamo Studio or Revit. Some \gls{ad} software tools, such as Luna Moth \citep{alfaiate2017luna}, Möbius \citep{janssen2016mobius} and PIM \citep{maleki2013programming}, also rely on ad hoc programming languages that support that paradigm.

On the right-hand side of Figure \ref{fig:caad-programming-languages}, the blue circle is for \textbf{declarative languages} i.e., languages that focus on the logic of the program without controlling the flow of execution. A common simplification of the principal difference between declarative and imperative languages is that the declarative paradigm focuses on \emph{what} to accomplish instead of \emph{how} to do so, as would be the case for the imperative paradigm approach. Within the declarative circle, we find \textbf{functional languages}, where programs are created by composing functions. That paradigm has its roots in lambda calculus \citep{church2016calculi}, a formal system that only uses functions for computation.

Often linked to functional programming (because it shares some common characteristics) is \textbf{dataflow programming} \citep{johnston2004advances}, where a program can be modelled as a directed graph that lets data flow along the edges after being processed by the graph's nodes. Although there are textual dataflow programming languages, their inherent capability to be represented as graphs has lead to the creation of many visual languages based on that paradigm. As for \gls{ad}, the main software tools in use today do rely on visual languages that integrate elements from both functional and dataflow programming. These include Grasshopper, depicted in Figure \ref{fig:gh}, but also GenerativeComponents and Dynamo Studio.

Very often, modern programming languages integrate features from multiple paradigms. One such example as far as \gls{ad} is concerned is AutoLISP and its enhanced version Vital LISP (later renamed Visual LISP), the AutoCAD scripting language that is a LISP dialect.
Figure \ref{fig:autolisp} shows an example of AutoLISP code that produces a spiral staircase.

It should be noted that most popular software tools integrate multiple programming languages. Most of the time, the core skeleton of an \gls{ad} model is a visual dataflow-based program, but sometimes specific components of that program can be written in another programming language supported by the software. This gives the designer access to multiple paradigms and representations, and he is free to choose the tool that best fit his design intent and/or abilities.

\section{Three-dimensional Human-Computer Interaction}
\label{sec:sota-3dui}

As made clear by its name, \textbf{\gls{hci}} is inherently an interdisciplinary domain, that combines knowledge from many disciplines. Our focus here will be on the technical (computer) side of \gls{hci}. The relevant literature in psychology, human factors and ergonomics therefore will be considered out of scope, although the interested reader may find good references on these subjects in \cite[chap.~3]{LaViola3duserinterfaces2017}, \citep{salvendy2006handbook}, \citep{mackenzie1992fitts}, and \cite[chap.~2]{mackenzie2012human}.

Interaction techniques are the necessary bridge between the user and the application's interface, they are how that user communicates intent to the system. 
In the world of two-dimensional applications, most interfaces rely on the \gls{wimp} metaphor and users typically interact with the system through a mouse and a keyboard. Since there is no such well-established standard for three-dimensional environments, interaction techniques and user interfaces adapted to that context can take many forms and rely on various input devices. 

This section covers such techniques, focused on the most basic manipulation tasks we need: selection, positioning and rotation. The section's structure is dictated by a common taxonomy \citep{LaViola3duserinterfaces2017} that classify techniques depending on whether they offer direct interaction with the target entity.

\subsection{Direct manipulation}

Amongst direct manipulation techniques, one way of categorising them is based on isomorphism: \textbf{isomorphic} approaches preserve a natural one-to-one mapping between input actions and their resulting effect, whereas \textbf{non-isomorphic} techniques afford non-realistic interactions and can even be based on ``magical'' or ``virtual'' tools. 

That being said, \textbf{direct} manipulation techniques very often either rely on \mbox(A) a \emph{touching} (grasping) metaphor or (B) a \emph{pointing} metaphor. The organisation of this section is therefore rather based on the metaphor being used.

\subsubsection{(A) Grasping metaphor}

As for techniques based on touching (or grasping), the user must reach the target object's position to interact with it. This is the most natural and intuitive way of interacting within three-dimensional environments since this is how human naturally interact in the real world.

\begin{figure}[h]
	\includegraphics[width=.9\textwidth]{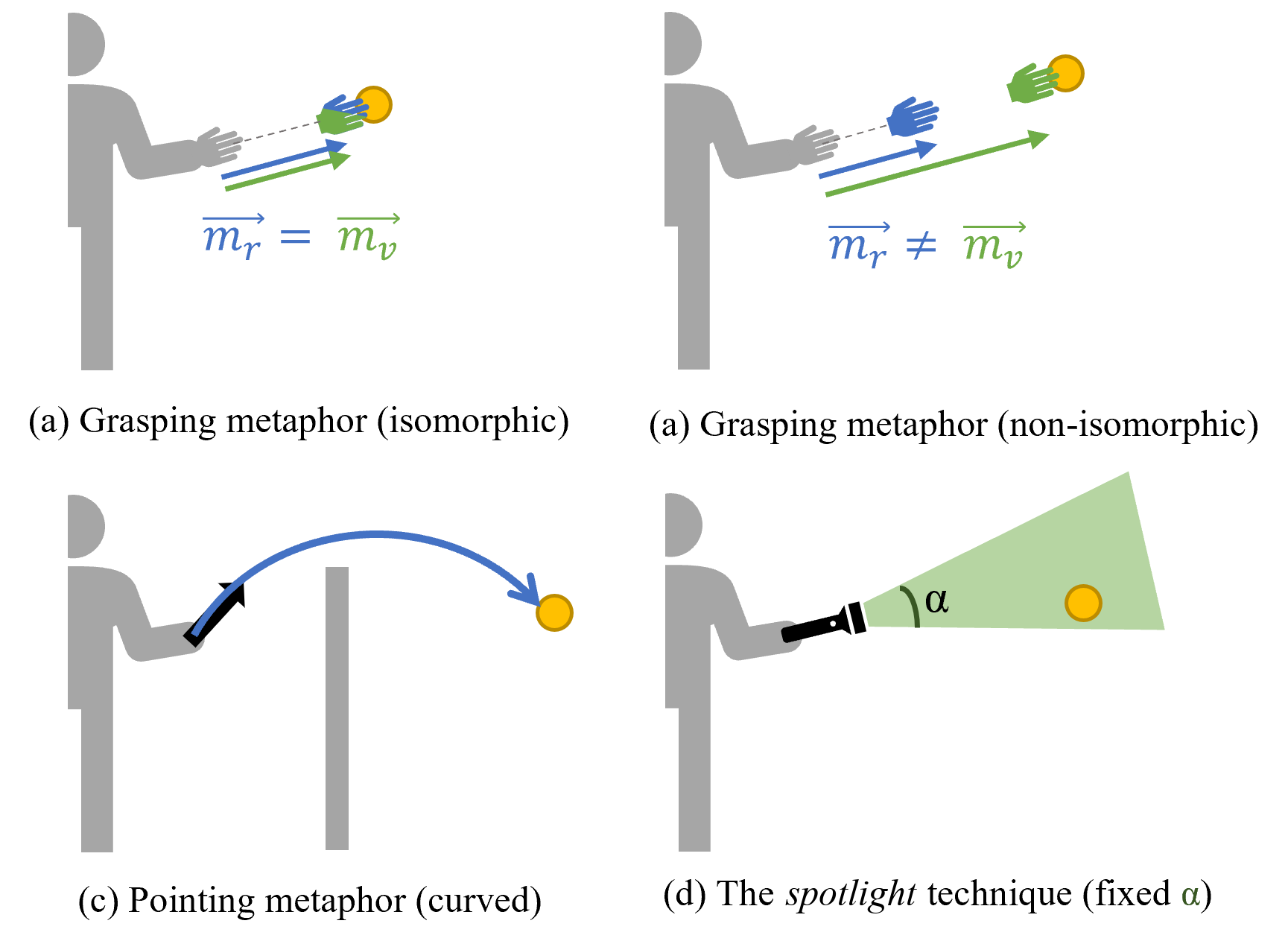}
	\centering
	\caption{\label{fig:direct-techniques} Examples of direct interaction techniques, relying on different metaphors to select a yellow ball.}
\end{figure}

Typically, grasping-based techniques rely on a virtual hand (or assimilated) that is mapped to a tracked object in the real world, that we will refer to as ``the real hand'' in the remaining of this section. A simplistic isomorphic example of such a setup would be to track a physical controller with six degrees of freedom and map its position and orientation to the virtual hand. If the physical controller has actionable buttons, they could be used to trigger the grasping of objects that collide with the virtual hand, so that they can be manipulated and then released (e.g., when releasing the button). Figure \ref{fig:direct-techniques}a depicts a similar example, with the user's real hand (in blue) overlapping the virtual one (in green) to touch a yellow ball.

When the tracking zone is limited in space compared to the ``interactable'' area, non-isomorphic mappings are typically used to mitigate these limitations. Figure \ref{fig:direct-techniques}b presents that idea, with the real hand's movement being amplified to change the virtual hand's position, so that it can reach the yellow ball even though the distance from the user would not allow such interaction in the real world (should the yellow ball have a physical counterpart in the same position).
The Go-Go~\citep{poupyrev1996go} technique is an example that only amplifies the virtual hand's motion when the real hand's distance to the user's body exceeds a predefined threshold.
On the other hand, different techniques such as PRISM \citep{frees2005precise} also rely on a non-linear mapping between the virtual hand and the real one's motion, but rather choose to scale down the virtual hand' movement when the real hand's motion is close to the user's body, to allow for more precise manipulations. Note that, in the case of that technique, an offset recovery is also present when the real hand's motion exceeds a given threshold, so as to allow for the virtual hand to catch up on the real hand. There is additionally a buffer zone in between those two threshold so that movements whose speed falls in that range can provide a direct (one-to-one) mapping between real and virtual motions.

\subsubsection{(B) Pointing metaphor}
The other metaphor commonly encountered is the pointing metaphor, that naturally mitigates space-related limitations, since aiming at an object becomes sufficient to start interacting with it. Similarly, the user is able to reposition that object by pointing towards a target position. Typically, the position and orientation of a tracked controller, that we will refer to as ``the wand'' in the remaining of this section, defines a ``laser beam'' that selects and manipulates objects. 

Techniques based on that metaphor differ in how the position and orientation of the wand affect the laser beam, and which object(s) are selected based on that beam. The most common version is often called ray-casting \citep{poupyrev1998egocentric}, where the wand defines a simple line segment and that segment's intersection with the environment defines the target object or position.
More complex techniques allow users to bend the line to mitigate limitations related to occlusion, as shown on Figure \ref{fig:direct-techniques}c, where selecting the yellow ball occluded by an object becomes possible thanks to the curve. There are many ways of bending the beam based on the wand's position and orientation, but most techniques rely on Bézier curves (e.g., the flexible pointer \citep{feiner2003flexible}).

To select distant objects more easily or in order to enable multi-objects selection, volumes can be used instead of simple rays. The selection volume size can be static (e.g., the spotlight technique \citep{liang1994jdcad} that relies on a selection cone, as pictured in Figure \ref{fig:direct-techniques}d) or dynamic e.g., the aperture technique \citep{forsberg1996aperture}, that expands on spotlight by allowing the user to control the spread angle of the cone by changing the distance from the eye to the wand (the angle gets bigger as the wand gets closer, as if the wand served as an open cylinder that would select everything the user can see through it). 

\begin{figure}[h]
	\includegraphics[width=.9\textwidth]{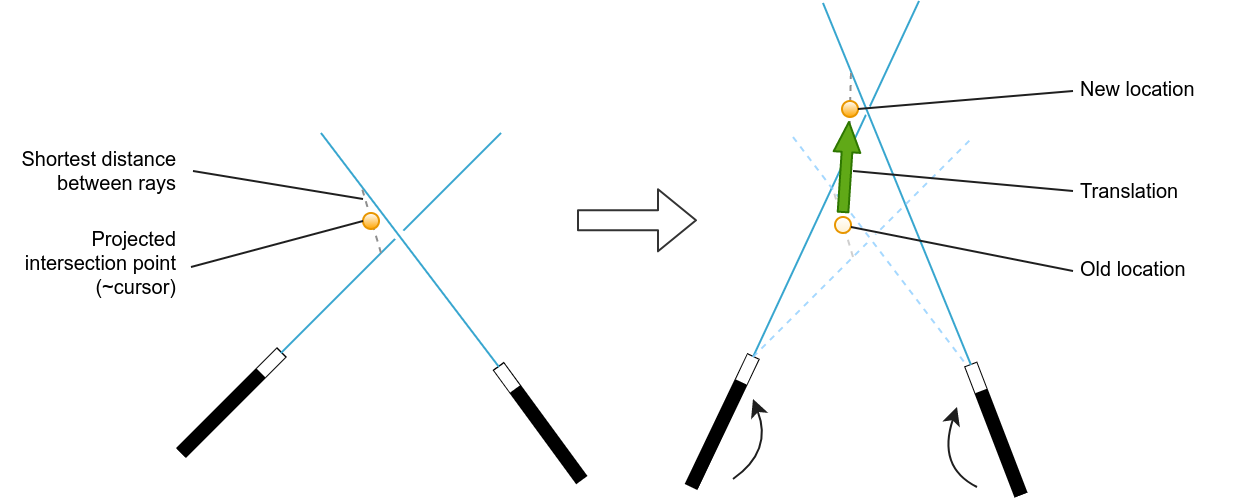}
	\centering
	\caption{\label{fig:isith} The idea behind the iSith bimanual interaction technique, with an example where a yellow ball is moved to a new location by rotating both wands.}
\end{figure}

More complex techniques that rely on two rays also exist, such as iSith \citep{wyss2006isith}, depicted in Figure \ref{fig:isith}. That technique uses a ``projected intersection point'' that is placed at the middle of the shortest line segment between both rays, that is then used as the interaction point (``cursor'') to select and move objects when pressing a button.

Other comparable techniques that add support for controlling more degrees of freedom have been proposed, ``Spindle + Wheel'' \citep{cho2015evaluation} that allows users to control with 7 degrees of freedom (3 positional axes, 3 rotational axes, and a global object scaling control) and is based on previous work \citep{mapes1995two} that proposed a similar approach that did not offer pitch control (one of the three rotational axes).

\subsection{Indirect manipulation}

Many other interaction techniques allow users to manipulate objects without directly interacting with them in the virtual environment. These may not be as natural as direct manipulation but provide alternative solutions to mitigate spatial and occlusion constraints so as to enable distant interaction.

One option is to rely on proxies, i.e., miniature representations of remote objects often located close to the user. Actions on a proxy are mapped to the proxy's full scale counterpart. This typically enables direct manipulation of the proxy, e.g., based on grasping techniques, but can produce an effect on a distant object.
A well-known technique that relies on this principle is WIM (World In Miniature) \citep{stoakley1995virtual}, that provides the user with a hand-held scale model of the virtual environment, similar to the maps or radars that are part of many game interfaces, but in three dimensions.

\begin{figure}[h]
	\includegraphics[width=.5\textwidth]{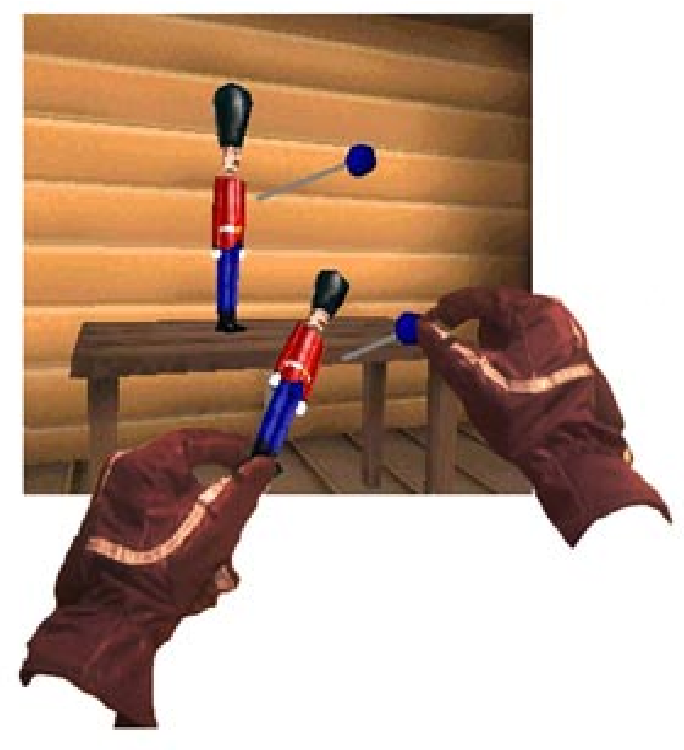}
	\centering
	\caption{\label{fig:voodoo-dolls} The Voodoo Dolls proxy-based indirect interaction technique. The blue pin can be moved relative to the nutcracker, since they were both selected and the pin was then grabbed by the user's dominant hand. Reproduced from \citep{pierce1999voodoo}.}
\end{figure}

A more advanced proxy-based technique is Voodoo Dolls \citep{pierce1999voodoo} that first asks the user to select one or multiple objects to interact with, through a pointing-based technique. Clones (dolls) of the selected objects are created and scaled down, then placed into the user's non-dominant hand. When the user grabs a doll with its dominant hand, the manipulation of the doll (and its full scale counterpart) can begin, with motions relative to dolls in the non-dominant hand. Figure \ref{fig:voodoo-dolls} shows that technique being used to interact with a nutcracker and a pin.

Another alternative to enable indirect manipulation is to make use of widgets. Such widgets are extremely common in desktop-based design tools for many domains (e.g., architectural design software or game engines), where the user can generally manipulate an object's position, rotation, or scale, based on handles placed in the target object's close environment. One handle typically only controls one degree of freedom but three-dimensional interfaces can also allow a widget to control two degrees of freedom at the same time. While controlling a third degree of freedom would be possible with 3D-tracked controllers, the widget is then only limited to a starting or anchoring role, and the technique therefore becomes more of a grasping-based one (potentially with an offset).

\subsection{Hybrid techniques}
As with tracking technologies, hybrid interaction approaches have been proposed to combine the advantages of multiple types of techniques and circumvent their individual shortcomings.
There are two ways to hybridise such techniques: either the user (or the system itself) can select the appropriate manipulation technique at any point in time, or specific techniques are assigned to specific tasks (or stages in the interaction).

\begin{figure}[h]
	\includegraphics[width=.8\textwidth]{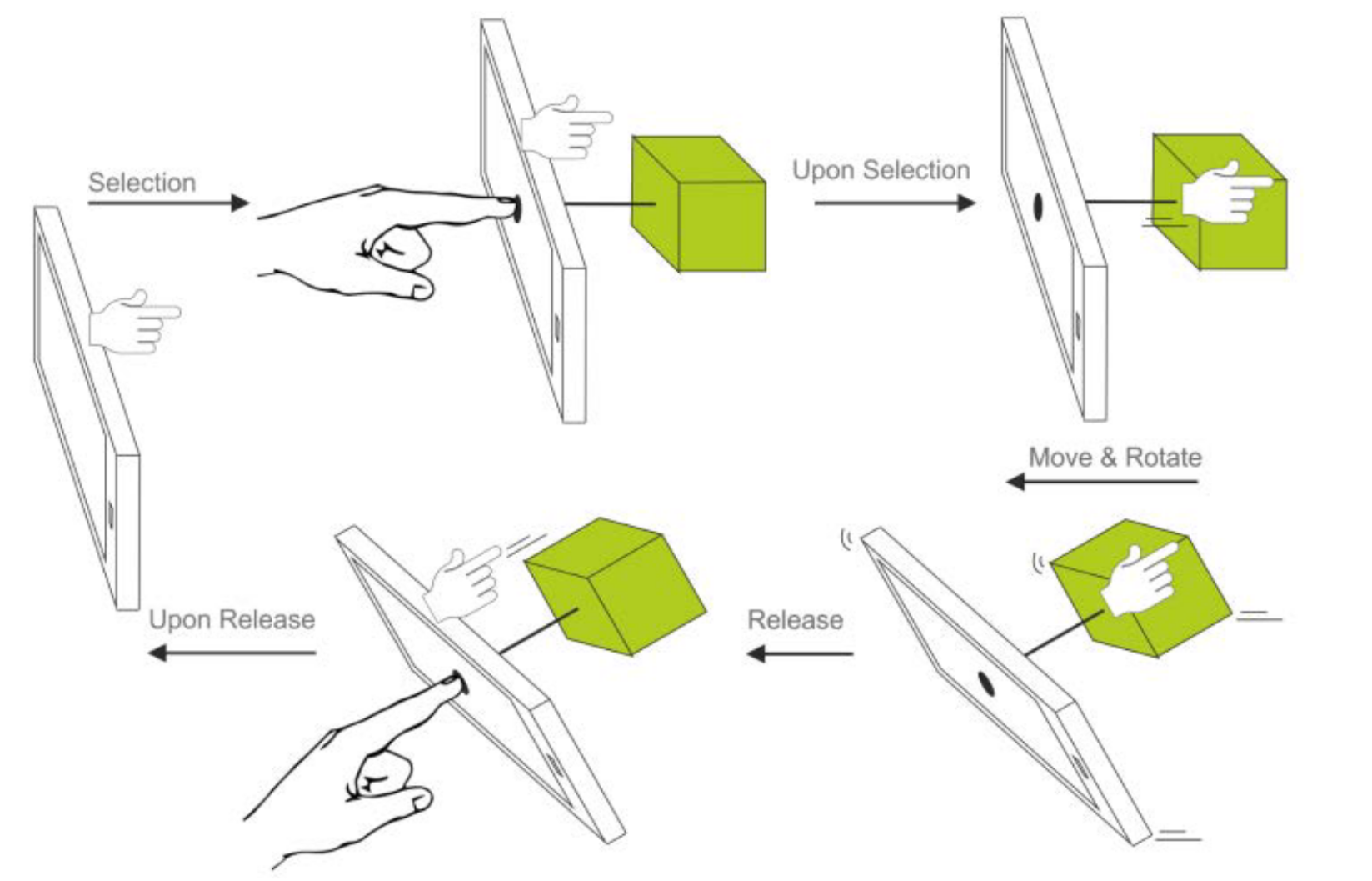}
	\centering
	\caption{\label{fig:homer} The HOMER-s hybrid manipulation technique, that combines both raycast and grasp-based approaches for handheld-AR interaction. Reproduced from \citep{mossel20133dtouch}.}
\end{figure}

An example of the latter is to perform the selection of an object through a raycast-based technique that teleports a virtual hand to the target location when the object is selected, before manipulating the target object using a grasping-based technique with scaled motion, similar to the previously mentioned Go-Go technique. \citep{bowman1997evaluation} proposed such a combination of techniques with HOMER (Hand-centered Object Manipulation Extending Ray-casting). That method has been adapted to handheld mobile \gls{ar} with HOMER-S \citep{mossel20133dtouch}, whose concept is depicted in Figure \ref{fig:homer}.

\section{New ways of interacting with architectural designs}
\label{sec:new-ways-archi}
As discussed in Section \ref{sec:sota-cad}, recent advances in various technologies have provided \gls{cad} software with new opportunities to interact with models. Architectural design is no exception and the advent of innovations such as 3D printing technologies have led to academic and industrial adoptions. Researchers have explored different uses of the technology, ranging from small-scale educational prototyping \citep{greenhalgh2016effects} to large-scale industrial constructions \citep{gardiner2011exploring, wu2016critical}.

Another recent technology that is being used extensively in architectural practice is photogrammetry, which can be used to document specific buildings for cultural heritage purposes \citep{hanan2015batak} or to provide a basis on which construction planning activities can be performed \citep{liu2001review}.

While the previous examples described potential enhancements to the design and construction processes that were ``external'' to the modelling tools themselves, work has been done to increase interactivity with existing software tools, as predicted in \citep{MonederoParametricdesignreview2000}.

This is the case for various stages of the design process. As for the conceptual design phase, where designers typically rely on sketching tools, the integration of generative design techniques has lead to systems such as DreamSketch \citep{kazi2017dreamsketch} that generates three-dimensional geometries based on free-from sketches. When sketching with that system, the designer creates objects that can be labelled as \textit{interface}, if they are part of the geometry, or \textit{obstacle}, if they are not to be included in the generated solution. The user can then define \textit{variables} by moving an object along the edges of the polygon that encloses that object's possible positions, and the system uses that information to generate potential solutions. Similar automated generation techniques have also been used for more advanced modelling activities, often through add-ons or plug-ins extending the functionality of popular modelling tools.

Section \ref{sec:dse} discusses how exploratory and optimisation-based methods have been applied to architectural design, while Section \ref{sec:sota-cad-immersion} overviews the current use of immersive technologies in the field.

\subsection{Design Space Exploration and Optimisation of parameter values}
\label{sec:dse}

\begin{figure}[h]
	\includegraphics[width=.8\textwidth]{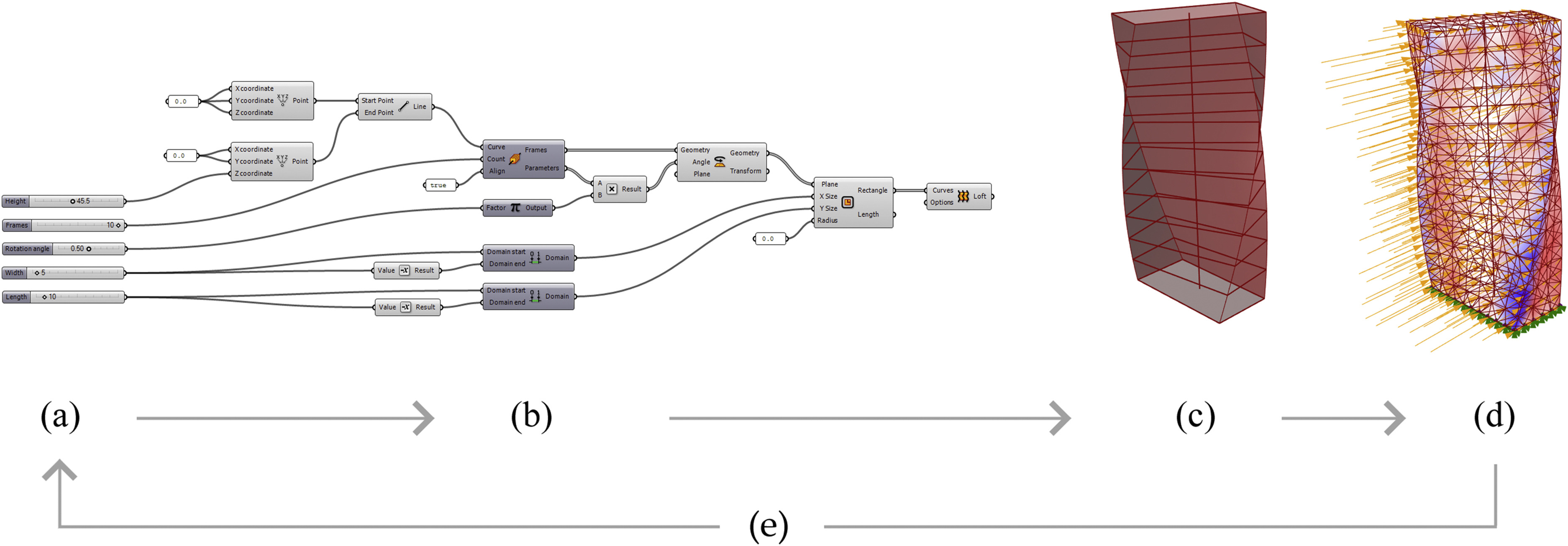}
	\centering
	\caption{\label{fig:parametric-modelling-cycle} A typical performance-based design exploration process where parameters (a) drive an \gls{ad} definition (b), that generates a geometrical representation (c) on which performance assessment tools can be run (d), so as to adjust parameter values (e) and iterate over the design. Reproduced from \citep{harding2017meta}.}
\end{figure}

The general concept of the \gls{ad} paradigm, consisting of generating geometry through an algorithm, already leads to a fundamental change in the way architects interact with their design, since they mentally switch from ``modelling an object'' to ``modelling the logic that generates an object''. The parametric nature of \gls{ad} definitions allows the designer to explore different solutions by adjusting these parameters. By integrating performance assessment tools into \gls{ad} software, the architect becomes able to iterate over parameter values until satisfied by both the generated design itself and its computed performance (e.g., via thermal or airflow analyses). Figure \ref{fig:parametric-modelling-cycle} represents a typical instance of such a performance-based design exploration process, likely for an airflow analysis on a building, created with Grasshopper.

It is even possible to go one step further and rely on artificial intelligence and optimisation techniques to close the loop, i.e., to let such techniques fix or improve at least part of the algorithmic (logic) or parametric (values) definitions. The process of using such techniques to explore solutions and find the best designs is called \textbf{\gls{dse}}.

The output of such techniques are typically optimised according to one or more target metrics, and it would consequently be easy to assume that the geometries that have been generated that way would be used as the final result. However, while this may indeed be the case for certain design teams, others only use these ``optimal'' models as starting points over which they can then iterate \citep{bradner2014parameters}.

It is important to realise that, even when only parameter values are being adjusted by such techniques, the design space very quickly becomes immense due to its combinatorial nature, since $n$ parameters with $m$ possible values each leads to $m^n$ possibilities. Even if $m$ is finite, $n$ leads to an exponential growth and it therefore rapidly becomes impossible to explore the whole design space in a reasonable time, when many parameters are involved.

For this reason, \gls{dse} and optimisation-based approaches often rely on approximation algorithms, such as metaheuristics \citep{talbi2009metaheuristics} (general procedures attempting to find a good solution without exploring the whole solution space). While a large body of work on \gls{dse} is linked to microprocessors and integrated circuits in general \citep{xie2006design}, similar approaches have been used for architectural design activities, and can help focus on a given criterion or a combination of criteria, using multi-objective optimisation techniques.

\subsubsection{Space layout}

\begin{figure}[h]
	\includegraphics[width=.8\textwidth]{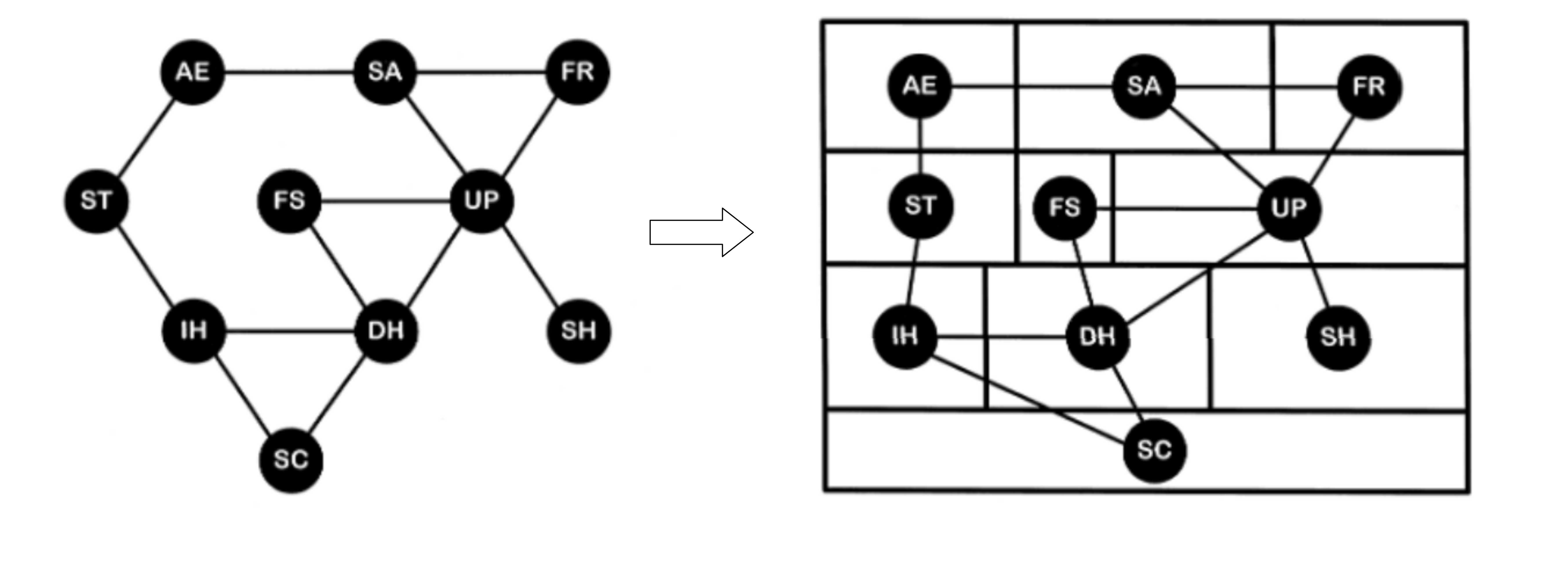} 
	\centering
	\caption{\label{fig:graph-layout} Generation of a block layout based on a technique using planar graph. Reproduced from \citep{liggett2000automated} as per \citep{goetschalckx1992interactive}.}
\end{figure}

A common application of these techniques is in space layout, where they produce automated facility, office, or housing arrangements. Methods based on single-parameter optimisation techniques (e.g., to minimise the cost of product flow between departments \citep{buffa1964allocating}) or graph theory (e.g., constructing a planar graph with a hexagonal structure to generate a rectangular block plan \citep{goetschalckx1992interactive}, as shown on Figure \ref{fig:graph-layout}) produce satisfying results on some instances of the space layout problem, but more complex techniques are necessary when multiple constraints or criteria are involved \citep{liggett2000automated}.

More recently, evolutionary algorithms have gained a lot of popularity for space layout problems, especially \textbf{\glspl{ga}} \citep{calixto2015literature}. This is likely due to two main factors working in their favour: their ability to handle multiple objectives and their inherent capacity to generate multiple solutions that the designer can choose from. In fact, an evolutionary algorithm naturally generates a new population of solutions at each iteration. That population results from crossovers and mutations on the best solutions (according to a fitness function that can accommodate for many parameters) from the previous generation.

Many examples of uses for \glspl{ga} solving space layout problems can be found in the previously cited \citep{calixto2015literature} as well as in \citep{turrin2011design} or even in commercial solutions such as Spacemaker\footnote{\url{www.spacemakerai.com}}.
Such tools can accommodate a plethora of criteria, such as energy efficiency \citep{wright2001simultaneous}, thermal comfort \citep{chen2008study}, or adjacency preference \citep{nagy2017project}. An extension to the latter has incorporated survey results to the generative process \citep{villaggi2018survey}. In fact, they integrated user satisfaction based on a questionnaire to that process, in order to enable certain occupant-level goals, that are usually disregarded, to be used.

\begin{figure}[h]
	\includegraphics[width=.9\textwidth]{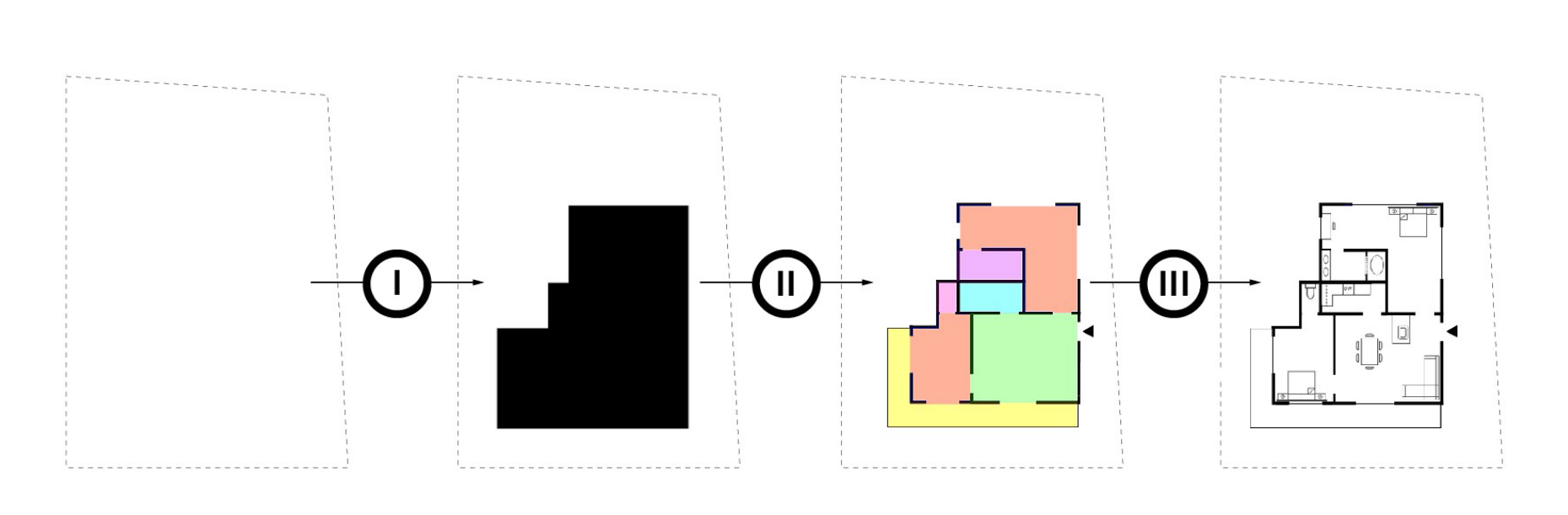}
	\centering
	\caption{\label{fig:floorplan-three-steps} Generation of a floor plan using a three-stage pipeline. Reproduced from \citep{chaillou2019ai}.}
\end{figure}

Other techniques based on machine learning, and neural networks in particular, have been used for similar purposes, often based on \textbf{\glspl{gan}}, a technique first proposed in \citep{goodfellow2014generative} that involves two competing neural networks: one that generates candidate solutions while the other evaluates them. \glspl{gan} have been used to generate bedroom configurations \citep{radford2015unsupervised} as well as floor plans, based on coloured drawings that represent areas with different functions \citep{huang2018architectural}, or simply based on a given parcel used as input for a three-stage pipeline \citep{chaillou2019ai} (building footprint outlining, room splitting, furnishing) represented in Figure \ref{fig:floorplan-three-steps}.

\subsubsection{Morphogenesis}
Another modelling application for \gls{dse} techniques is morphogenesis, i.e., the generation of form. 
For instance, \citep{caldas2003genetic} proposed to use \glspl{ga} to produce building envelopes, based on factors such as ventilation, lighting and construction costs.
Similar work includes envelope generation that searches for the optimal shape of a tower, to maximise solar radiation while satisfying geometric constraints \citep{besserud2008architectural}, or to provide the best possible views \citep{doraiswamy2015topology}.

Morphogenic approaches based on deep learning have also been proposed, with \citep{as2018artificial} exploring the use of different types of neural networks to generate houses.

Smaller parts of a design are also covered by GA-based approaches, with \citep{turrin2010performance} addressing the generation of roofs that offer good daylight and thermal comfort, \citep{choi2014evaluation} designing louvres based on thermal performance, or \citep{ercan2015performance} exploring shading devices to be placed on facades to optimise daylight while avoiding excessive solar heat gain.

Although generally rather linked to engineering than architectural design, many researchers have integrated these techniques to structural design (i.e., the creation of stable structures capable of resisting loads and various forces being applied on their parts). For example, \citep{kaveh2008structural} relied on an ant colony algorithm (a metaheuristic that is inspired by the behaviour of ants using pheromones to converge towards better solutions) to generate stiff structures given a certain quantity of materials. 

A more interdisciplinary work was proposed in \citep{miles2001conceptual}, whose authors created a system to be used by both engineers and architects. That system uses a \gls{ga} to produce floor plans and determine the layout of columns based on criteria such as lighting or ventilation, but also structural performance. Similarly, \citep{mueller2015combining} combined structural performance with designer preferences using an evolutionary algorithm.

The latter example integrates an interactive aspect, since the designer selects the best solutions for each iteration of the evolutionary algorithm, thereby ``piloting'' the exploration of the design space. This is also the case for \citep{marin2012creativity} who have used so-called Interactive \glspl{ga}, to produce buildings that both rely on energetic performance optimisation and designer preferences. A similar approach has also been proposed for urban landscape design \citep{koma2017research}. Commercial solutions integrating such concepts are also available, such as One Click LCA\footnote{\url{www.oneclicklca.com/fr/parametric-and-generative-carbon-optimisation}} that can allow Grasshopper designers to minimise the carbon footprint of a building \citep{apellaniz2021holistic}.

\subsection{Use of immersive technologies}
\label{sec:sota-cad-immersion}

In addition to these automated approaches to generate designs, innovative ``input'' interaction devices have also increasingly been integrated to \gls{cad} software tools.
This includes hand or body tracking devices that enable gestural interfaces for conceptual design \citep{khan2019gesture} or even more advanced modelling activities \citep{dave2013gesture}, but tangible user interfaces have also been created to facilitate collaboration as part of architectural design activities \citep{gu2011technological}, for various purposes such as daylight simulation control \citep{nasman2013evaluation}, or simply to move a geometry \citep{abdelmohsen2007tangicad}. 

This section will rather focus on advanced ``output'' devices and immersive displays in particular.

As mentioned in Chapter 1, the current use of \gls{ar} and \gls{vr} technology in the \gls{aec} sector is mostly limited to visualisation purposes \citep{blach1998flexible}, often with marketing goals in mind, i.e., to help convince and sell a product to a client \citep{juan2018developing}. This ability of immersive displays to facilitate the understanding of three-dimensional volumes for laymen is indeed one of the main benefits that has been pointed out. 

It has in fact been demonstrated that such technologies improve the understanding of objects \citep{louis2020high} and space \citep{schnabel2003spatial}. Additionally, several studies \citep{johnson1963sketchpad, paes2017immersive} have shown that the spatial perception is even enhanced as the immersion deepens. Therefore, as envisioned and described in \citep{brooks1987walkthrough}, immersive walkthroughs using these technologies favour the dialogue between stakeholders \citep{CoroadoVIARMODESVisualizationinteraction2015, lodesigning} (e.g., involving designers, constructors and clients), leading to better decision-making and easier detection of issues \citep{RoupeImmersivevisualizationBuilding2016}, and helping to move towards user-centred design, leading to better (accepted) designs \citep{mobach2008virtual}.

Design software usually comes with features to export the geometry being worked on. Most major CA(A)D tools in fact support exporting to the OBJ and/or FBX formats, as they are amongst the most common options found on the market. Both formats can be used as a basis to create an immersive experience quite easily with popular game engines such as Unity\footnote{\url{www.unity.com}} and Unreal Engine\footnote{\url{www.unrealengine.com}}. With the growth of the demand for immersive productions, a number of tools were developed to facilitate and (semi-)automate the creation of such experiences, especially for \gls{vr}. Some of these tools allow users to have control over environmental factors (sun position, daylight intensity) or even provide access to extrinsic properties of the rendered design, such as texturing, scaling, or positioning options. However, most of the currently available tools do not enable changes on the form itself and should rather be seen as inspection or showcasing tools.

\begin{figure}[h]
	\includegraphics[width=.8\textwidth]{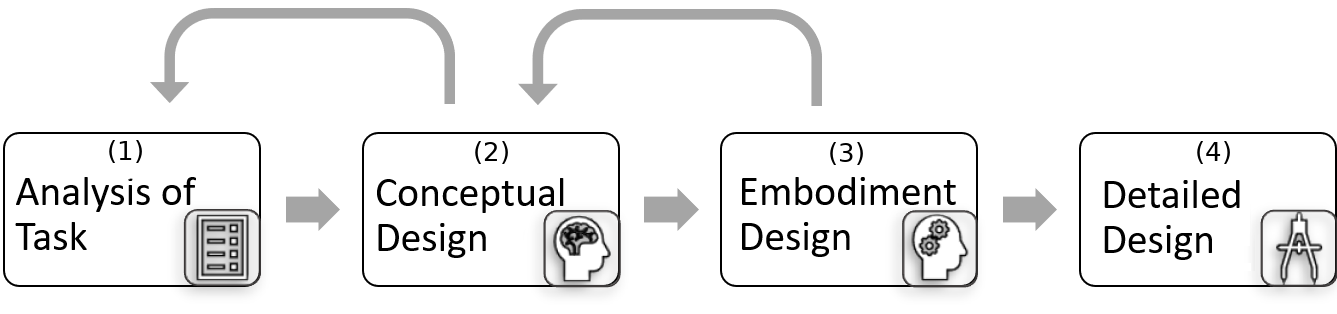}
	\centering
	\caption{\label{fig:howard-design-process} The stages of the design process, according to \citep{howard2008describing}.}
\end{figure}

Design activities tend to follow a well-known sequence of processes. Based on a literature review in both engineering design and cognitive psychology, \citep{howard2008describing} describes the usual design process as a 4-stage endeavour: (1) analysis of the task at hand, (2) conceptual design, (3) embodiment design, and (4) detailed design. That process, illustrated in Figure \ref{fig:howard-design-process}, also allows designers to go back to previous stages until they reach stage (4). The remaining of this section will discuss existing work on modifications to the model itself, at different stages of the design process.

\subsubsection{Sketching and sculpting}

During the conceptual phase, a designer explores different ideas and traditionally relies on hand drawings and scale models. Their digital and three-dimensional equivalents are free-form sketching and what is sometimes called sculpting, often based on voxels, that allow users to spawn or remove material at the pointer's location. The corresponding tools are not sufficient to produce actual \gls{cad} models, but provide the opportunity to explore design options in 3D environments during the early phases of the process.

Early examples include \citep{kameyama1997virtual} that is based on a virtual clay metaphor allowing the designer to add or remove some of the clay, and the more advanced \citep{wesche2001freedrawer}, based on line drawings but with support for creating surfaces based on these lines. Both of these systems rely on 3D-tracked controllers and stereoscopic \glspl{hmd} used at the time.

Contemporary commercial applications following similar approaches but involving modern \gls{hmd} and input devices have been released, including Gravity Sketch\footnote{\url{www.gravitysketch.com}} and Google Tilt Brush\footnote{\url{www.tiltbrush.com}}, that was selected in \citep{AroraExperimentalEvaluationSketching2017} to evaluate sketching in \gls{vr} with a surface to support the drawing.

\begin{figure}[h]
	\includegraphics[width=.7\textwidth]{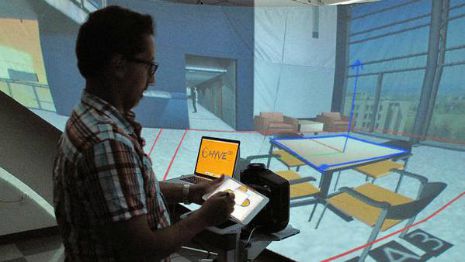}
	\centering
	\caption{\label{fig:hyve-3d} The Hyve-3D \gls{vr} sketching system. Reproduced from \citep{DortaHyve3D3DCursor2016}.}
\end{figure}

As for research contributions more focused on architectural design, \citep{donath1995vrad} presented voxDesign \citep{donath1995vrad}, a \gls{vr} sketching tool for early phases of the design process relying on a 3D-tracked controller, while \citep{DortaHyve3D3DCursor2016} more recently proposed Hyve-3D, another conceptual design system that suggests using a tablet to control a 3D cursor that moves around and enables sketching, as depicted on Figure \ref{fig:hyve-3d}.

\subsubsection{Direct modelling}
\begin{figure}[h]
	\includegraphics[width=.9\textwidth]{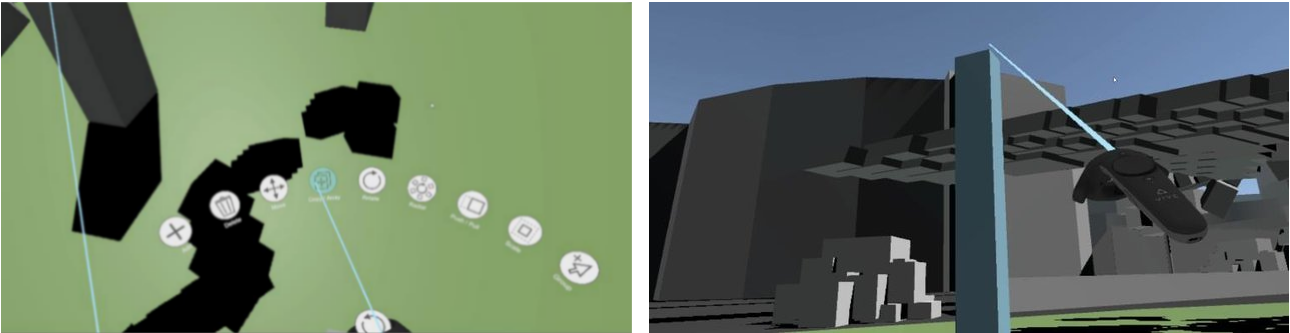}
	\centering
	\caption{\label{fig:innes} A \gls{vr} modelling system that allows the user to create basic shapes and manipulate them with a selection of tools. Reproduced from \citep{innes2017virtual}.}
\end{figure}

Slightly more advanced modelling capabilities were added to solutions such as \citep{Butterworth3DMthreedimensionalmodeler1992}, \citep{Billinghurst3DPalettevirtual1997} and \citep{ChuMultisensoryuserinterface1997}, that all enable basic shapes to be created from a \gls{vr} application. Other systems were proposed more recently with architectural modelling in mind during development, including \citep{mine2014making} and \citep{innes2017virtual}, illustrated in Figure \ref{fig:innes}.

Commercial applications replicating such an approach are Microsoft Maquette\footnote{\url{www.maquette.ms}} and Google Blocks\footnote{\url{arvr.google.com/blocks/}}, that both offer the essential functionality to create low-complexity geometries from a VR-based environment.

However, the aforementioned tools lack support for some usual \gls{cad} features and cannot truly be considered as the \gls{vr} equivalents or at least the companions to modern desktop-based tools, and are therefore confined to a conceptual design role.

More advanced \gls{vr} modelling systems that integrate with popular \gls{cad} software have been developed, including the MARUI\footnote{\url{www.marui-plugin.com/marui4}} plug-in for Maya\footnote{\url{www.autodesk.com/products/maya}}, and Mindesk\footnote{\url{https://mindeskvr.com/}}, compatible with Rhinoceros\footnote{\url{www.rhino3d.com}}. Besides enabling interoperability between a \gls{vr} interface and a desktop \gls{cad} application, these solutions offer more complex design features than the systems we previously mentioned, e.g., to manipulate \gls{nurbs} curves.

\subsubsection{Algorithmic Design}
\label{sec:sota-immersive-ad}

Algorithmic Design is also covered by immersive solutions, to a certain extent. Generating a \gls{vr} (or even \gls{ar}) experience can be done quite easily using a game engine but, in order to do so, one must typically first export the design to a specific format to allow the game engine to load the corresponding file and import it into a virtual world.
Some tools facilitate that process, such as Twinmotion\footnote{\url{www.unrealengine.com/twinmotion}} and IrisVR Prospect\footnote{\url{www.irisvr.com/prospect}} by providing ``one-click'' exports from well known 3D modellers, such as Rhinoceros and Revit\footnote{\url{www.autodesk.com/products/revit}}), to a \gls{vr} environment that includes the up-to-date geometrical representation. They therefore do smooth out the burden of creating such experiences, but still provide limited control over the design artefact itself (what we called extrinsic properties in Section \ref{sec:sota-cad-immersion}). Such tools can be used with some \gls{ad} editors, since they generally rely on --- or easily connect to --- general modelling systems that these tools support. But their usage remains limited to visualisation purposes. 

\begin{figure}
	\includegraphics[width=.7\textwidth]{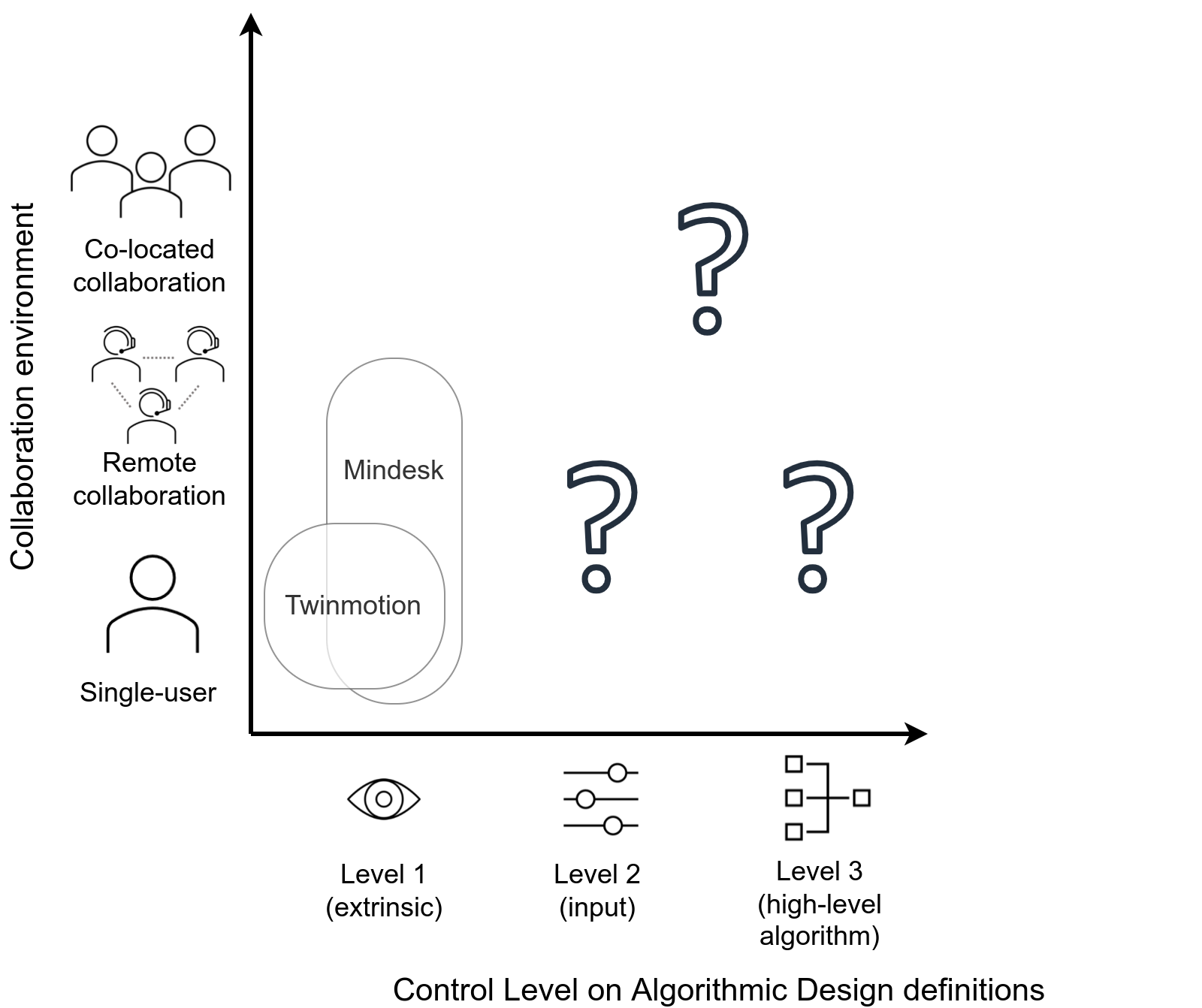}
	\centering
	\caption{\label{fig:prototypes-comparison-control-collab-questions} Immersive \gls{ad} landscape in 2017: collaboration scenarios and levels of control (defined in details in Section \ref{sec:ad-requirements}).}
\end{figure}

These types of tools only provide extrinsic control over the designed model, but deeper control over \gls{ad} definitions is possible and will be discussed in the remaining of the present section, then further described in Section \ref{sec:ad-requirements}. Figure \ref{fig:prototypes-comparison-control-collab-questions} clarifies what the state of the practice was at the start of our research work (in 2017), with regards to immersive control over \gls{ad} models, and immersive collaboration scenarios.

Some proposals however enable manipulation of key control points employed by an \gls{ad} algorithm, e.g., \citep{kwiecinski2017participatory} that uses tangible items as part of an \gls{ar} experience. Other systems allow \gls{vr} users to remove elements from a structural analysis calculation, from a \gls{vr} environment that highlights geometry parts that do not meet pre-specified requirements \citep{Moubilerealtimeparametricstructural2018}. These types of interactions can be subsumed as input control, since they do not interact with the internals of the algorithm but allow for changes to the entities it uses.

\begin{figure}[h]
	\includegraphics[width=.6\textwidth]{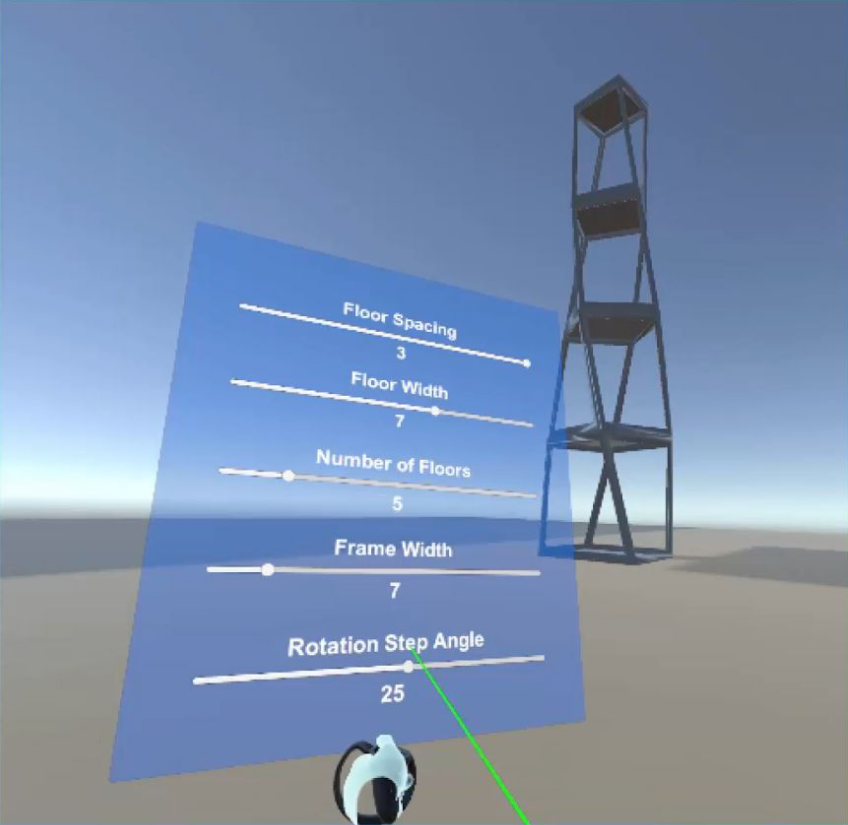}
	\centering
	\caption{\label{fig:hawton} A \gls{vr} application that allows for parameter adjustments to an \gls{ad} definition. Reproduced from \citep{hawtonshared}.}
\end{figure}

Another set of tools that fall into the same category lets users adjust an \gls{ad} algorithm's parameter values. Such immersive adjustments are made possible in a prototype presented in \citep{hawtonshared}, depicted in Figure \ref{fig:hawton}, that relies on the Oculus Rift\footnote{\url{www.oculus.com/rift}} \gls{vr} \gls{hmd}. Such a feature is also covered by \citep{Moubilerealtimeparametricstructural2018} that works with the same headset. In both cases, the user is presented with a panel, attached to a standard \gls{vr} controller (tracked in 3D). The panel contains a list of parameters, coming from a Grasshopper definition, whose values can be changed. They both support number parameters, tweaked through the manipulation of sliders. Changes made to the parameter values are sent back to Grasshopper so as to modify the generated geometry whose updated version is, in turn, relayed back into the \gls{vr} environment.

A commercial solution, Fologram\footnote{\url{www.fologram.com}}, includes the same functionality but targets \gls{ar} displays such as Microsoft's Hololens\footnote{\url{www.microsoft.com/hololens}} and can even work with simple smartphones. 
The user can therefore visualise a geometrical representation while manipulating the parameters driving the (re)generation of that geometry, with no need to leave the augmented experience to do so.
Note that these tools are directly related to the prototype we will describe in Section \ref{sec:ecaade}; they were developed simultaneously yet result from entirely independent works.

\begin{figure}[h]
	\includegraphics[width=.6\textwidth]{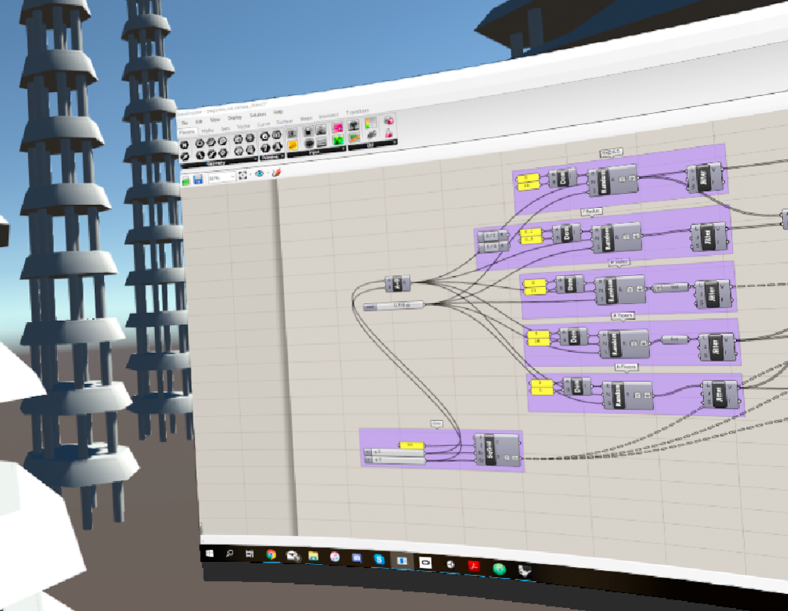}
	\centering
	\caption{\label{fig:castelo} A mirror view of the desktop's display, incorporated into a VR-based environment. Reproduced from \citep{castelo2020inside}.}
\end{figure}

The aforementioned solutions cannot be used to add or remove components, nor do they allow to interact with the links between these components. To overcome that limitation, researchers have recently started working on adding control over the model itself. A \gls{vr} proposal described in \citep{castelo2020inside} relies on desktop mirroring i.e., the \gls{vr} user has access to a ``window'' that mirrors the view of the computer running Grasshopper. In order to interact with that window, one of the \gls{vr} controllers ``simulates'' a standard desktop mouse, with a ``point and click'' approach. Figure \ref{fig:castelo} shows that mirror view integrated into the \gls{vr} environment.
Mindesk, a commercial software that we mentioned previously, was updated to provide the same functionality following the same approach of mirroring the computer screen.

Providing a mirror view of the desktop interface coupled with simulated mouse and keyboard input gives the user access to the same feature set as within the desktop tool itself. Nevertheless, interacting that way does not take full advantage of the 3D-tracked controllers and leads to difficulties when small items have to be selected or more generally when precision is required.
Despite predating the aforementioned work, the prototype we describe in Chapter \ref{chap:immersive-design} is an attempt at tackling that interaction problem.

\subsection{Survey on the potential of Virtual Reality for architectural design}
\label{sec:survey}
Based on the state of the art we discussed in this chapter, we posit that immersive technologies should be integrated as part of the \gls{caad} process, and for \gls{ad} in particular. To verify these claims, we conducted an online survey on the potential of using \gls{vr} for architectural design.

The survey was shared in January 2020 to practitioners on the Rhinoceros fora\footnote{\url{https://discourse.mcneel.com/t/your-opinion-on-vr-for-architectural-design/95111}}, researchers on the official eCAADe LinkedIn and Facebook pages, as well as architecture students from three French-speaking universities, and was made available in two languages (French and English) in order to target and solicit a broader audience.

This section discusses the results we got out of that survey, even though one should keep in mind that only 36 complete responses were received. This is probably good enough to provide anecdotal evidence but is not sufficient to capture statistically significant evidence. A copy of the complete questionnaire and the results we gathered can be found online\footnote{\url{zenodo.org/record/4696074)}}. The code we used to analyse that data and produce figures is also available\footnote{\url{zenodo.org/record/4696071)}}. We also published a paper that expands on the results we obtained \citep{coppens2021integrating}.

\subsubsection{Population}

\begin{figure}
	\centering
	\begin{subfigure}[t]{0.48\textwidth}
		\centering
		\includegraphics[width=\textwidth]{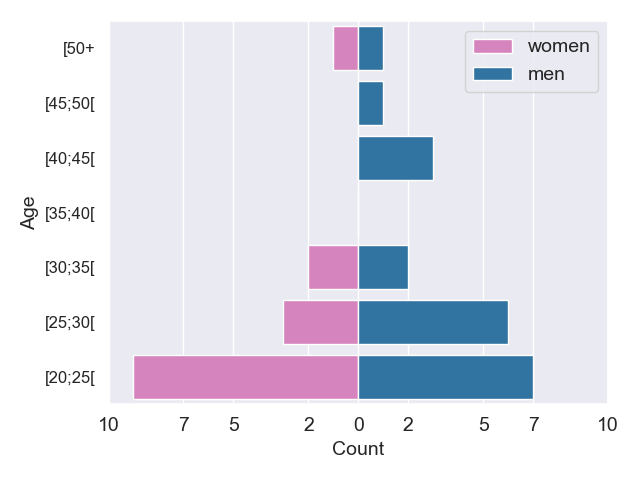}
		\caption{\label{fig:cib-age} Respondents' age pyramid.}
	\end{subfigure}
	\hfill
	\begin{subfigure}[t]{0.48\textwidth}
		\centering
		\includegraphics[width=\textwidth]{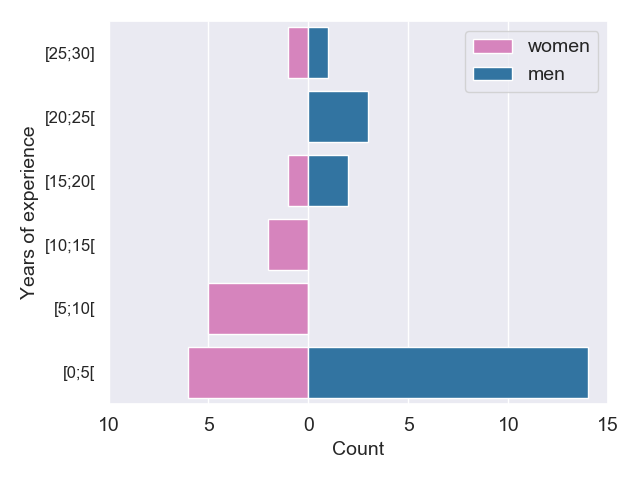}
		\caption{\label{fig:cib-xp} Distribution of experience amongst respondents.}
	\end{subfigure}
	\caption{Profile of survey respondents.}
\end{figure}

By targeting diverse venues, we managed to obtain a population of respondents with different profiles. Figure \ref{fig:cib-age} shows the age distribution amongst the respondents, highlighting that 25 out of 36 participants (69\%) were between 20 and 29 years inclusive (median: 25, interquartile range: 9.25). Figure \ref{fig:cib-xp} presents the distribution of experience amongst respondents through a population pyramid reporting the number of years of experience in architecture or related fields (median: 3, interquartile range: 8.5). These two figures seem to suggest that a significant part of the surveyed population is comprised of students and recently graduated practitioners.

Based on their reported architectural experience and diplomas, we classified respondents into four distinct architectural profiles: 2 \emph{uninitiated} respondents, 11 \emph{novices}, 11 \emph{competent} participants, and 12 \emph{experts}. We observed a balanced grouping of respondents in the three categories with the most qualified profiles, as well as a good gender balance across these categories.

\begin{figure}[h]
	\includegraphics[width=.6\textwidth]{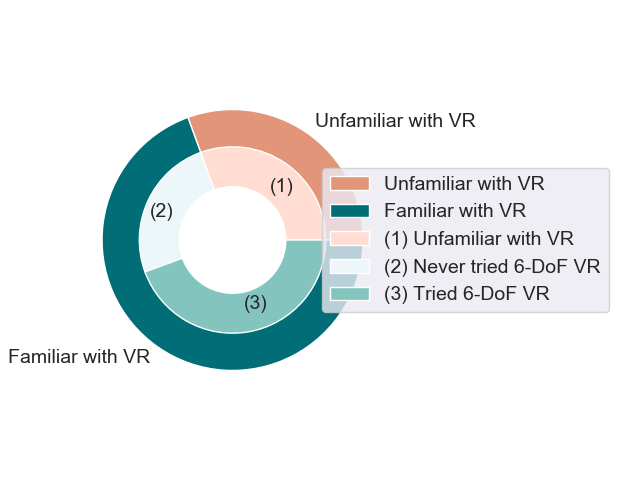}
	\centering
	\caption{\label{fig:cib-vr} Familiarity of respondents with different quality levels of \gls{vr}.}
\end{figure}

We also wanted to assess the respondents' familiarity with \gls{vr} as well as their experience with the technology, by asking what devices they had used. This allowed us to separate between three types of respondents: (1) those without \gls{vr} experience; (2) those with prior exposure to ``low-quality'' \gls{vr} devices (rotational tracking only); and (3) those that had been able to try fully tracked (\gls{6dof}) \gls{vr} experiences. Figure \ref{fig:cib-vr} depicts the corresponding distribution, with 11 respondents stating they were not familiar with \gls{vr} at all (1), 9 belonging to type (2) since they were familiar with \gls{vr} but had not tried a \gls{6dof} experience, and 16 with \gls{6dof} \gls{vr} experience (3).

\subsubsection{Potential of Virtual Reality for architecture}

The core focus of the questionnaire was to assess the current use and the potential of using \gls{vr} for architectural design practices. Respondents that had indicated prior exposure to \gls{vr} were asked whether they had tried to use such devices for architecture-related activities. Among respondents that indicated having tried \gls{vr}, more than half (11 out of 21) had also done so for architecture-related activities. For them, we queried which tools they had used and what limitations they encountered when using them. 

Only 6 participants answered the question about the perceived limitations of architecture-specific \gls{vr} tools, but they mentioned tooling complexity (R18: ``work-intensive transition from regular \gls{cad} model to VR''; R17: ``difficult to set a proper scale for the imagery''), issues with user interfaces or interactions (R27: ``lack of easy-to-use interface''; R56: ``limited interactions''), as well as collaboration needs (R27: ``it gets kind of lonely in VR [...] on projects with multiple stakeholders, it takes a long time to present, because everybody wants to go in'') and hardware costs. That being said, more than half (6 out of 11) of these respondents mentioned game engines in the list of tools they used for architectural activities in \gls{vr}, which leads us to think they may not reasonably be aware of more ad-hoc and easy-to-use software solutions.

\begin{figure}[h]
	\includegraphics[width=.7\textwidth]{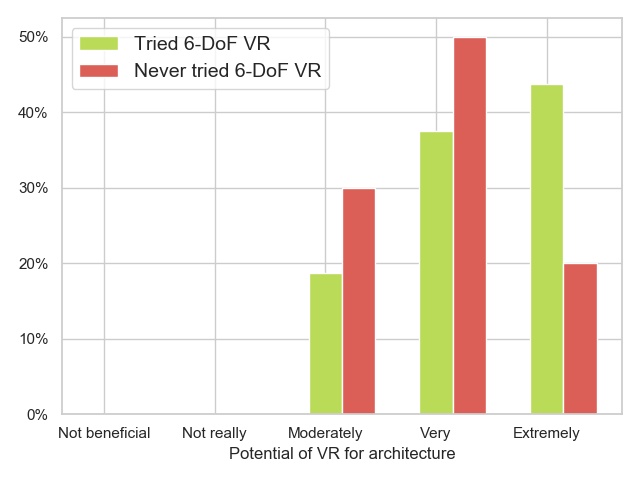}
	\centering
	\caption{\label{fig:cib-archi-potential} Perceived potential of \gls{vr} for architecture, depending on whether respondents had prior exposure to \gls{6dof} \gls{vr}.}
\end{figure}

We also asked all respondents about the potential of \gls{vr} for architecture-related activities, regardless of their previous experience with such tools and devices.
Figure \ref{fig:cib-archi-potential} shows the results we obtained for that question, contrasted with the respondents' prior experience with \gls{vr}. One can observe that as much as 75\% (25 out of 36) respondents indicated that they considered the technology could be at least very useful to the field, while the remaining 25\% still consider the technology to be moderately useful.

In addition to these general results, we drilled down into specific stages and actors of the architectural design process that are perceived as the most suitable targets for embracing \gls{vr}. We proposed four uses of the technology as possible answers for a multiple-choice question: (1) after the design process; (2) during the design process, for informing stakeholders other than the designers; (3) during the design process, to be used by the designer himself; and (4) right from the start and all along the design process.

Stage 1 covers what we identified as the most common contemporary use of \gls{vr} technology in the architectural context, i.e., to show a finished design to a client. Stage 2 suggests the potential to show stakeholders a work-in-progress in order to gather early feedback that can be taken into account for subsequent iterations. Stage 3 encompasses a workflow where the designer sporadically checks on a design in \gls{vr}, e.g., in order to evaluate the design at full scale and from a human point-of-view. Finally, Stage 4 represents the extreme case where \gls{vr} is fully integrated into tools that support all steps of the architectural design process, potentially replacing non-VR solutions.

\begin{figure}[h]
	\includegraphics[width=.6\textwidth]{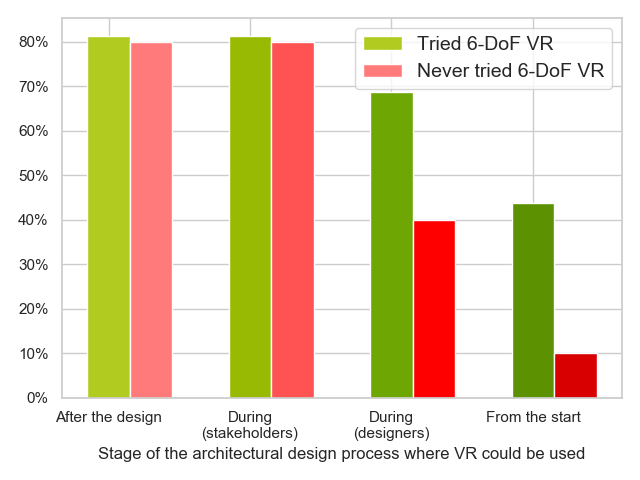}
	\centering
	\caption{\label{fig:cib-archi-stages} Stages of the Architectural Design process suitable for \gls{vr} integration.}
\end{figure}

Figure \ref{fig:cib-archi-stages} presents the answers received for that question, compared against prior \gls{vr} exposure. A large proportion of respondents (29 out of 36) consider \gls{vr} technology suitable for presenting a finished project, which supports the assumptions about the current use of the technology we derived from our literature review. More surprisingly, there are equally many respondents that indicate it could be used to involve stakeholders during the design process. Slightly more than half of the respondents (19 out of 36) believe designers themselves could use the technology during their architecture-related activities, while 9 respondents indicate they believe \gls{vr} tools could be used right from the start and all along the design process.

\subsubsection{Potential of Virtual Reality for Algorithmic Design}
The questionnaire then focused on how \gls{vr} could be used to support \gls{ad}. For this part of the survey, we only considered respondents that were aware of algorithmic design or parametric modelling tools. The vast majority of them (34 out of 36) signalled an awareness of such tools. Given the vague and confusing term ``parametric'', and in order to better appreciate the level of understanding of what the paradigm entails, we asked respondents ``How would you define parametric modelling/design?''. 

In a similar way to \citep{stals2018immersion}, we classified respondents based on their answer into three distinct categories: ``wrong definition'', ``correct definition'', and ``unclear definition''. While the first two categories are rather straightforward, the latter contains respondents that did mention the parametric aspect but whose proposed definition did not contain any reference to algorithms, programming or interlinked components, and did not mention that variations of parameter values produce chain reactions. 
Since such an incomplete definition means the presence of parametric objects (e.g., parametrised primitive shapes in traditional \gls{cad}) is enough to fill the bill, it would also apply to non-algorithmic software.

Out of 34 respondents that answered this question, only 12 gave a definition that we considered to be correct, while 18 persons provided a definition that we classified as being unclear.

\begin{figure}[h]
	\includegraphics[width=.65\textwidth]{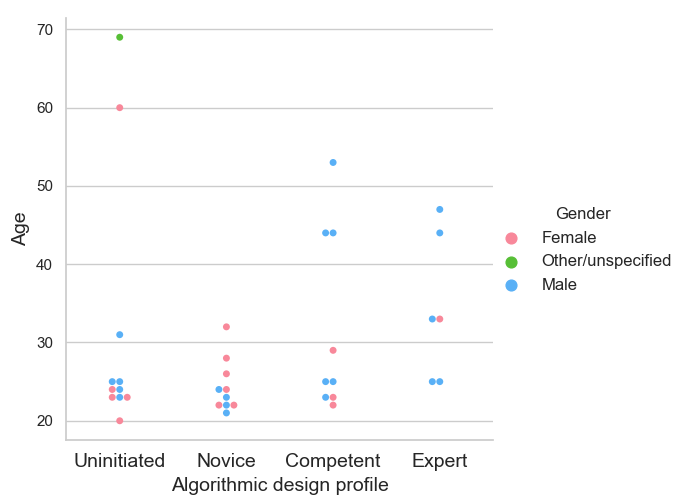}
	\centering
	\caption{\label{fig:cib-pmdef} Algorithmic Design profile of the respondents.}
\end{figure}

By combining the respondents' answer with their architectural profile, we classified them into four categories based on their \gls{ad} profile: \emph{uninitiated}, \emph{novice}, \emph{competent} and \emph{expert}. Figure \ref{fig:cib-pmdef} shows the distribution of these categories across respondents, in relation to their age.

Next, we queried respondents on the usefulness and desirability of integrating \gls{ad} software and \gls{vr} hardware. We first showed them an online video of the prototype we will describe in Chapter \ref{chap:immersive-design}. This prototype demonstrates the feasibility of editing a Grasshopper algorithm in \gls{vr} with an interface adapted to the three-dimensional immersive context. It should however be noted the responses are therefore based on the respondents' subjective perception derived from a video that demonstrates a prototype that does not intend to provide a consumer-grade experience. Exacerbated by the fact that the video has to be watched on a 2D screen that does not generate the same immersion as an actual VR-based test, some of the responses may have been negatively impacted.

\begin{figure}
	\centering
	\begin{subfigure}[t]{0.48\textwidth}
		\centering
		\includegraphics[width=\textwidth]{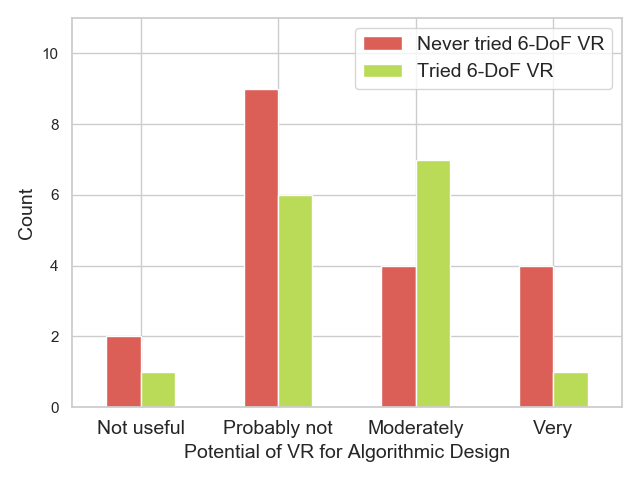}
		\caption{Usefulness of \gls{vr} for \gls{ad}, broken down by respondents' prior exposure to \mbox{\gls{6dof}} \gls{vr} devices and experiences.}
		\label{fig:survey-ad-vs-tried-vr}
	\end{subfigure}
	\hfill
	\begin{subfigure}[t]{0.48\textwidth}
		\centering
		\includegraphics[width=\textwidth]{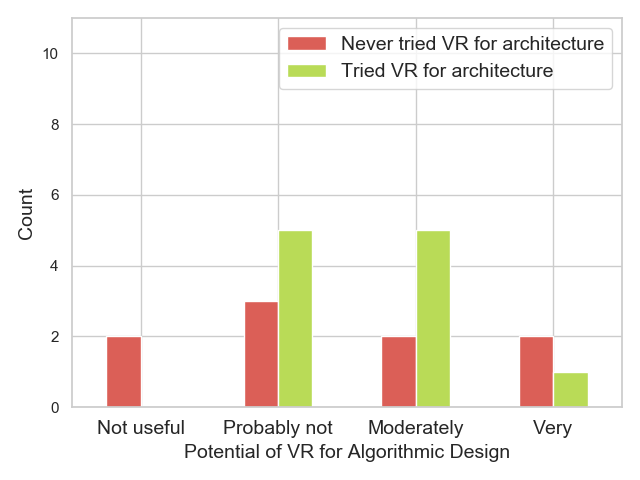}
		\caption{Usefulness of \gls{vr} for \gls{ad}, broken down by respondents' prior exposure to \gls{vr} tools for architecture (if they were familiar with \gls{vr}).}
		\label{fig:survey-ad-vs-tried-vr-archi}
	\end{subfigure}
	\caption{Usefulness of \gls{vr} for Architectural Design.}
	\label{fig:survey-ad}
\end{figure}

Nevertheless, after viewing the video, each respondent was presented with a question on the usefulness of similar VR-enabled functionality if it were combined with a 3D visualisation of the geometry being worked on. We identified an interest in having \gls{vr} tooling adapted to the \gls{ad} paradigm, but not nearly as pronounced as the one for \gls{caad} in general. In fact, as illustrated on Figure \ref{fig:survey-ad}, almost half of the respondents (16 out of 34) consider \gls{vr} functionality as moderately or very useful for \gls{ad}. The two sub-figures compare these answers with prior exposure of respondents to \gls{6dof} \gls{vr} on the one hand (\ref{fig:survey-ad-vs-tried-vr}) and to \gls{vr} tools for architecture on the other hand (\ref{fig:survey-ad-vs-tried-vr-archi}). Note that the latter only takes 20 answers into account, since we did not include respondents that indicated they were not familiar with \gls{vr} or had never experienced the technology. Figure 14 does take those respondents into account.

The proposed prototype was slightly better received amongst respondents with prior exposure to \gls{6dof} \gls{vr} and \gls{vr} tools for architecture, respectively with 54.5\% (6 out of 11)
and 53.3\% (8 out of 15) of the answers that consider it to be at least ``moderately useful'' within these subgroups of the population. Although we notice that slight difference, the number of respondents is too low to draw any statistically significant conclusions.

\subsubsection{General outcome}
As a whole, the survey results confirm our assumption that the \gls{vr} technology should be used earlier in the design process. A very large number of respondents see potential for a more user-centric approach that would involve stakeholders during the design process. Although the demand for \gls{vr} editing tools is not as high (for architecture in general and for \gls{ad} more specifically), it still is there. 

Despite the threats to validity we have to take into account (especially the population size that prevents us from drawing statistically significant conclusions), we consider the findings we derived from this survey to provide sufficient anecdotal evidence of the need for better immersive tooling in architectural practice that goes beyond the current usage. Architects need solutions to be able to edit and evaluate designs from an immersive context, with adapted user interfaces and interactions.
\chapter{Immersive parameter adjustment for algorithmic co-design}
\label{chap:immersive-adjustment}
\epigraph{
	``A display connected to a digital computer gives us a chance to gain familiarity with concepts not realizable in the physical world. It is a looking glass into a mathematical wonderland.'' }{Ivan Sutherland}

As stated in Chapter 1 and as supported by the survey discussed in Section \ref{sec:survey}, we posit that immersive technologies should be integrated into the architectural design process and to \gls{ad} in particular. To address this need, we follow a prototype-based process described in Section \ref{sec:methodology} and therefore develop research prototype tools that show the potential of using immersive technologies for \gls{ad} activities.

Based on our literature review and the state of the art of the various domains our research covers, we start working on how to bring immersive technologies to \gls{ad} in architecture. In this chapter, we explore different display options (\gls{vr} and \gls{ar}) and interaction modalities (based on desktop-like interfaces adapted for \glsfirst{6dof}, but also using \textbf{\glspl{tui}}) and focus on immersive geometry visualisation and immersive value adjustment of parameters shared with an \gls{ad} definition.

\pagebreak
\section{Requirements for creating immersive Algorithmic Design tooling}
\label{sec:ad-requirements}

The most essential aspect to take advantage of immersive technologies is to provide a way for an \gls{ad} project to be visualised in an immersive manner, preferably with a navigation system so as to be able to examine the rendered architectural geometry from different angles.

Since we do not aim to replace desktop-based tools such as Grasshopper entirely, our goal is to interact with existing \gls{ad} definitions created with such tools. Two options are available to do so. 

The first one is implementing a way to read from and write to a specific \gls{ad} format (e.g., Grasshopper's file format) from an immersive application, to process a given definition. We can then develop a module that generates the geometrical representation based on that definition. 

Another option is to create a ``bridge'', between an immersive application and the desktop-based tool, through a custom component for that desktop tool. The component would simply share geometry and parameter data to a companion immersive application. Generating the geometrical representation based on the definition remains the desktop tool's responsibility in that case; the custom component simply sends the result to the immersive application.

We opted for this latter option, since it does not force us to dive into the logic of how to generate geometries based on an algorithm. Moreover, it allows the resulting immersive application to be tool-agnostic, in the sense that any \gls{ad} software would be able to work with the immersive application, as long as a bridge plug-in is created for that specific software.

Ideally, the process of transferring the geometry from a \gls{caad} desktop-tool to the immersive application should satisfy two main properties: it should be fast so that the user benefits from near-instant updates on a modified geometrical representation, but also automatic, so that the designer does not have to go through a cumbersome process of exporting a geometry each time a new version is to be evaluated. One might also be tempted to aim for the update process to be incremental (working with modifications to the current state of the geometry instead of resending the whole geometry). Unfortunately, doing so would require a geometry-adapted diff tool \citep{dobovs20123d} that would inherently be imprecise (because the rendering-adapted representation of the geometry is itself imprecise) and would likely be computationally expensive. In the specific context of \gls{ad}, a particular change (e.g., a value change on a parameter) may impact the whole geometry, rendering such diff tool even less appropriate.

To go further than visualisation only, it is needed to add different kinds of interactions with the design, that we classify in 4 groups corresponding to different levels of control, as shown in Table \ref{tab:control-levels}.

We first differentiate between \textbf{extrinsic} and \textbf{intrinsic} control. \emph{Extrinsic} modifications to a project only impact the rendered geometry or its surrounding environment (e.g., lighting or weather). This correspond to a first level of control (\emph{Level 1}) and includes modifications to textures, or changes in zoom level. 

On the other hand, \emph{intrinsic} control actually allows for modifications to the \gls{ad} model itself. We further split intrinsic control into three more levels of control: input control (\emph{Level 2}), high-level algorithm (\emph{Level 3}) and low-level algorithm (\emph{Level 4}). These three levels allow designers to have control over the actual \gls{ad} definition, impacting the way the geometrical representation is generated. 

Input control is limited to the data used by the algorithm (e.g., parameter values and key or anchor points for certain components), while both algorithm control levels allow for modifications to the algorithm itself. 

High-level algorithm control provides the designer with the ability to manipulate the (often visual) algorithm that drives the geometry generation. This includes adding or removing components in a visual algorithm (as in Grasshopper) and making changes to the links between them. 

Low-level algorithm control allows designers and potentially software developers to modify the (often textual) code inside the components used by the high-level algorithm. This includes implementing custom components to perform project or company-specific geometric operations, or connecting to an external service to retrieve or share data.

A designer clearly needs at least \emph{Level 3} control to create an architectural project, but can be confined with \emph{Level 2} control when tweaking an existing project (therefore on a temporary basis). \emph{Level 4} is only really needed when the built-in components in the \gls{ad} tool do not suffice for specific needs, and when no custom component was created or shared by a third-party to cover these needs.

\begin{table}[]
	\resizebox{\textwidth}{!}
{
	\begin{tabular}{>{\centering\arraybackslash}p{.06\textwidth}|>{\centering\arraybackslash}p{.125\textwidth}|>{\centering\arraybackslash}p{.15\textwidth}|>{\centering\arraybackslash}p{.34\textwidth}|>{\centering\arraybackslash}p{.35\textwidth}}
		& Type of control & Target users & Description & Examples\\\hline
		Level 1 & extrinsic & designer and others & Making modifications to the rendered geometry only & textures, zooming, geometry positioning, lighting\\\hline
		Level 2 & input & designer only &  Making changes to data used by the algorithm generating the geometry & adjustment of parameter values, key point positioning\\\hline
		Level 3 & high-level algorithm & designer only & Editing the (generally visual) algorithm used to generate the geometry & adding links (edges), components (nodes), removing or grouping them\\\hline
		Level 4 & low-level algorithm & designer and developer & Developing (generally textual) code for custom components to be used for Level 3 control & performing geometric operations or connecting to external services
	\end{tabular}
}

\caption{\label{tab:control-levels} Types of control over Algorithmic Designs.}
\end{table}

We already discussed in Section \ref{sec:sota-cad-immersion} that \emph{Level~1} can be achieved with many existing tools. The current chapter presents our work towards enabling \emph{Level~2} control over an \gls{ad} project from within an immersive experience. Chapter \ref{chap:immersive-design} will present how we manage \emph{Level~3} control.

\emph{Level 4} is however outside of the scope of this dissertation. This is due to two reasons: (1) it typically involves textual code and typing and immersion do not currently mix very well \citep{grubert2018text}, and (2) custom component development is usually done through specialised development software, outside of Grasshopper. (2) means that integrating immersive technologies for \emph{Level 4} control would heavily complexify the development of a solution, since it would have to communicate with these specialised development tools and have them process changes to produce new versions of the custom component being modified. Even then, Grasshopper would need to be restarted to reload all components and thereby take into account changes made to the specific component that was just modified. (1) further implies that overcoming these obstacles are not even guaranteed to provide a truly useful interface.

To clarify some usage scenarios where \emph{Level~2} control is likely to be beneficial to the design process, we provide the following two hypothetical use cases, that we will use as an inspiration to develop our research prototypes.

\begin{usecase}{1}[] 
	\label{usecase-1}
	An architectural firm is in charge of the construction of a modern-looking clubhouse for a golf club. Based on the club's demands, the architects have designed a first version of the building that will welcome the 300 registered members. The design process in quite advanced and the building is expected to be close to its final form. The architects would like to get input from the clients (the club's board members), so they invite them to a \gls{vr} session where both the lead architect and board members visualise the virtual building in its current state. The \gls{vr} visualisation helps the clients understand what the actual clubhouse will look like. The architect guides the clients around the virtual environment and adjusts parameter values along the way to accommodate the remarks from the clients.
\end{usecase}

\begin{usecase}{2}[]
	\label{usecase-2}
	The municipal council of a city wants to create a public space in a disused area of 1,000 square meters, nestled amid a few apartments and shopping buildings. 
	They appointed an architectural firm that works with Algorithmic Design to do so and asked them to include a big sculpture to be placed somewhere in the centre of that space. The project is already well advanced in Grasshopper and the architects now want the opinion of the council to adjust parameter values. They therefore invite council members for a meeting at their office, that themselves invite representatives from commercial and residential buildings in the project's vicinity with them. Using the immersive and interactive system at the architects' office, all these stakeholders are able to visualise the project from different angles and can take an active part in the remaining (parameter adjustment) design decisions. The \gls{ar} and \gls{vr} visualisations help non-architects to better appreciate the dimensions of the sculpture and how it integrates with the surroundings.
\end{usecase}

\section{Prototype for adjusting models in Virtual Reality}
\label{sec:ecaade}

In order to enable \emph{Level~2} control over an \gls{ad} project, we developed a proof-of-concept research prototype, that serves as a bridge between a Grasshopper definition and a \gls{vr} headset. We therefore decided to call it \textbf{GHVRBridge}. We chose Grasshopper because it is very popular in both industry and research \citep{cichocka2017optimization}, and supports the creation of custom components through a \Csharp \, \gls{sdk}. As for the \gls{vr} headset, we chose the HTC Vive \gls{hmd} since it was, at the time, one of the very few options to offer good quality (including \gls{6dof} tracking) at an affordable price. To ensure generalisability of the approach, we rely on a cross-headset toolkit that should work with other devices as well, so as to handle different settings should the need arise.

\begin{figure}
	\includegraphics[width=1.45\textwidth, angle=90]{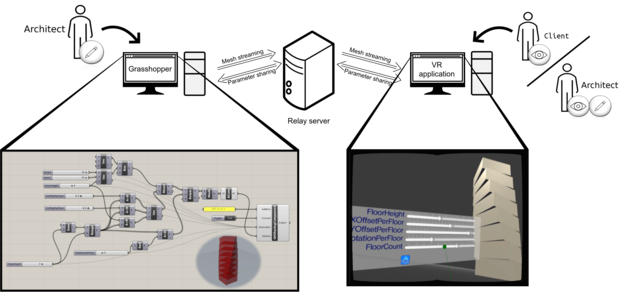}
	\centering
	\caption{\label{fig:ecaade-schema} GHVRBridge: a proof-of-concept research prototype that enables VR-based visualisation and adjustments of Grasshopper definitions.}
\end{figure}

A video demonstrating GHVRBridge is available online\footnote{\url{http://informatique.umons.ac.be/staff/Coppens.Adrien/?video=eCAADe2018}} and our codebase is hosted as open-source software on a GitHub repository\footnote{\url{https://github.com/qdrien/GHVRBridge}}, available under the MIT\footnote{\url{https://opensource.org/licenses/MIT}} license. The prototype was also described in a paper we published \citep{coppens2018parametric}.

Figure \ref{fig:ecaade-schema} presents the overall concept and structure of GHVRBridge, involving three separate software components in addition to Grasshopper itself: a Grasshopper component that shares geometries and parameters, a \gls{vr} application for the HTC Vive, and a relay server to forward the information that needs to be exchanged. The next two sections focus on the first two components only, since the relay server's functionality is rather self-explanatory. It allows for \gls{vr} users to be in a different location than the computer running Grasshopper, and will facilitate the later extension to a collaborative setting.

\subsection{Grasshopper custom component for external parameter value adjustment}
\label{sec:ecaade-gh}
\begin{figure}[h]
	\includegraphics[width=.9\textwidth]{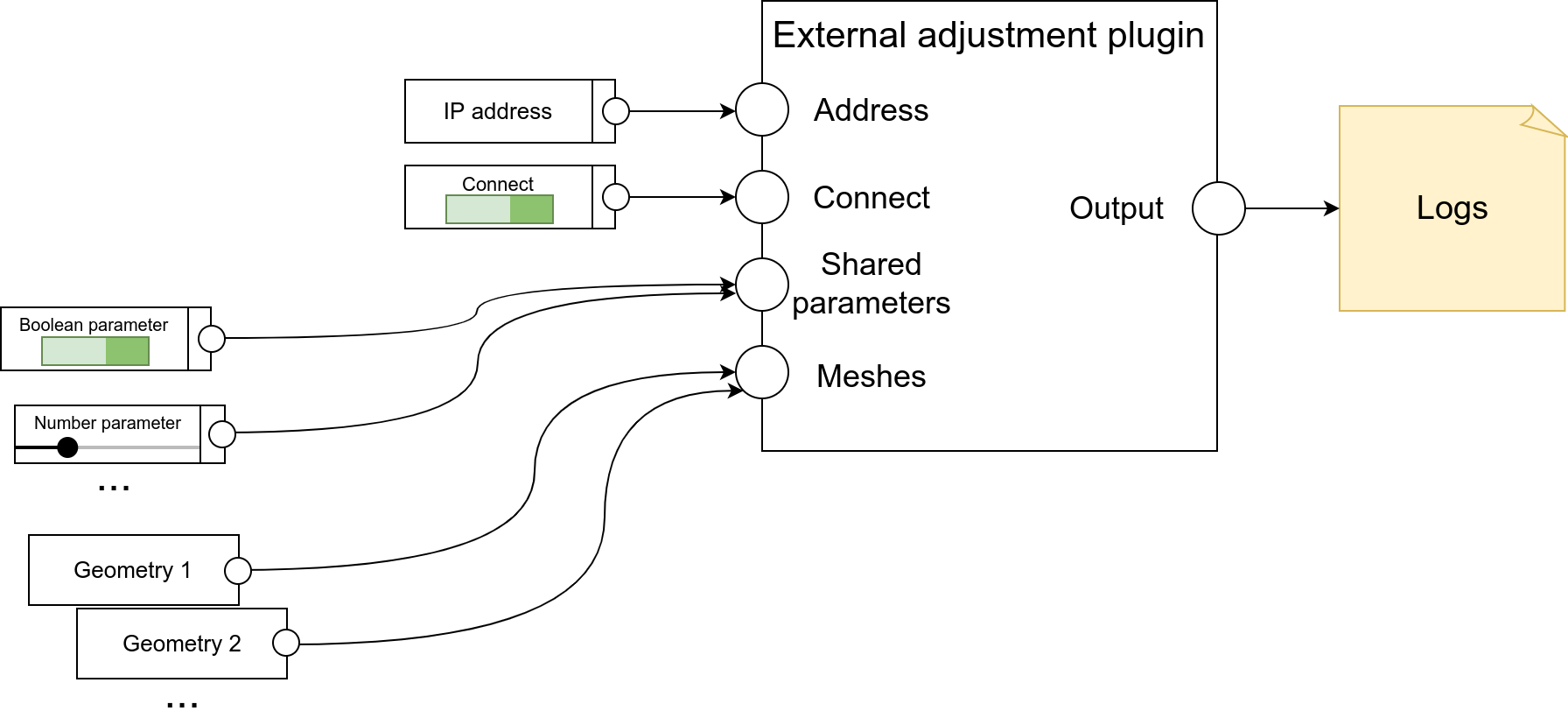}
	\centering
	\caption{\label{fig:ecaade-gh} The GHVRBridge Grasshopper custom component for external geometry sharing and parameter adjustment.}
\end{figure}

We developed a Grasshopper sharing component, written in \Csharp. It should be placed as any other Grasshopper component inside the \gls{ad} definition in order to interact with other components.
The component contains 4 main input ports: \emph{Address}, \emph{Connect}, \emph{Meshes} and \emph{Shared parameters}. The corresponding representation in Grasshopper is outlined in Figure \ref{fig:ecaade-gh}. An actual concrete example is shown in the green component in the bottom-left picture of Figure \ref{fig:ecaade-schema}.
When activated via a boolean switch linked to the \emph{Connect} input port, the component connects to the relay server indicated by the \emph{Address} port, via the WebSocket protocol \citep{fette2011websocket}. It then iterates over the input parameters that the designer has linked to the \emph{Shared parameters} port, stores their properties (name, type, range, values and a unique identifier) and sends this information to the relay server that forwards it to the VR applications. The geometries linked to the \emph{Meshes} port are then shared as well.

The component listens to incoming parameter adjustment messages, coming from a VR application and relayed by the WebSocket server. When one such message is received, the component iterates over the list of value changes it contains and replicates them on the corresponding input parameters in Grasshopper (identifying them thanks to the previously mentioned unique identifiers).

To ensure that messages are properly sent and received, we require a transmission protocol that provides reliable transfers; otherwise, geometries or parameter updates may never reach their destination. WebSocket is a well-established protocol that enables bidirectional communication over a \gls{tcp} \citep{postel1981tcp} connection, providing such reliable transfers. 
While removing the complexity of handling plain \gls{tcp} connections manually thanks to well-supported libraries, WebSocket does not consume more network traffic than plain \gls{tcp}, except for the initial handshake (at the start of the connection) \citep{skvorc2014performance}. We therefore chose to rely on that protocol for implementing GHVRBridge.

In order to optimise the content (payload) of the messages exchanged by GHVRBridge, the data is structured via Flatbuffers\footnote{\url{https://google.github.io/flatbuffers/}}, a serialisation library that can easily convert mesh and parameter data to a binary format that minimises the size of the (binary) payload. Compared to other very popular exchange formats such as JSON \citep{json}, XML \citep{xml} and YAML \citep{yaml}, Flatbuffers (de)serialises faster and produces smaller transmittable messages\footnote{\url{https://google.github.io/flatbuffers/flatbuffers_benchmarks.html}}, reducing the traffic to be sent over the internet in the case of a remote communication. 

The Flatbuffers library requires that we define the schema (format) of the data in a \texttt{.fbs} file, before we can build converters from and to that format. This schema is presented in Listing \ref{lst:fbs}. In it, we define a \texttt{Components} message that can contain multiple instances of a \texttt{Component}, which can either be a \texttt{BooleanToggle} or a \texttt{NumberSlider}. As required by the syntax, both are encapsulated into a \texttt{GenericComponent}. Depending on the type of component, different fields are listed. The \texttt{value} field has a different type depending on the component's type, but a \texttt{NumberSlider} also requires information about the range of possible values and it therefore contains additional fields to store that information.

Based on a given schema, Flatbuffers creates a set of classes to be used in the project's codebase. The behaviour of the component we use to transfer geometries and parameters is defined in a class called \texttt{GHSharingComponent}, that instantiates \texttt{GH\_Component}, so that it can be integrated into a Grasshopper (visual algorithm) definition.

\begin{figure}
	\includegraphics[width=.9\textwidth]{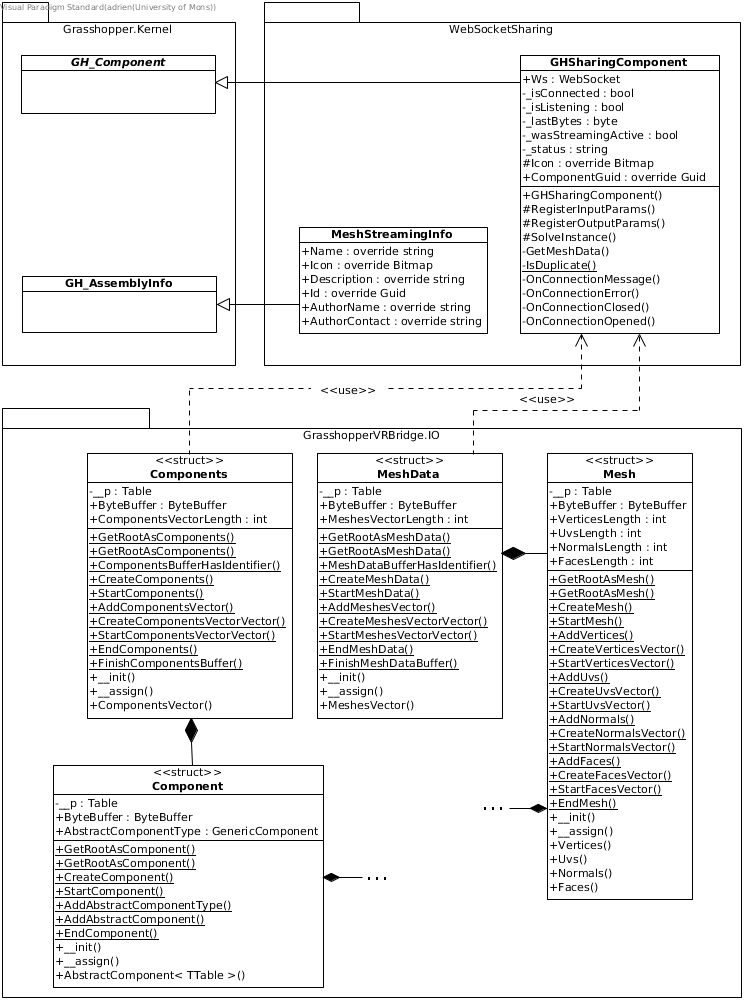}
	\centering
	\caption{\label{fig:uml-ghvrbridge-gh} The class diagram for GHVRBridge's custom Grasshopper component. Only the ``entry classes'' (\texttt{Components} and \texttt{MeshData}) of the \texttt{GrasshopperVRBridge.IO} package are shown so as to keep the diagram readable on a single page.}
\end{figure}

Figure \ref{fig:uml-ghvrbridge-gh} displays the class diagram of the \Csharp \, project for the Grasshopper custom component we developed, including part of the \texttt{GrasshopperVRBridge.IO} package that was automatically generated using Flatbuffers (the dots on the Figure accounts for the other classes that were automatically generated based on the schema).

\noindent
\begin{lstlisting}[float, label=lst:fbs,caption={The Flatbuffers Schema definition for exchanging parameter data.}]
	namespace GrasshopperVRBridge.IO;

	enum Accuracy:byte {
		Float = 0,
		Integer = 1,
		Even = 2,
		Odd = 3
	}
	
	table BooleanToggle {
		name:string;
		guid:string;
		value:bool;
	}

	table NumberSlider {
		name:string;
		guid:string;
		value:float;
		accuracy:Accuracy;
		min:float;
		max:float;
		epsilon:float;
		decimal_places:short;
	}
	
	union GenericComponent {
		BooleanToggle,
		NumberSlider
	}
	
	table Component {
		abstractComponent:GenericComponent;
	}
	
	table Components {
		componentsVector:[Component];
	}
	
	root_type Components;

	file_identifier "PARA";
\end{lstlisting}

\subsection{Virtual Reality application}
\label{sec:ecaade-vr}


In order to create the \gls{vr} part of GHVRBridge that will consume the data provided by the custom Grasshopper component described in the previous section, we rely on Unity, a game engine that facilitates the creation of interactive three-dimensional applications. Additionally, Unity also helps us to create \gls{vr} experiences in particular, through an easy access to SteamVR, the platform that handles the communication with the HTC Vive headset. Since we develop Grasshopper custom components using \Csharp , we define the behaviour of the \gls{vr} application with \Csharp \, scripts. This facilitates code reuse between different components of the GHVRBridge system and lowers the risk of incompatibility issues.

The \gls{vr} application supports two types of users. Simple clients can visualise the rendered geometry in \gls{vr} and receive updates whenever a new one is generated in Grasshopper. On the other hand, designers additionally see parameters and can modify their values so as to edit the Grasshopper definition. Depending on how the application is started (configured for a certain type of user), it listens to a WebSocket service that only shares the necessary information for that type of user. 

When receiving a geometry update, the \gls{vr} application reads the Flatbuffers payload that contains what is called a mesh.
This is a simplified representation of the surface of an object using polygons (typically triangles), defined by vertices, edges, and faces.
From the mesh data, the \gls{vr} application creates a three-dimensional object (such as the egg-shaped tower in the bottom right picture of Figure \ref{fig:ecaade-schema}) that uses the corresponding polygons, with two peculiarities to take into account. 

\begin{figure}[h]
	\includegraphics[width=.8\textwidth]{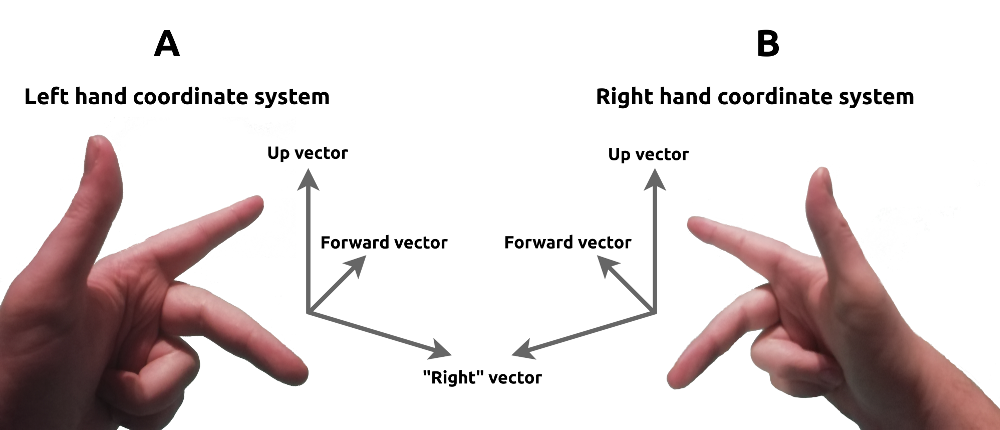}
	\centering
	\caption[]{\label{fig:coordinate-systems} Left-handed and right-handed coordinate systems.}
\end{figure}

The first one is that Grasshopper uses a right-handed $z$-up coordinate system, while Unity is based on a left-handed $y$-up system. To clarify, Grasshopper uses a system such as the one depicted on the right (B) side of Figure \ref{fig:coordinate-systems}, with the $z$ axis pointing upwards, the $y$ axis pointing forward, and the $x$ axis pointing to the left. On the other hand, Unity's coordinate system is based on the left side (A) of Figure \ref{fig:coordinate-systems}, with $y$ pointing up, $z$ pointing forward, and $x$ pointing right. This implies that $y$ and $z$ values need to be switched, and $x$ values must be modified to -$x$.

The second need is the computation of normals, i.e., vectors perpendicular to mesh surfaces and pointing outwards. Normals allow Unity's lighting system to properly project light on the generated objects and compute shadows; there is one normal to be computed for each vertex in the mesh.

As for parameter updates, the \gls{vr} application reads the information according to the format presented in Listing \ref{lst:fbs}, and generates the appropriate widgets. Boolean parameters are mapped to switch toggles, while numbers lead to sliders whose possible values depend on the received settings data.
These widgets are placed on a panel attached to the user's non-dominant hand (left hand by default but the user can switch hands by interacting with a button), that is shown on the bottom-right picture of Figure \ref{fig:ecaade-schema}.

When the user makes changes to a parameter value, the application sends a modification update to the relay server that then forwards it to the Grasshopper component. That update contains the unique identifier of the parameter that was modified, as well as the new value, which is then reflected onto the actual Grasshopper parameter.

\begin{figure}
	\includegraphics[width=.95\textwidth]{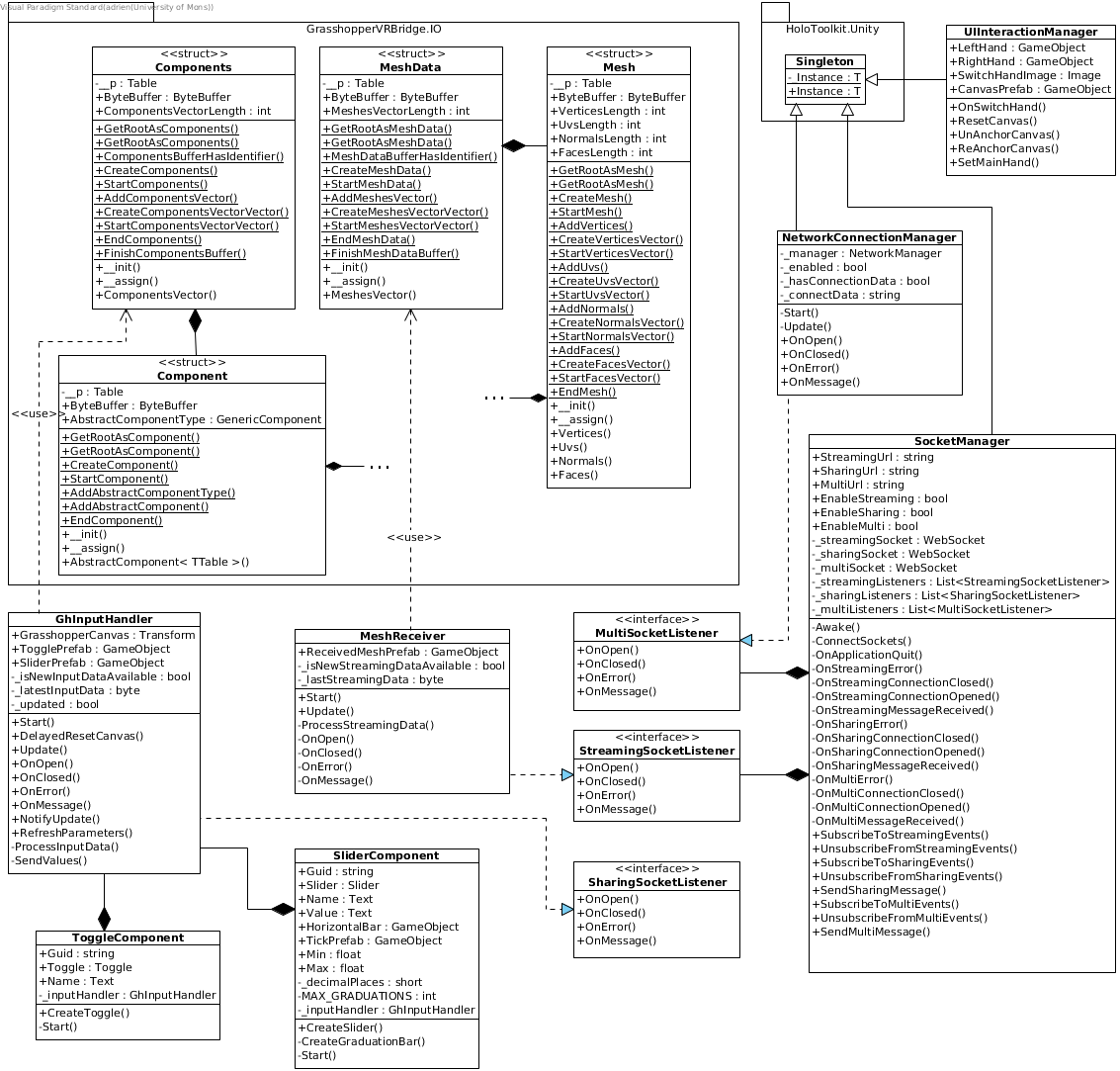}
	\centering
	\caption{\label{fig:uml-ghvrbridge-vr} Part of the class diagram for GHVRBridge's \gls{vr} application, clarifying some of the networking and input handling aspects.}
\end{figure}

To clarify some of the networking and input handling aspects, Figure \ref{fig:uml-ghvrbridge-vr} shows a portion of the class diagram for GHVRBridge's \gls{vr} application. On the Figure, the same automatically-generated \texttt{GrasshopperVRBridge.IO} package can be found, since the data receiving classes of the \gls{vr} application rely on them to read from (and create) update messages. The initiated reader will recognise the Observer and the Singleton design patterns \citep{gamma1995elements}. The former is used to allow subscribing objects to listen to updates efficiently, while the latter ensures only one instance of a given class exists. Due to Unity's usage of ``game object'' entities, on which ``scripts'' that communicate through messages are attached, the Singleton design pattern is also a more efficient way to enable communication between scripts (e.g., to notify the \texttt{SocketManager} that a parameter update has to be sent) than the standard messaging system.

\section{Enabling immersive collaboration}
\label{sec:cdve}

The GHVRBridge proof-of-concept prototype we described thus far allows for two different modes that can accommodate designers that want to tweak parameters, and simple clients that only want to visualise the geometrical representation inside a \gls{vr} environment. 

But the architectural design process is a collaborative endeavour; it is therefore desirable to allow that collaboration to continue inside the \gls{vr} application. We want to allow multiple designers to work on the same model at the same time \citep{dagit1993establishing}, from the same shared \gls{vr} experience. Furthermore, in order to involve clients during the design activities and thereby move to a \gls{vr}-based user-centred process that increases quality and performance in architectural projects \citep{Bullingerusercentreddesign2010}, we need to allow both types of users (designers and simple clients) to participate in the same (shared) virtual experience.

\begin{figure}[h]
	\includegraphics[width=.95\textwidth]{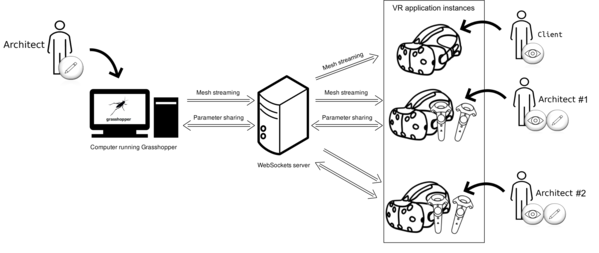}
	\centering
	\caption{\label{fig:cdve-schema} The GHVRBridge prototype adapted for multi-user immersive adjustment of Grasshopper parameter values, with two possible modes: mesh streaming (i.e., geometry visualisation only) or parameter sharing (i.e., control over shared parameters).}
\end{figure}

The way we designed the GHVRBridge system makes it possible to connect multiple instance of the \gls{vr} application to the relay server, to facilitate the sharing of the experience by several \gls{vr} users that potentially have different roles (viewer only or designer with control over the parameter values), as illustrated in Figure \ref{fig:cdve-schema}. We note that, as seen on the Figure, only one Grasshopper instance is running; the collaboration we enable is focused at the \gls{vr} level. Desktop-level collaboration is indeed out of scope of the present research since we work with immersive technologies, and would likely need to be implemented as part of the core code of Grasshopper by the vendor itself (and not as a custom third-party component).

A video demonstrating the collaborative extension of GHVRBridge is available online\footnote{\url{http://informatique.umons.ac.be/staff/Coppens.Adrien/?video=CDVE2018}} and our codebase is hosted as open-source software on a GitHub repository\footnote{\url{https://github.com/qdrien/GHVRBridge}}, available under the MIT\footnote{\url{https://opensource.org/licenses/MIT}} license. We also published a paper describing this collaborative extension and the challenges we encountered \citep{coppens2018towards}.

\subsection{Virtual co-presence}
\label{sec:copresence}
Regardless of the exact collaboration scenario (e.g., a designer showing adjusting parameter values with a client), \gls{vr} users need to see one another and be able to observe the actions of other users to collaborate efficiently \citep{Nguyensurveycommunicationawareness2014}. While \gls{vr} headsets typically occlude the wearer's view of the real world, it is possible to replace a physical co-presence by a virtual one, using avatars that mimic what the real person is doing.

\begin{figure}[h]
	\includegraphics[width=.4\textwidth]{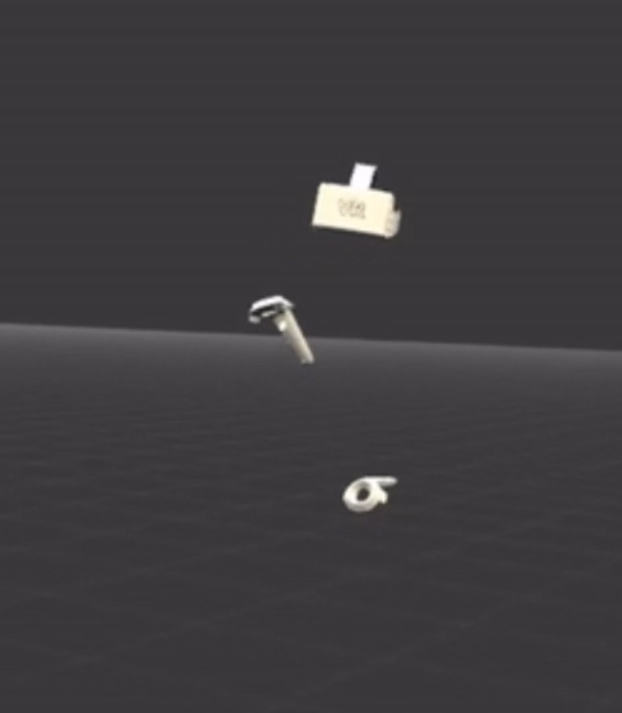}
	\centering
	\caption{\label{fig:cdve-avatar} A collaborator waving towards another \gls{vr} user, as seen from that user's point-of-view. Both the collaborator and the user are co-located in the same virtual environment, with the motions of each individual's head and controllers being tracked and shown to others.}
\end{figure}

Depending on the exact immersive devices being used, different tracking options could be available. With the HTC Vive, it is only possible to know the position and rotation of the headset and the controllers. This means we can only portray these items' motions accurately. Under the reasonable hypothesis that the user is wearing the headset and holding the controllers in his hands, we can even depict the user's head and may approximately picture his hands (although he may hold the controllers or press buttons in a different way than the one we expect). We chose to limit ourselves to the simple factual depiction of the \gls{hmd} and the controllers, as shown in Figure \ref{fig:cdve-avatar}. We therefore only depict objects that are actually tracked, without trying to infer elements (gaze, fingers, etc) that may then be incorrectly represented (e.g., if the user does not grasp the controllers the way we expect him to) even if that would have lead to a stronger level of immersion \citep{lin2019effect}.

In order to enable a shared experience, head and controller positions and orientations have to be sent to all VR application instances sharing the collaborative experience. This information does not have to transit through the relay server nor reach the Grasshopper component, as it is only valuable to \gls{vr} instances. Our approach therefore is to let \gls{vr} instances communicate ``co-presence information'' only between them. When initially connecting to the relay server, a \gls{vr} instance indicates that it wishes to join the shared experience, to which the server either replies with the current host's IP address, or simply by ``you'', thereby informing that particular instance that it will be the session's host. In the latter case, the IP address of the machine on which the \gls{vr} application is running is stored by the server to be provided to other \gls{vr} instances that may join the session later. In case the host is disconnected, a new one is elected by the server and broadcasted to all \gls{vr} instances.

The system to decide and advertise the host allows \gls{vr} instances to communicate directly with that host, that will forward (position and orientation) updates to other instances. 
Instances will have to exchange messages at a high rate to provide a somewhat smooth experience. Because of that and since losing a packet (i.e., missing a position update) is not really a problem as newer information will quickly arrive, we should this time rely on the User Datagram Protocol (UDP) \citep{postel1980rfc0768} to transfer the data. In fact, that protocol does not provide mechanisms similar to \gls{tcp} to ensure reliable transfer of data, but offers better speed. 
To implement the co-presence feature that was just described, we relied on Unity's built-in multiplayer High Level \gls{api} (HLAPI\footnote{\url{https://docs.unity3d.com/Manual/UNetUsingHLAPI.html}}), that is implemented on top of UDP. 

\subsection{Towards collaborative parameter value adjustment}
\label{sec:conflicts-params}

While the previous section explains how we minimised message size and latency for remote collaboration, another challenge not directly related to network aspects arises when several users are allowed to modify parameter values: handling concurrent and potentially conflicting modifications. Safeguards may be put in place to avoid a frustrating user experience. 

Many researchers have worked on issues related to concurrent editing and collaboration conflicts. There are in fact even venues (both conferences and journals) specifically focused on such topics, such as the Computer-Supported Collaborative Work (CSCW) conference\footnote{\url{https://cscw.acm.org/}}. An extensive literature can be found on methods that handle concurrency for database manipulations in particular \citep{munson1996concurrency}. 

Concurrent modification of parameters is not exactly similar to reading and writing database entries since methods used in that context need to handle transaction dependencies (modifications of elements that depend on each other are not allowed). The programming context of \gls{ad} means it is tempting to look at solutions used for collaborative software development. But, in that case, semantic understanding is key and software merging techniques \citep{mens2002state} are necessary since changes to the code are typically done without conflict checking mechanisms, and are only merged together later (with modern version control systems). 

Since \gls{ad} is a particular paradigm for \gls{caad}, it would be equally tempting to develop upon general collaborative design tools. But these tools typically do not rely on algorithms and therefore need to deal with different issues than the ones we are being faced with. For instance, collaborative direct modelling tools must make sure that instructions sent by different collaborators do not create an invalid result (incompatible changes, potentially related to the order in which instructions are processed) \citep{hepworth2014model}.

It therefore is more appropriate to assess the potential of techniques used in other contexts, including collaborative database management. Another field that we can draw inspiration from is multi-processors systems with shared resources. The problem of avoiding conflicts between tasks that must run in disjoint time intervals is called mutual exclusion \citep{baker1996mutual}.

By reading papers from that literature and reasoning about the way to apply them in the concurrent parameter adjustment context, we ended up with a few options, listed hereafter: 

\begin{itemize}

	\item \textit{Overwriting updates}: 
	
	Whenever the system receives an update of a parameter value, the previous one is overwritten. This default behaviour is the most tolerant, but does not provide any safeguard to concurrent modifications. In addition, latency has to be minimal for this approach to work, otherwise the system will appear to be ignoring certain user changes. This solution can be acceptable if designers rarely modify parameters simultaneously, and if the rate of change is not too high.
	
	\item \textit{Reactive locking:}

	Whenever a designer starts modifying a parameter value, his collaborators are notified (e.g., through some visual clue) and they can no longer modify that parameter until the first designer releases it. Conflicts could still happen due to latency, if another designer tries to modify the same parameter before receiving the ``lock notification''. In that case, the system should notify the user that his modification request was rejected.

	\item \textit{Preemptive locking:}

	Preemptive locking is an even more conservative approach that prevents the previous problem from happening. By default, parameters cannot be edited. If a designer wants to modify a parameter value, he first needs to request access to that parameter. This is similar to the voting mechanism in place in many systems tackling the mutual exclusion problem. An example that we could draw inspiration from would be an algorithm proposed in \citep{maekawa1985n}, that only requires $\sqrt{n}$ messages to coordinate $n$ nodes in a decentralised system.
	
	\item \textit{Privilege strategy:} 

	In addition to the locking strategy, a mechanism based on user privileges could be put into place. Users with higher privileges might be granted the ability to take control of a parameter that is being modified concurrently by someone with lower privileges. In that case, the less-privileged user should be notified that he can no longer control the said parameter. 
	
	\item \textit{Parameter layers:}

	We could consider grouping parameters so that users could get access to different groups of parameters. This could reduce the concurrent modification problem for instances where users mostly get access to different groups of parameters, or even remove that problem entirely when all users are assigned to different groups. We should however note that grouping parameters should in general induce more conflicts, since two users willing to interact with two different parameters could be stopped if these two parameters are part of the same group. That solution, if implemented, implies that grouping elements should be made with caution.

\end{itemize}

Collaborative testing and evaluation sessions using the GHVRBridge prototype will typically involve a limited number of designers with concurrent access to the same parameters, because of hardware and space constraints, but also because studies suggest that working groups should be rather small to be effective (e.g., at most four as per \citep{steiner1972group}).
The limited number of active participants should result in very few conflicts. Furthermore, immersive sessions involving multiple designers only really make sense if these designers are co-designing a project and discussing changes to be made to the geometry, based on the immersive visualisation they are experiencing. In that case, they will probably naturally avoid concurrent modifications since they will be focusing on one change at a time (the one they are discussing). 

For these reasons, we chose to stick with the simple \emph{overwriting} approach and simply added an ``update rate limiter'' to make sure one application cannot send too many messages to the system. This helps in reducing the risk of conflict, but also the burden on Grasshopper, that needs to recompute, regenerate and resend the geometry after each update.

\section{Evolving the concept for Augmented Reality and Tangible User Interfaces}
\label{sec:list}

While the GHVRBridge prototype (presented in Section \ref{sec:ecaade}) and its adaptation to a collaborative setting (presented in Section \ref{sec:cdve}) enable parameter sharing and geometry visualisation in \gls{vr}, their simple user interface still resembles the classical \gls{wimp} approach, with a \gls{vr} controller acting as a three-dimensional version of a mouse.

To provide a more natural collaboration environment within the same physical room, we worked with the Luxembourg Institute of Science and Technology to develop another prototype that combines different kinds of interaction devices. It in fact relies on a \gls{vr} headset but additionally utilises \gls{ar} glasses (Microsoft Hololens) and Tangible User Interfaces (a display table with tangible items on it) so as to create a multi-modal system. We call that multi-modal prototype system \textbf{GHXR}, for Grasshopper for \gls{xr}.

\subsection{Overview of the system}

\begin{figure}
	\includegraphics[width=.75\textwidth]{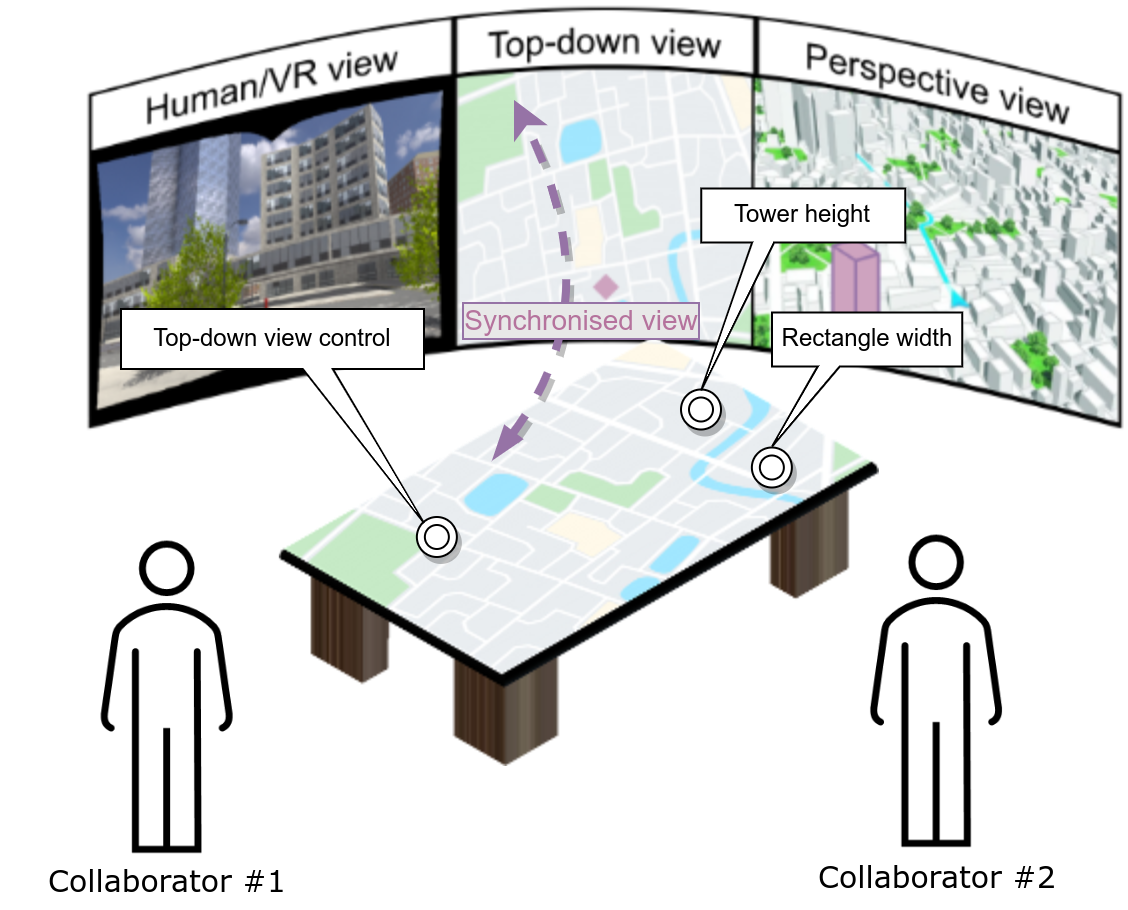}
	\centering
	\caption{\label{fig:ghxr-mockup-base} Mock-up of the GHXR system, with a table display on which users interact and collaborate. They are surrounded by about 300\degree \, circular screens, displaying different views. The example shows a building project in a city centre, with two Grasshopper parameters available to the users.}
	
	  \vspace*{\floatsep}
	
	\includegraphics[width=.75\textwidth]{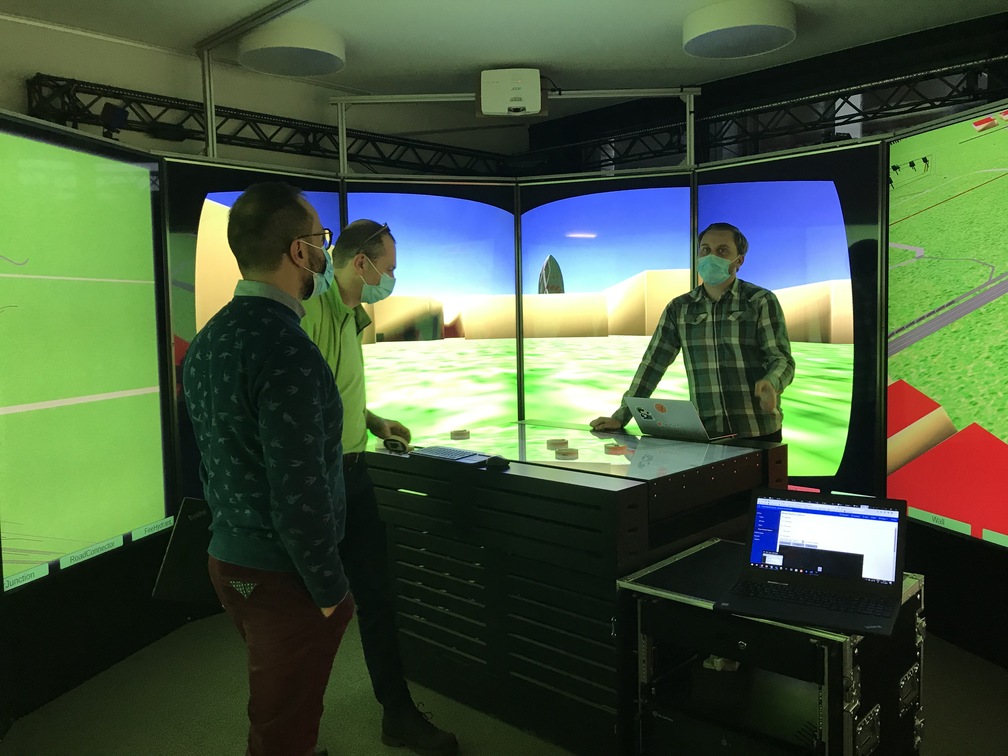}
	\centering
	\caption{\label{fig:ghxr-realpic} The GHXR system, with three users discussing over an \gls{ad} model, using the different views offered by GHXR.}	
\end{figure}

\begin{figure}
	\includegraphics[width=1.45\textwidth, angle=90]{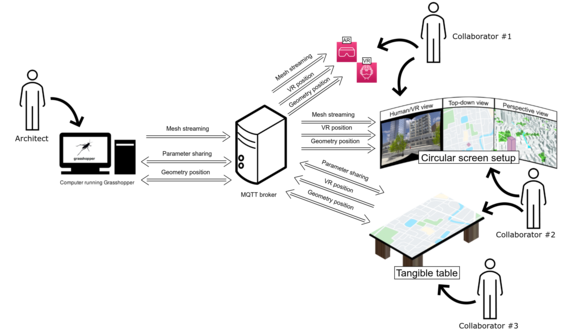}
	\centering
	\caption{\label{fig:ghxr-devices} Constituents of the GHXR prototype and data transfer between these constituents.}
\end{figure}

The general idea, depicted in Figure \ref{fig:ghxr-mockup-base}, is that a table display is placed in the middle of a room in which collaborators can work on a joint project. These collaborators are surrounded by a circular screen setup (covering about 300\degree \, around them), that displays three different views of the geometrical representation being worked on and its surroundings: a first-person point-of-view, a top-down (plan) view that is also reproduced on the table display, and a 3D perspective view. The designer is therefore free to choose the appropriate representation depending on the specific aspect or problem being discussed, at any point in time. The constituents of the system and the data transfers between such constituents are depicted in Figure \ref{fig:ghxr-devices}.

\begin{figure}
	\includegraphics[width=.9\textwidth]{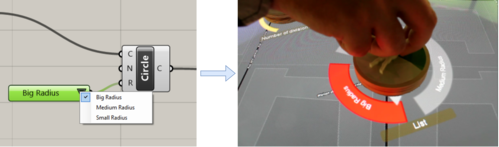}
	\centering
	\caption{\label{fig:ghxr-widget} A Grasshopper list component and its assigned tangible item in GHXR's table application, with the corresponding software widget around it.}
\end{figure}

A tangible item (that we may simply refer to as ``a tangible'' hereafter) that roughly looks like a hockey puck is placed on the table, as shown on Figure \ref{fig:ghxr-widget}, for each Grasshopper parameter that is shared.
Such a tangible is tracked via a paper marker (see Section \ref{sec:sota-tracking} and Figure \ref{fig:paper-markers} in particular) placed on its bottom and a camera watching through the display to recognise this marker, so that its position and orientation on the surface of the table are known. When a designer rotates a tangible mapped to a parameter, the value of that parameter is adjusted and sent to Grasshopper to update the generated geometry.

In addition to the tangibles representing Grasshopper parameters, two more tangible items are placed on the table: one to control the top-down view, allowing to zoom in or out by rotating the item as well as to reposition the view by sliding the item, and one to move the human (first-person) point-of-view (to teleport the view to the target location by moving the item and to change the view angle by rotating the item).

\begin{figure}
	\includegraphics[width=.75\textwidth]{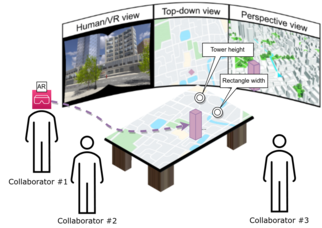}
	\centering
	\caption{\label{fig:ghxr-mockup-ar} The additional 3D hologram of the building that \gls{ar} users can see, placed at the correct location on the table's plan view.}
\end{figure}

The human point-of-view is also visible through the \gls{vr} headset, but in visualisation mode only. Write access to parameters is not allowed since they are being controlled by the tangibles on the table. In addition to the \gls{vr}, table, and circular screen displays, a user wearing an \gls{ar} headset can see a 3D hologram of the geometry being worked on, projected at the correct location on the table, as pictured in Figure \ref{fig:ghxr-mockup-ar}.

Figure \ref{fig:ghxr-realpic} depicts three users collaborating through GHXR. The system naturally enhances collaboration compared to GHVRBridge, since users can see and talk with each other, and interact with the same physical tangible items. A video demonstrating GHXR is available online\footnote{\url{https://youtu.be/L5dqMx7rnmM}} and our codebase is hosted as open-source software on a GitHub repository\footnote{\url{https://github.com/qdrien/GHXR}}, available under the MIT\footnote{\url{https://opensource.org/licenses/MIT}} license. 

The next 4 sections discuss the necessary constituents to produce GHXR: a Grasshopper custom component, a table display application, an application for a circular screen setup that surrounds the users collaborating through the table application, and two immersive applications (for the \gls{vr} and the \gls{ar} headsets, respectively).

\subsection{Grasshopper custom component}

Similar to the Grasshopper custom component for GHVRBridge we described in Section \ref{sec:ecaade-gh}, we develop another Grasshopper component to exchange geometry and parameter data with a server, that itself relays that data to the applications that run on the table display, the computer controlling the circular screens, as well as on the \gls{vr} and \gls{ar} headsets. 

\begin{figure}
	\includegraphics[height=.9\textwidth, angle=90]{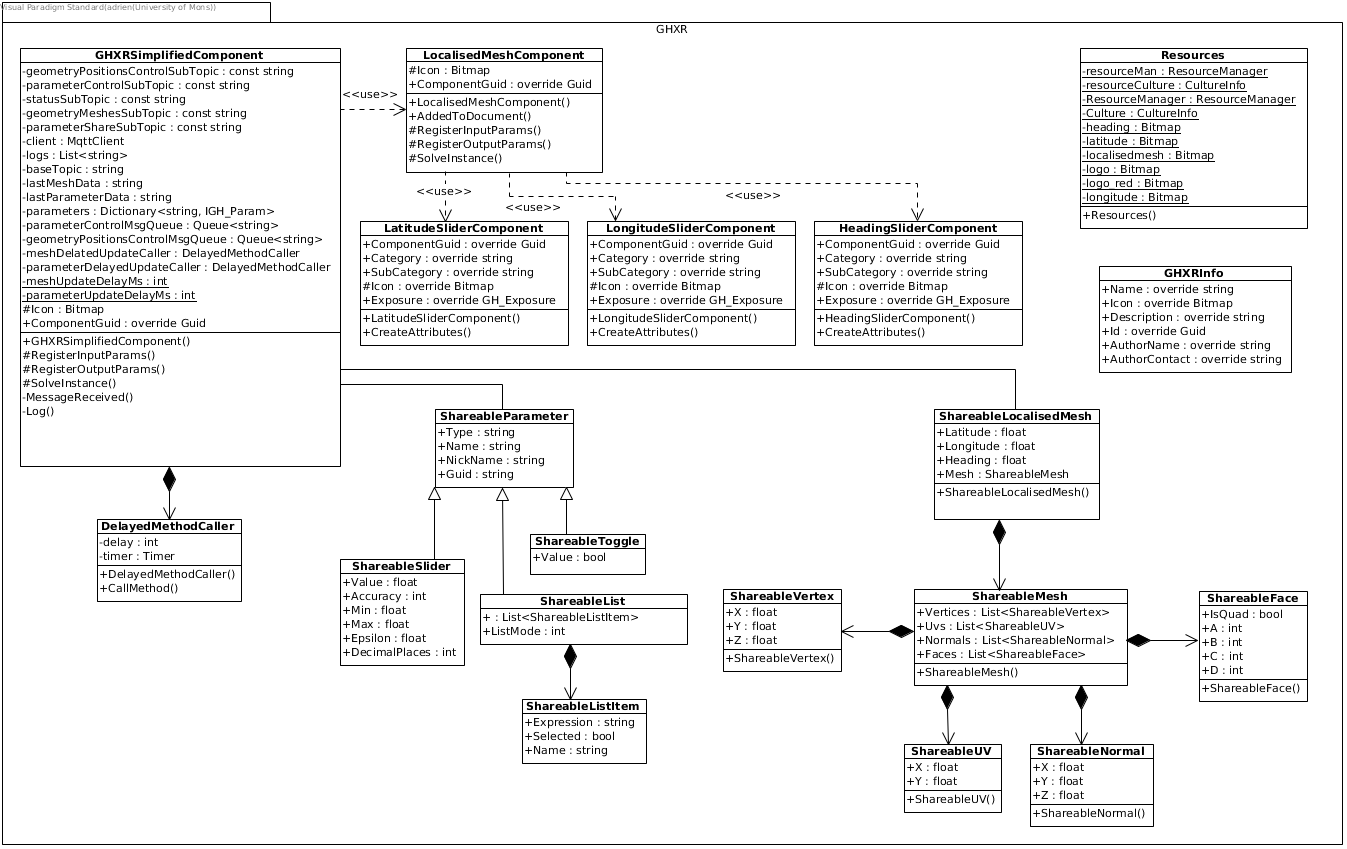}
	\centering
	\caption{\label{fig:ghxr-class} The class diagram for GHXR's Grasshopper custom component.}
\end{figure}

The ``Shareable'' classes on the class diagram of Figure \ref{fig:ghxr-class} are data classes used to easily convert from and to the transfer format (JSON in this case since the system is here supposed to exchange data only on a local network). On the same Figure, one can notice the \texttt{GHXRSimplifiedComponent}, handling most of the logic behind the Grasshopper custom component (its name contains ``Simplified'' because it requires less configuration on the user's end than a previous version). 

A \texttt{DelayedMethodCaller} class can also be seen on the class diagram ; it serves to delay parameter and geometry updates, so as to limit the frequency of such updates that are sent to ``consuming'' applications. If another change generates a newer update within that \texttt{DelayedMethodCaller}'s timeframe, only the newest update will effectively be sent.

Instead of using WebSockets as for GHVRBridge, we here rely on \gls{mqtt} \citep{light2017mosquitto} as the communication protocol, since it is already supported by the framework we need to use for the table display.

Since the geometry shared from Grasshopper is localised for GHXR, we include the GPS position (manually indicated by the user) and heading information together with the mesh data (that itself has the same structure as described in Section \ref{sec:ecaade-vr}). As for parameters, we support number and boolean values (as for GHVRBridge), but we also add support for lists of textual values, that are themselves mapped to any values within Grasshopper.

In order to develop the Grasshopper custom component that shares the geometries and parameters, we had to overcome a few obstacles related to concurrent access to software objects. 

The first one is that modifications to Grasshopper objects (e.g., parameters) cannot be made from anywhere else than the main application thread and it is therefore necessary to rely on a message queuing system to save incoming updates for later processing.

Another difficulty we faced is that the messages queues, that are necessary to store the exchanged information in this queuing system, needs to be manipulated by both the main and background threads. We therefore have to make sure to avoid what is called race conditions, i.e., concurrent modification of an object, leading to application bugs or crashes.

\begin{figure}
	\includegraphics[width=.9\textwidth]{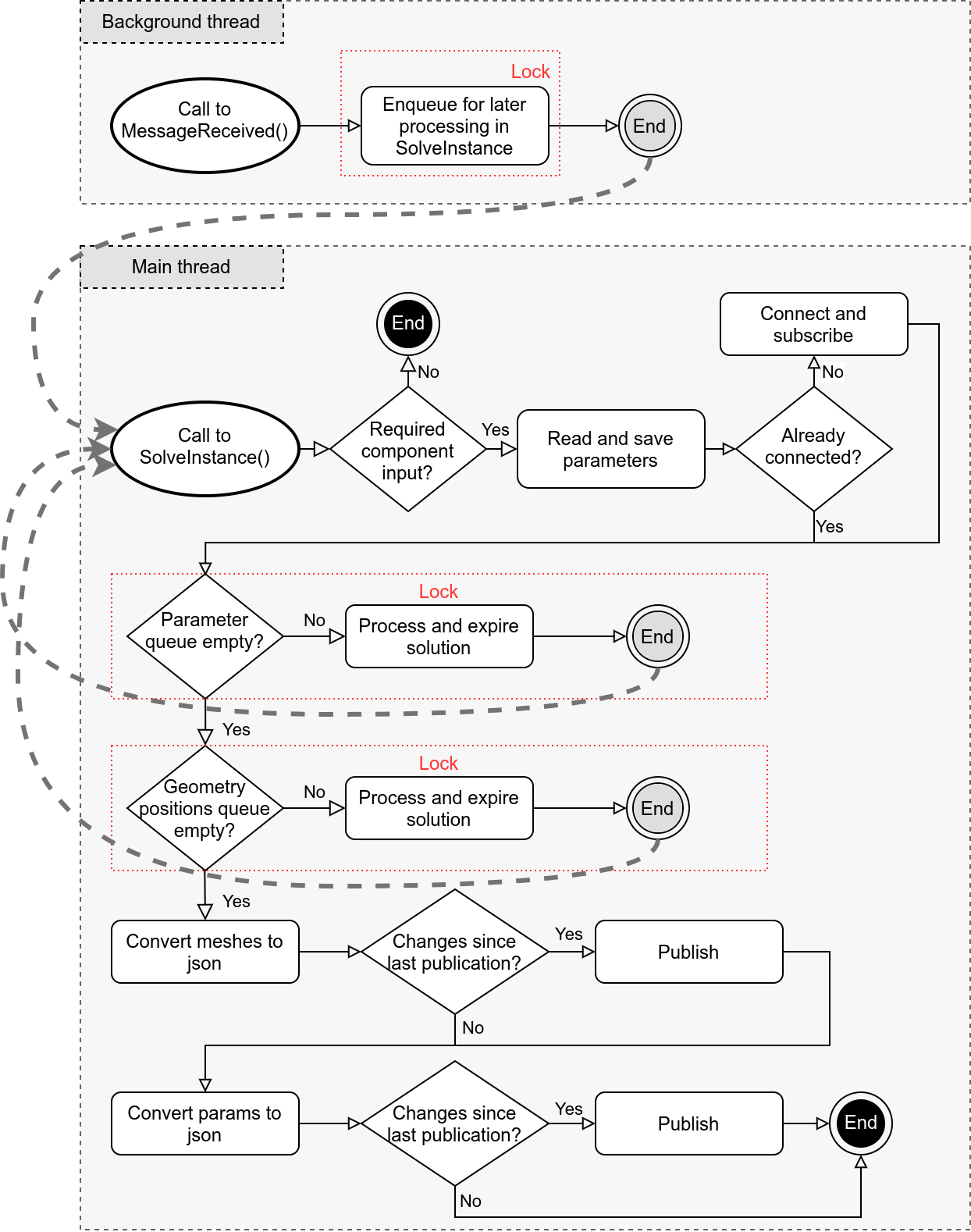}
	\centering
	\caption{\label{fig:ghxr-plugin-logic} The underlying logic behind GHXR's custom component for Grasshopper, to send geometries and share parameters while handling concurrency issues. The black ``End'' circles indicate simple terminations of the \texttt{SolveInstance()} method, while their grey counterparts represent terminations that are preceded by the expiration of the Grasshopper solution, leading to a new call to \texttt{SolveInstance()}, hence the inclusion of dashed arrows.}
\end{figure}

Figure \ref{fig:ghxr-plugin-logic} describes the logic we rely on to handle these issues.

When receiving incoming updates from an application, the GHXR Grasshopper component is notified through the \verb|MessageReceived()| method that runs in a background thread. After enqueueing the message, the \verb|MessageReceived()| method ``expires'' the Grasshopper solution, forcing it to be recomputed and leading to a call to \verb|SolveInstance()|, that itself runs from the main thread.

After making sure that the custom component is properly setup and connected, the \verb|SolveInstance()| method can dequeue messages since it is running on the main thread and the corresponding data can be processed on it. Once changes contained in these messages are reproduced in the Grasshopper definition, it still is necessary to expire the solution once again (or at least the corresponding components and the ones that depend on them), so that Grasshopper recomputes the (part of) the visual algorithm to produce the new geometrical representation.

In order to avoid race conditions, we use the locking mechanism available in \Csharp \, to make sure the queues cannot be manipulated at the same time by different threads. Using that mechanism, when a thread A wants to get access to a variable that is currently locked by another thread B, A's execution is paused until B releases that variable.

Note that a similar logic was applied to solve the same sort of problems for the component described in Section \ref{sec:ecaade-gh}.

\subsection{Table application}

In order to display a plan view and handle interaction with tangibles placed on the tangible table, we rely on an existing framework called TULIP \citep{tobias2015tulip}. 
This open-source framework, custom-built at the Luxembourg Institute of Science and Technology, is fully integrated with the tangible table.
TULIP is a modular framework, in the sense that it allows for modules to be integrated to a project. For instance, we here use the GIS module to retrieve data from OpenStreetMap\footnote{\url{www.openstreetmap.org}} to display the plan view, as well as the IOT module to handle the connection with the MQTT relay server (called a broker in MQTT terminology).

The TULIP framework works with scenarios described in XML files, that feed the core software, itself in charge of creating widgets. Widgets are interactable software entities placed on tangible items, as shown in Figure \ref{fig:ghxr-widget}, where we see a widget for value control over a parameter simply named ``List'', that can take three values: small, medium, or big radius

The specific scenario we use for the table application tells TULIP to display a plan view of a geographical location that therefore appears on the table. The exact location is specified inside the scenario's XML file. The table application then connects to the \gls{mqtt} broker and awaits a parameter sharing message. Once received, it uses that data to map each parameter to a tangible item with an appropriate widget (depending on its type and possible values) so that changes to the widget value through the tangible item are sent back to Grasshopper via the broker.

The scenario also creates additional widgets to control the top-down view and the first-person views, and shares the GPS coordinates of the corners (bounding box) of the table's plan view, so that other visualisation applications know what is being displayed on it.

\subsection{Circular screen setup and Virtual Reality applications}
Since a significant part of the code used for the \gls{vr} experience is shared with the application that runs on the circular screen setup (hereafter called the visualisation application), we describe both of them in this section.

They both parse an OBJ file representing the surroundings (based on OpenStreetMap data) of the building that designers are working on and generate 3D shapes that replicate that data inside the virtual environment. They then place the geometry coming from Grasshopper at the correct location inside that environment, thanks to the geolocation data contained in the message. The inclusion of these surrounding buildings provides users with the option to consider the vicinity of the designed project when working on it.

\begin{figure}
	\centering
	\begin{subfigure}[t]{.24\textwidth}
		\centering
		\includegraphics[height=.14\textheight]{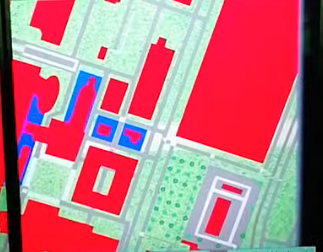}
		\caption{Top-down view.}
		\label{fig:ghxr-topdown}
	\end{subfigure}\hfill%
	\begin{subfigure}[t]{.417\textwidth}
		\centering
		\includegraphics[height=.14\textheight]{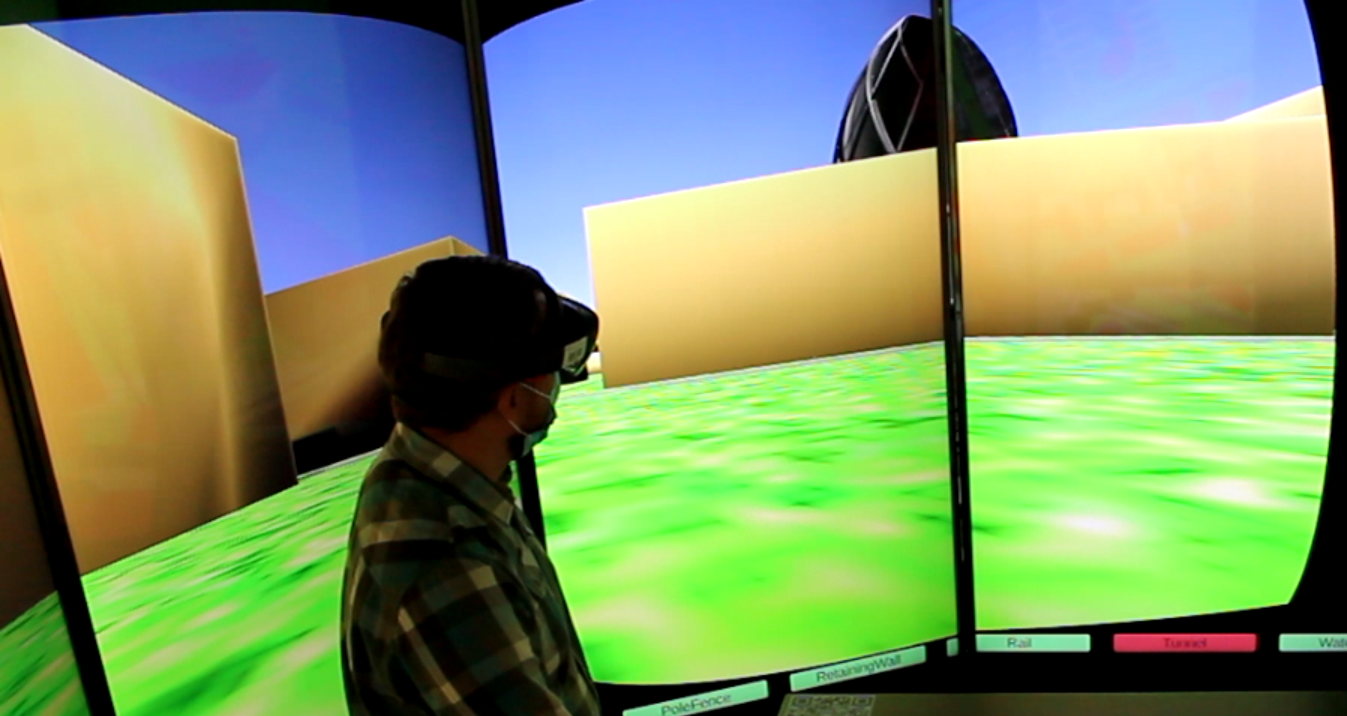}
		\caption{First-person view.}
		\label{fig:ghxr-human}
	\end{subfigure}
	\begin{subfigure}[t]{.3\textwidth}
		\centering
		\includegraphics[height=.14\textheight]{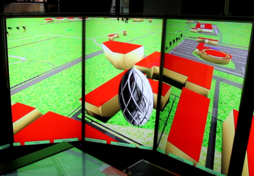}
		\caption{Perspective view.}
		\label{fig:ghxr-perspective}
	\end{subfigure}\hfill%
	\caption[]{The three views available to the user of the GHXR system, on the circular screen setup.}
	\label{fig:ghxr-threeviews}
\end{figure}

The visualisation application displays three views, presented in Figure \ref{fig:ghxr-threeviews}, that are based on three cameras placed in that environment. First, there is a top-down view, based on an isometric camera placed above the target location, that replicates the bounding box of the table display as shown on Figure \ref{fig:ghxr-topdown}). The second camera is a standard perspective camera looking at the geometrical representation, therefore providing a perspective view, as seen on Figure \ref{fig:ghxr-perspective}. Finally, a first-person view is provided by another perspective camera placed at human height, as shown in Figure \ref{fig:ghxr-human}.

When users interact with the tangible item that controls the human view, the corresponding camera is teleported at the given location and rotates according to the given angle.

The VR application only displays the first-person view but also listens to first-person view position updates so as to teleport the user to the right location. We do not apply these updates in a continuous manner to avoid cybersickness issues, and instead use an approach where an update is only applied if no other update quickly follows that one. This introduces a small delay but ensures that only stable data is processed.

In between position updates coming from the table widget, the \gls{vr} user is free to explore the environment by teleporting himself using \gls{vr} controllers (pointing metaphor with a curved beam) and looking around (by moving his head).

\subsection{Augmented Reality application}

\begin{figure}
	\includegraphics[width=.5\textwidth]{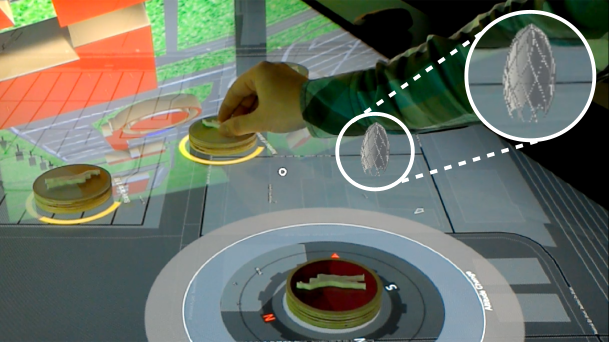}
	\centering
	\caption{\label{fig:ghxr-ar} The point of view of an \gls{ar} user interacting with a widget mapped to a parameter, while visualising a hologram of the geometrical representation at the target location on the plan view displayed on the table.}
\end{figure}

We additionally developed a Hololens application that also connects to the \gls{mqtt} broker, in order to retrieve geometry data, so as to create a 3D hologram to be placed on the table, as shown in Figure \ref{fig:ghxr-ar}. Due to the limited processing power available on the Hololens headset, it was not reasonably possible to show the surrounding buildings in \gls{ar}, even though the code would have easily been adapted.

The GPS data included in the geometry messages, is enough to properly locate the geometrical representation relative to the ``virtual'' bounding box, but it is still necessary to somehow recognise the physical table's position to map its surface to the bounding box and place the hologram on it at the correct scale.

To do so, we rely on paper markers, recognised by Vuforia\footnote{\url{http://vuforia.com}}, and placed in two opposite corners of the table. These corners are then mapped to the GPS-based bounding box and the hologram can therefore be placed relative to them.

An additional particularity that we handled was the conversion of GPS coordinates between the EPSG:3857\footnote{\url{https://epsg.io/3857}} and the EPSG:4326\footnote{\url{https://epsg.io/4326}} formats. Both formats are based on the approximation that the earth is close to a reference ellipsoid, that serves as the basis for a coordinate system. 

In the case of EPSG:4326, coordinates are given in latitude and longitude, i.e., degrees of deviation from a reference meridian and parallel (respectively). This is the type of system humans generally refer to when talking about a GPS position.

EPSG:3857, also called the pseudo-Mercator system, uses a projection of the previous coordinate system onto a plan, resulting in a rectangular world map. It is used internally by most computer-based map software, even though such software often offer a user interface that supports EPSG:4326.

The surrounding buildings we get from OpenStreetMap are enclosed in a bounding box given in the EPSG:4326 format, while the tangible table sends its own bounding box in the EPSG:3857 format. As mentioned earlier, since humans tend to use EPSG:4326, we also expect to receive coordinates in that format for geometry position data. This means that we need to be able to convert coordinates from one system to the other.

The following formulas can be used to do so, with $EH$ representing half of the circumference of the Earth at the Equator, $\ell$ a latitude, and $L$ a longitude:


EPSG:4326 $\rightarrow$ EPSG:3857:
$\begin{cases}
\ell_{3857} = \frac{\log_{10}(\tan(90 + \ell_{4326}) * \frac{\pi}{360})}{\frac{\pi}{180}} * \frac{EH}{180}\\
L_{3857} = L_{4326} * \frac{EH}{180}
\end{cases}$

EPSG:3857 $\rightarrow$ EPSG:4326:
$\begin{cases}
	\ell_{4326} = \frac{\arctan(\exp(\ell_{3857} * \frac{\pi}{EH})) * 360}{\pi - 90}\\
	L_{4326} = L_{3857} * \frac{180}{EH}
\end{cases}$

Parts of the code also compute distances between two GPS positions (e.g., to place the geometrical representation at the specified location, relative to the virtual environment's boundaries). To do so, we base ourselves on the haversine formula that is widely used in navigation. It computes the distance between two points on a sphere, therefore assuming the Earth is spherical but still manages to keep the error below 0.1\%, even for long distances{\interfootnotelinepenalty10000 \footnote{\url{https://docs.microsoft.com/en-us/dotnet/api/system.device.location.geocoordinate.getdistanceto}}}.

The exact formula we use to calculate the distance $d$ between two points $(\ell_1, L_1)$ and $(\ell_2, L_2)$ is as follows: 

$d\Big((\ell_1, L_1), (\ell_2, L_2)\Big) = ER * 2 * \text{arctan2}(\sqrt{\alpha}, \sqrt{1 - \alpha})$

with: 
$\begin{cases}
ER \text{, the Earth's radius (since the formula considers it is a sphere)}\\
arctan2 \text{, computing the arctan and returning an angle that can be} \\
\hspace{4.13em}\text{in any of the four quadrants of the trigonometric circle}
\end{cases}$

and where:
$\begin{cases}
\alpha=\sin(\frac{\widetilde{\ell_2}-\widetilde{\ell_1}}{2})^2 + \cos(\widetilde{\ell_1}) * \cos(\widetilde{\ell_2}) * \sin(\frac{\widetilde{L_2} - \widetilde{L_1}}{2})^2\\
\widetilde{\ell_2}=\ell_2 * \frac{\pi}{180}\\
\widetilde{\ell_1}=\ell_1 * \frac{\pi}{180}\\
\widetilde{L_1}=L_1 * \frac{\pi}{180}\\
\widetilde{L_2}=L_2 * \frac{\pi}{180}
\end{cases}$

\section{Validation}
\label{sec:validation-adjustment}

The GHVRBridge prototype was tested at various events, including several workshops at the Faculty of Architecture and Urban Planning (UMONS) and through demonstrations at the National School of Architecture in Nancy. While the participants were mostly enthusiastic about the prospect of modifying an \gls{ad} project in \gls{vr}, some of them mentioned the need to go further than parameter adjustment, with a few of them suggesting to move away from the panel-based interface (seen as mimicking standard desktop software with a point and click interaction). Both of these types of suggestions led us to develop the other 2 prototypes, with GHXR pushing the interface towards a more natural interaction and GHVRGraph (described in Chapter \ref{chap:immersive-design}) enabling control over the visual algorithm itself.

As for the GHXR prototype, we intended to follow a more rigorous validation, and planned on inviting potential users to validate the system on site. We wanted to reach participants with different profiles (both professional practitioners and individuals with a more academic background, including students, researchers and professors) and varying experience with \gls{ad} tools. 

Unfortunately, due to the sanitary situation (COVID-19 pandemic), these evaluations had to be moved to an online setting and were therefore delayed. For that purpose, we implemented an alternative version of the system, that runs on a standard desktop computer and relies on a simulator for the tangible table. 

We designed an evaluation protocol that lasts for around 40 minutes and includes questions on the participant's background, a presentation of the GHXR system, a task to perform and post-task interview. The complete protocol can be found in Appendix \ref{apx:ghxr-protocol}, including the post-task interview that mostly consist of open-ended questions on different aspects of the system. It also mentions a questionnaire, available in Appendix \ref{apx:ghxr-sus}, based on the System Usability Scale (SUS) \citep{brooke1996sus} that evaluates the usability of a system based on 10 criteria. The questionnaire we used includes an additional question about the overall user-friendliness of the system (through a 7-point scale from ``worst imaginable'' to ``best imaginable'') at the end of the survey, based on \citep{bangor2008empirical}.

We were only able to convince 3 architects, with limited experience outside of the University context, to take part in these evaluations at the time of writing this dissertation so it is hard to draw conclusions on the otherwise excellent average SUS score we achieved (86/100) with the limited number of participants, but it at least matches the way they rated the system's overall user-friendliness (2 chose ``excellent'' and 1 settled for ``good''). 

The open-ended questions brought their own light to the evaluation process. For instance, all 3 participants indicated that the surroundings were useful and that their level of detail (surrounding volumes instead of a photo-realistic representation) was sufficient for early design stages. They also all mentioned how the system's tangible interface was intuitive and how important that was for an interactive setup; they therefore see potential in the system to help with discussions between architects and other stakeholder, including clients, urban planners and construction companies.

There was however no consensus on what immersive technology was the most appropriate or useful. While 2 participants indicated the \gls{ar} holograms appearing on the table were not essential to the system with a \gls{vr} view being available, the other participant stated the \gls{ar} visualisation was the most interesting of the available views, because it allows for face-to-face discussions while visualising a project in 3D. This tends to indicate that systems that enable both \gls{ar} and \gls{vr} visualisations should be preferred, so as to allow users to choose the most appropriate medium for their specific situation.

\section{Discussion}

The two prototypes (GHVRBridge and GHXR) presented in this chapter enable \gls{vr}-based visualisation of geometries, so that designers and other stakeholders can see full-scale versions of such geometries through an immersive virtual medium. They can therefore experience what the final project will look like from different realistic angles before it is built.

Both proof-of-concept research prototypes go further than simple visualisation tools and even allow for more than extrinsic modifications to the geometries since they both address the adjustment of parameter values for \gls{ad} definitions within immersive environments. The designer therefore does not need to remove the immersive headset to change these values, and changes to the geometry are automatically forwarded to the immersive experience so that the new version can immediately be evaluated. This is a game changer compared to simple ``\gls{vr} exporting'' tools, resulting in a smoother integration of the technology into \gls{ad} practice, making it possible to move towards a user-centred design approach that involves stakeholders during the design process.

The two hypothetical use cases introduced at the end of Section \ref{sec:ad-requirements} indeed are likely to benefit from the developed prototypes. GHVRBridge is particularly suited for Use Case \ref{usecase-1}, while GHXR better correspond to Use Case \ref{usecase-2}.


The two developed prototypes differ in many ways, as summarised in Table \ref{tab:ecaade-vs-ghxr}.
We would be tempted to say GHXR is more advanced in general, since it offers all the features of GHVRBridge, and adds to it the ability to see the surrounding buildings, the option to change the geometrical representation's position, and the possibility to share an additional parameter type (lists). It also is more accomplished, with more visualisation options (\gls{ar} and non-immersive displays) as well as more adapted user interactions (through \glspl{tui}) that naturally favour collaboration. This overall superiority is quite logical considering we developed GHVRBridge in 2017-2018, and GHXR in 2021. GHXR could therefore benefit from prior developments made for GHVRBridge and was built upon them.

\begin{table}
	\begin{tabular}{>{\centering\arraybackslash}p{.19\textwidth}|>{\centering\arraybackslash}p{.35\textwidth}|>{\centering\arraybackslash}p{.35\textwidth}}
		& Prototype 1: GHVRBridge & Prototype 2: GHXR \\
		\hline
		Control & parameter values, navigation & parameter values, navigation, geometry position \\\hline
		Interaction paradigm & 3D-adapted \gls{wimp} & \gls{tui} \\\hline
		Geometry visualisation & at scale & at scale, perspective, top-down \\\hline
		Surroundings visualisation & not available & OpenStreetMap buildings data \\\hline
		Visualisation modalities & \gls{vr} & \gls{vr}, \gls{ar}, non-immersive \\\hline
		Hardware cost & $\sim$2,000\euro, including a computer & $\sim$30,000\euro \\\hline
		Transfer protocol & WebSockets & \gls{mqtt} \\\hline
		Collaboration & local or remote & mostly local, but \gls{vr} remote visualisation is possible \\
	\end{tabular}
	\caption{Comparison of the GHVRBridge and GHXR proof-of-concept research prototypes, that enable parameter adjustment of \gls{ad} definitions within immersive environments.}
	\label{tab:ecaade-vs-ghxr}
\end{table}

This does not mean that GHVRBridge has become irrelevant since the costs linked to the hardware requirements of GHVRBridge are much lower than those of GHXR. Another consideration is that the collaborative extension of GHVRBridge was designed with \emph{remote collaboration} in mind, meaning it is optimised for that purpose as explained in Section \ref{sec:ecaade-gh}. GHXR was instead built for a \emph{local setup} and works with a simple JSON format, leading to additional delays when system components have to communicate over the internet.
Even though the format could be changed to a more optimised one, the table display is where the control over the system is located, so remote collaboration would be limited to visualisation for off-site collaborators.

These characteristics mean that the two developed prototypes target different use cases, projects, and users. 
Individuals and small companies should be able to afford a system such as GHVRBridge (considering they can afford Grasshopper and therefore Rhinoceros in the first place), while only larger companies and organisations can potentially allow themselves to buy the necessary equipment for GHXR. With the many displays and views as well as the more natural interaction and collaboration, GHXR is inherently better suited for discussions that involve more collaborators than GHVRBridge, and for bigger projects (e.g., at the urban scale, to discuss with public authorities). The ability to see the surrounding buildings in GHXR also implies that it can be used to assess how a designed geometry integrates with its environment, meaning the prototype may be used for projects where the vicinity is of particular importance.

It is also worth noting that both prototypes can only handle a certain number of simultaneously modifiable parameters due to their ``client'' applications. While both Grasshopper custom components can virtually handle any number of parameters, GHVRBridge's \gls{vr} application uses a panel that is only large enough for 6 simultaneous parameters and GHXR's table device can only track 32 tangible items (about 25 of those being available for parameters, but even that number would induce significant clutter on the table, leading to an unpleasant experience). 

While GHVRBridge's panel could be extended (e.g., made larger or scrollable when more parameters are shared), the limit on GHXR's tangible items is hardware-related and adding support for additional parameters would only be possible through a rotating approach (allowing a dynamic mapping of parameters to tangible items e.g., through a side panel on the table display that would allow users to select which parameter to assign for a particular tangible item). At the same time, the need for a larger number of adjustable parameters is questionable, since both systems enable rapid \gls{vr}-based collaborative iterations and it seems more appropriate to limit the flexibility to a few crucial parameters.

As reported in Section \ref{sec:sota-immersive-ad}, since the development of GHVRBridge in 2018, other companies and research teams have commercialised and published similar tools, such as Fologram\footnote{\url{www.fologram.com}} and the previously mentioned prototype \citep{hawtonshared}. Our point of view on the integration of immersive technologies for \gls{ad} activities therefore is not an isolated opinion, as confirmed by the informal feedback we received from architects, students and laymen during the visits, workshops, and exhibitions we took part in.

While we mostly received positive feedback when presenting GHVRBridge and GHXR to architects (as stated in the previous Section), some mentioned the need to go further than mere parameter values and allow users to modify the \gls{ad} visual algorithm itself (components and links, what we called \emph{Level 3} earlier) since this would provide greater flexibility in terms of possible adjustments to the design. This pushed us to explore that possibility, which will be presented in Chapter \ref{chap:immersive-design}.

\chapter{Immersive visual programming for Algorithmic Design}
\label{chap:immersive-design}
\epigraph{
	``The ultimate display would, of course, be a room within which the computer can control the existence of matter.'' }{Ivan Sutherland}

According to the prototype-based iterative research process described in Section \ref{sec:methodology}, the proof-of-concept applications enabling \emph{Level 2} control over \gls{ad} definitions (as per Table \ref{tab:control-levels} in Section \ref{sec:ad-requirements}), that we presented in Chapter \ref{chap:immersive-adjustment}, were subject to evaluations. 
When presenting our \gls{vr}-based solutions for parameter value adjustment to architects, some of them requested to have more control over the \gls{ad} visual algorithm. Parameter value adjustment was indeed often considered to be too restrictive. This lead us to explore \gls{vr}-based \emph{Level~3} control, i.e., the ability to edit the visual algorithm, by adding or removing components or links between them.

The present chapter discusses how we achieved that goal, through the implementation of a research prototype called \textbf{GHVRGraph}.
We again develop a \gls{vr}-based editing tool but explore different interaction modalities, by relying on different techniques borrowed from Section \ref{sec:sota-3dui} that we implement with \gls{6dof} controllers as well as a hand-tracking device, but also with speech-based interaction. We then reflect on that prototype and provide insights as to how such features should be implemented in an immersive context, with adapted interaction mechanisms.

\pagebreak
\section{A graph representation of Algorithmic Design definitions}
The GHVRGraph research prototype tool enables \gls{vr}-based \emph{Level 3} control over \gls{ad} definitions. Architects using it are therefore able to manipulate components and links from the visual algorithm, from an immersive environment.

Before diving into the implementation of the GHVRGraph prototype, we had to reflect on how to represent \gls{ad} visual algorithms using an appropriate formalism that leads to a convenient data structure to be used for the prototype's implementation. 

\begin{figure}[h]
	\includegraphics[width=.9\textwidth]{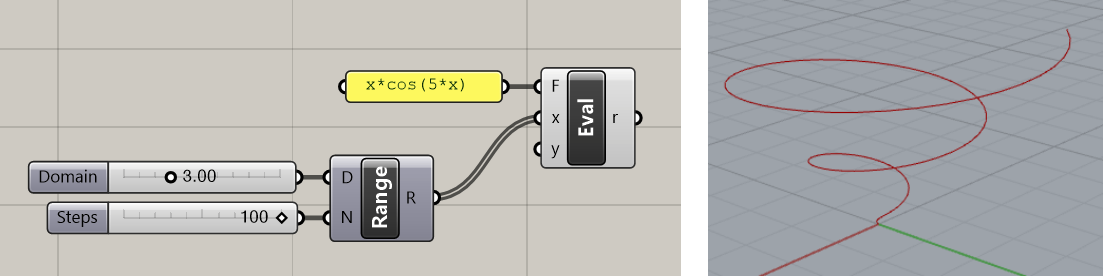}
	\centering
	\caption{\label{fig:graph-gh} Example of a Grasshopper definition, part of the more complete definition from Figure \ref{fig:gh} generating the conical spiral displayed on the right side of this figure.}
\end{figure}

We use part of the Grasshopper definition shown in Figure \ref{fig:gh} as a running example to discuss the representation of visual algorithms with that formalism. The sub-part of the definition that we keep for this purpose is shown in Figure \ref{fig:graph-gh} and includes two number components (sliders), one text panel containing a formula, and two components with different input and output ports.

As discussed in Section \ref{sec:intro-ad}, \gls{ad} algorithms, such as those created with Grasshopper, can be seen as a type of dataflow models. Dataflow models can be represented with directed graphs, that we formally define in Definition \ref{def:dg}.

\begin{definition}[]
	\label{def:dg}
	\textbf{A directed graph (digraph)} is an ordered pair $(V,E)$ such that:
	
	$\begin{cases}
	V & \text{is a set of vertices}\\
	E \subseteq V \times V & \text{is a binary relation over } V \text{ that defines a set of directed edges}
	\end{cases}$
\end{definition}

Since Grasshopper definitions cannot contain a loop (i.e., it is not possible to have a link from a component $A$ to a component $B$ if there exists a succession of links that already connect $B$ to $A$), the corresponding directed graphs would be acyclic. We therefore formally define \textbf{\glspl{dag}} in Definition \ref{def:dag}.

\begin{definition}[]
	\label{def:dag}
	\textbf{A (vertex-labelled) Directed Acyclic Graph (DAG)} is a triple $(V,E,vlabel)$ such that:
	
	$\begin{cases}
	(V,E) & \text{is a digraph}\\
	E & \text{is an acyclic relation}\\
	vlabel: V \rightarrow Label & \text{is a total function that assigns a label to each vertex}
	\end{cases}$
	where $Label$ represents the set of all possible labels.
\end{definition}

We would be tempted to convert \gls{ad} definitions to \glspl{dag}, representing components as vertices in such graphs, and links between components as edges.
However, Grasshopper algorithms also require to specify input and output ports on components. Examples of such ports in Figure \ref{fig:graph-gh} include the ones from the \texttt{Range} component, that has two input ports, respectively for the \texttt{Domain}, \texttt{Steps} parameters, as well as one output port returning the resulting \texttt{Range}).

\begin{figure}
	\includegraphics[width=.5\textwidth]{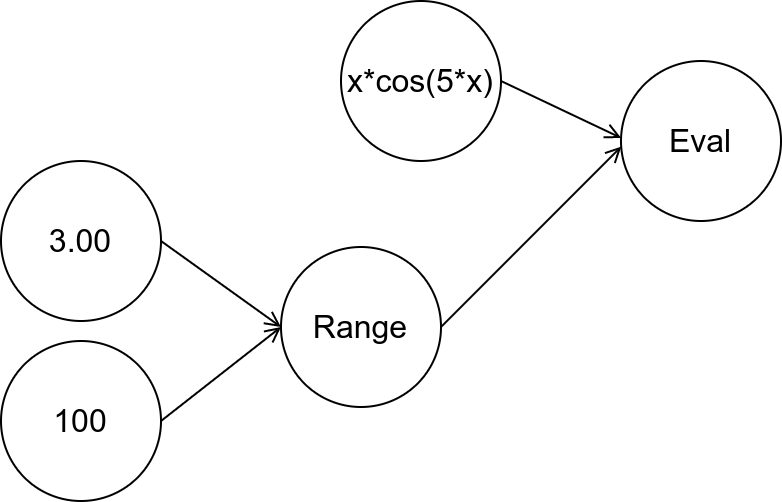}
	\centering
	\caption{\label{fig:graph-dag} An oversimplified \gls{dag} representation for the example previously given in Figure \ref{fig:graph-gh}. Ports are not represented and the corresponding information is lost.}
\end{figure}

For that reason and as exhibited on Figure \ref{fig:graph-dag}, a basic (vertex-labelled) \gls{dag} is not enough to convey that information: it is no longer possible to know which ports are used by the edges based on that representation alone, and that information would consequently be lost. 

A solution to that problem would be to additionally label the edges with the ports they are connected to, but we discarded that option because of implementation details. It would in fact be impractical since we would have to iterate over all the edges going from or to a component, in order to find the edges that only concern a specific port from that component. 
Navigating the graph and converting it to the Grasshopper file format when saving changes would then become more complex and may lead to clumsy code. We therefore opted for another solution representing ports as vertices, that we describe hereafter.

We chose to represent \gls{ad} definitions using a specific graph-based formalism that is well-known in graph transformation theory \citep{ehrig2004fundamental}. To be more specific, we decided to make use of a \textbf{type graph} in order to formally specify what a valid \gls{ad} definition is composed of, and \textbf{typed graphs} to represent such definitions. 

This is similar to the representation of software design models (e.g., statecharts and class diagrams) in \citep{mens2005use}, where typed graphs are used in order to specify model refactorings using graph transformations.

\begin{figure}
	\includegraphics[width=.32\textwidth]{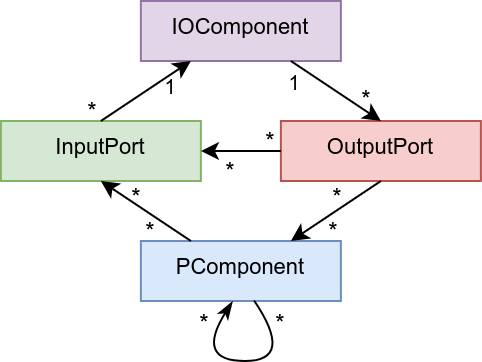}
	\centering
	\caption{\label{fig:graph-typegraph} The typegraph for the graph structure. An arrow from one element $A$ to another element $B$ indicates that it is possible for an edge to connect an element of type $A$ to an element of type $B$.}
	
	\vspace*{\floatsep}
	
	\includegraphics[width=.97\textwidth]{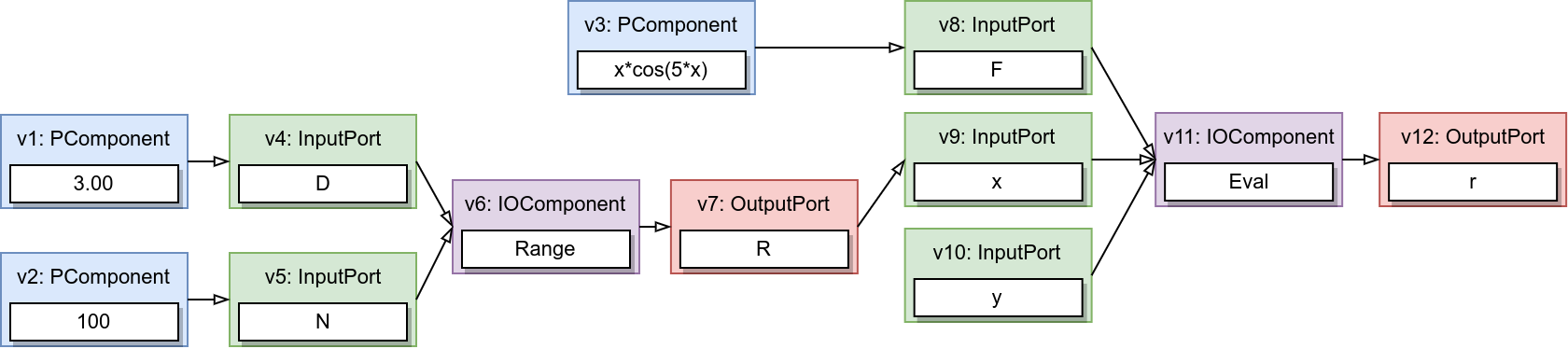}
	\centering
	\caption{\label{fig:graph-typedgraph} The vertex-labelled, typed, acyclic directed graph for the example given in Figure \ref{fig:graph-gh}, converted to the typed graph representation.}
\end{figure}

Definition \ref{def:typedgraph} formally defines the notion of typed graph, as we used it in the context of visual \gls{ad}.

\pagebreak 

\begin{definition}[]
	\label{def:typedgraph}
	\textbf{A typed graph} $G$ is a 5-tuple $(V,E, vlabel, TG, vtype)$ such that:
	
	$\begin{cases}
	(V,E,vlabel) & \text{is a (vertex-labelled) DAG}\\
	TG = (TV,TE) & \text{is a digraph}\\
	vtype:V \rightarrow TV & \text{is a total function that assigns a type to each vertex}\\
	\forall (v_i,v_j) \in E: & \big(vtype(v_i), vtype(v_j)\big) \in TE
	\end{cases}$
	
\end{definition}
We call $TG$ the type graph and $G$ the typed graph, and we say that $G$ is typed over $TG$.

In the case of \gls{ad} visual algorithms, the digraph that defines valid definitions is depicted in Figure \ref{fig:graph-typegraph}, and formally described in Example \ref{ex:typegraph}.

\begin{example}[]
	\label{ex:typegraph}
	The digraph $TG = (TV,TE)$, that is depicted in Figure \ref{fig:graph-typegraph}, describes valid \gls{ad} graphs, with:
	
	$\begin{cases}
	TV= &\{\text{IOComponent},\text{InputPort},\text{OutputPort},\text{PComponent}\}\\
	TE= &\{(\text{InputPort},\text{IOComponent}), \allowbreak(\text{IOComponent},\text{OutputPort}), \\
	& \hspace{0.5em} (\text{OutputPort},\text{PComponent}), \allowbreak(\text{PComponent},\text{InputPort}), \\
	& \hspace{0.5em} (\text{OutputPort},\text{InputPort}), \allowbreak(\text{PComponent},\text{PComponent})\}
	\end{cases}$
\end{example}

Figure \ref{fig:graph-typedgraph} depicts the typed graph that corresponds to the running Grasshopper definition example from Figure \ref{fig:graph-gh}.

This representation of Grasshopper definitions resembles the \gls{gpm}, proposed in \citep{janssen2015types}, which also includes a reflection on how to represent \gls{ad} visual definitions using graphs. The \gls{gpm} distinguishes data nodes from operation nodes (geometric or computational processing of incoming data), and considers ports as data nodes. The model therefore relies on an extended version of a \gls{dag}, where these two types of nodes are allowed. We however need to differentiate predefined ports from input data and output channels at the interaction level, to prevent invalid modifications to the \gls{ad} definition (e.g., removing a link from an input port to its associated component).

While conceptually interesting, the idea of combining ports and data into a single notion of ``data node'' is therefore impractical for implementation purposes. The authors behind \gls{gpm} indeed state that they only intend to provide an analytical device, not discuss actual implementation \citep{janssen2015types}.
We therefore rely on the presented typed graph formalism to represent \gls{ad} visual definitions. 

That formalism also has the advantage that it comes with tools such as AGG \cite{taentzer1999agg}, allowing automated checking of graph transformations and constraints, so we could have relied on such tools to validate manipulations made to the graph representation in our application. Unfortunately, \gls{ad} visual algorithms come with specific constraints (e.g., most components have a set of input and output ports that is predefined and cannot be changed by the user) and concepts (e.g., number parameters come with specific settings restricting the range of possible values) that make it hard to fully rely on such a formalism. We therefore opted to constrain and check the validity of manipulations directly in our code.

In order to prevent the aforementioned invalid modifications from occurring, GHVRGraph must for instance prevent users from adding or removing an edge (link) between a port and its component. It should also not allow them to add or remove a port on a component.

\section{Interoperability with Grasshopper}
\label{sec:ghvrgraph-interoperability}

In order to replicate the representation discussed in the previous section and to allow users to manipulate the resulting graph, we rely on the Unity-specific version\footnote{\url{https://github.com/davidgutierrezpalma/quickgraph4unity}} of the QuickGraph\footnote{\url{https://github.com/YaccConstructor/QuickGraph}} library. This allows us to preserve a certain degree of genericity since other graph-based representations could rapidly be adapted to work with the \gls{vr}-based environment and interactions we developed. The library also enables us to use graph algorithms such as topological sorting and leaf-finding functions.

\subsection{Defining vertex objects}
The QuickGraph library works with graphs whose vertices can only be of a single concrete type i.e., the corresponding class cannot be abstract. We therefore circumvent this limitation by defining a concrete \texttt{Vertex} class that contains a \emph{chunk} field. That field corresponds to an abstract \texttt{Chunk} class, from which we can define a hierarchy of vertex types to allow us to handle the different types of vertices required for representing \gls{ad} definitions.

\begin{figure}
	\includegraphics[height=.8\textwidth, angle=90]{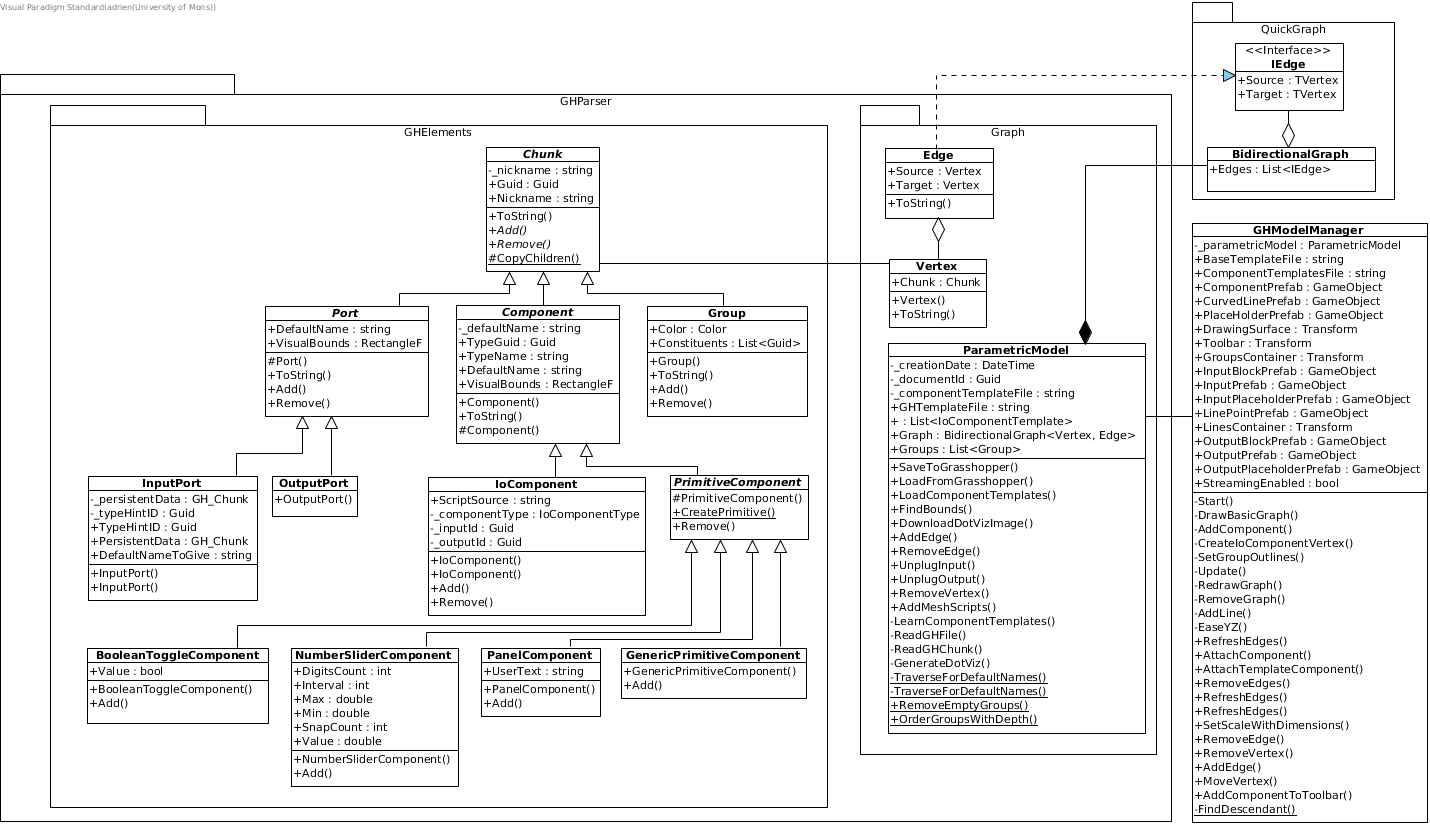}
	\centering
	\caption{\label{fig:ghvrgraph-class-split01} A section of the class diagram for GHVRGraph, showing the hierarchy of classes we use to represent Grasshopper ``chunks'' in the research prototype, as well as how the graph structure is stored using QuickGraph.}
\end{figure}

The hierarchy of classes we use for the GHVRGraph prototype tool is shown on the \texttt{GHElements} package in the class diagram of Figure \ref{fig:ghvrgraph-class-split01}. It is mostly based on the internal representation of the corresponding elements in the Grasshopper desktop tool. All vertex types indeed inherit from the aforementioned \texttt{Chunk} abstract class, and a vertex can be of type \texttt{Port}, \texttt{Component}, or \texttt{Group}.

Except for the \texttt{Group} class that will be discussed in Section \ref{sec:gh-groups}, these classes are also abstract, and concrete objects must therefore be of a more specific type. An \texttt{IoComponent} is a component that has at least one input or output port, and can have both types of ports, while a \texttt{PrimitiveComponent} correspond to a vertex that contains input data or is an output channel (a \texttt{PanelComponent} can be both). 

We additionally created a \texttt{GenericPrimitiveComponent} to handle primitive components that do not fall into the previous three categories. This allows us to successfully read from and write to Grasshopper files that contain such components.

\subsection{Converting Grasshopper files to a graph-based representation}

While the exact implementation details are out of scope of this document, we will briefly discuss the main ideas as well as the Grasshopper file format and a challenge we encountered when parsing such files. When saving a definition within the Grasshopper desktop application, the user is offered two format options: the default binary \texttt{.gh} format, or a XML-based \texttt{.ghx} format. 

A Grasshopper definition is essentially stored as a list of XML \citep{xml} entities called \emph{chunks}. In addition to ``header'' chunks that contain the likes of versioning information, timestamps and some user preferences, most of the content in such a file is within the \texttt{DefinitionObjects} chunk, that contains ``sub-chunks'' that correspond to interactable entities such as components and input data. The corresponding XML element contains various information, such as a type identifier, a name, the display boundaries within the definition, and an instance identifier that uniquely identifies an element in the definition. Depending on the particular type of element, sub-chunks may be present, for instance to give information about a component's ports. 

Links are indirectly stored within these sub-chunks, via \texttt{Source} items, that indicate the instance identifiers of entities that link to the current entity. 
In order to obtain the vertex-labelled typed graph, we must process these \texttt{Source} items and therefore construct the graph backwards (with regards to the direction of the edges).

Since there is no guarantee that a unique identifier referenced by a \texttt{Source} item corresponds to an element that has already been processed, we need to be able to create temporary placeholder components, to construct the graph with a ``placeholder'' for the expected component until we reach that component's description in the file. 




\subsection{Groups and clusters}
\label{sec:gh-groups}

In Grasshopper, there is a grouping concept that allows designers to divide and structure their work in sets of components, similarly to how a software developer would separate some code into multiple classes and files. Examples of \textbf{groups} can be seen on Figure \ref{fig:gh-example}, where they are represented by coloured rectangles that enclose all their components.
In the Grasshopper file format, a group is represented as a chunk that contains a list of one or more \texttt{ID} items to reference the constituents' unique identifiers. 

There is no operational semantics linked to that group concept, in the sense that placing components in a group cannot modify the generated output geometry.
Since groups only constitute a visual aid for the designer, we do not include them into the typed graph representation but still create a \texttt{Group} class to match the grouping concept and handle such groups in GHVRGraph. This allows us to display them in the \gls{vr}-based representation and preserve the grouping information in the save files generated from GHVRGraph.


A related element in Grasshopper is the concept of a \textbf{cluster}, that also bundles components together. Unlike groups however, it serves as a black box that summarises its constituents' input and output ports. This black box is reusable, in the sense that multiple copies of a cluster can be included in the definition.

A user can choose to look into a cluster that was created earlier and modify its constituents ; doing so will impact all copies of that cluster.
The cluster concept therefore goes beyond simple visual aid, but clusters are unfortunately stored as an encrypted item in the Grasshopper file format. We consequently cannot fully support clusters in GHVRGraph, because we have no way of allowing \gls{vr} users to ``open'' a cluster and interact with its constituents.


\subsection{Immersive visual representation}
\label{sec:ghvrgraph-representation}

In order to allow designers to manipulate Grasshopper definitions in \gls{vr}, we needed to represent the typed graph, that we construct from the imported Grasshopper definition, within the immersive environment. To provide \gls{vr} users with interaction mechanisms that are appropriate to the 3D context, the graph representation should be three-dimensional. That representation should allow \gls{vr} users to easily manipulate interactable objects such as components and links, through the corresponding vertices and edges. The 3D representation should also maintain a mapping with the Grasshopper desktop (2D) canvas, so that changes to the placement of components are preserved when switching between the desktop tool and the GHVRGraph application (e.g., moving an input parameter closer to a related input or output component).

\begin{figure}
	\includegraphics[width=.95\textwidth]{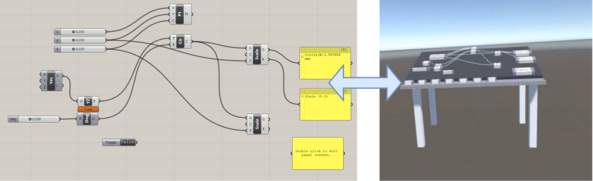}
	\centering
	\caption{\label{fig:ghvrgraph-table} A Grasshopper definition and its corresponding immersive visualisation in GHVRGraph, showing the table metaphor we use.}
\end{figure}

We decided to rely on a table metaphor to represent the typed graph in \gls{vr} within GHVRGraph. Figure \ref{fig:ghvrgraph-table} shows a side by side comparison of the same \gls{ad} definition, represented in the Grasshopper desktop tool on the left, and in GHVRGraph on the right.
In front of the \gls{vr} user's starting position, we place a virtual table on which we add boxes that roughly resemble the corresponding Grasshopper components, albeit in a three-dimensional form. 

Most of the developments for the GHVRGraph prototype happened during the one-month long 2019 edition of the eNTERFACE workshop\footnote{\url{http://web3.bilkent.edu.tr/enterface19/}}, where we gathered early feedback from workshop attendees (about 60 participants). Based on their reports and on the literature on ergonomics, that suggests the preferred display zone should be below the horizontal line of sight \citep{PheasantBodyspaceAnthropometryergonomics2018}, we opted for the metaphor of a virtual table instead of a board-based metaphor.

Another suggestion we received from the aforementioned workshop participants was to slightly incline the table, since this would allow the user to place himself in a position where the table is angled towards him. In fact, a flat and horizontal surface would require the user to incline his neck further in order to look down on the table. The suggestion to incline the surface is supported by the recommendation on viewing angle in \citep{PheasantBodyspaceAnthropometryergonomics2018} and is corroborated by a study specifically addressing neck pain \citep{yip2008relationship}, that concludes that the associated pain decreases as the user gets closer to a forward head posture (i.e., a 90\degree \, angle between the neck and the horizontal line of sight).


\section{Exploring interaction techniques for graph manipulation} 
\label{sec:ghvrgraph-interactions}

Displaying a 3D rendering of a graph-based \gls{ad} definition on a virtual table is only a first step. Designers need ways to manipulate components and links, not just visualise them. Based on the immersive interaction techniques identified in Section \ref{sec:sota-3dui} as well as popular and affordable devices, we explored different options to enable ``basic editing'' of \gls{ad} definitions and their underlying typed graphs in \gls{vr}. By ``basic editing'' we understand the ability to add and remove components and links (not individual ports since we want to preserve a valid mapping with Grasshopper). We additionally allow users to move components, i.e., position them differently, without changing their connections.

The set of interaction devices we used to explore \gls{vr}-based graph editing included: (1) the HTC Vive \gls{hmd} together with its associated \gls{6dof} controllers; and (2) the Leap Motion hand-tracking system. Using them, we explored different techniques as different variants of the same GHVRGraph prototype. Table \ref{tab:enterface} gives an overview of these prototype variants, named \textcolor{blue}{$P_1$}, \textcolor{cyan}{$P_{1}^{speech}$} and \textcolor{red}{$P_2$}, and how they cover various modalities and techniques, based on the categories we presented in Section \ref{sec:sota-3dui}.

\begin{table}[]
	\resizebox{\textwidth}{!}
	{
\begin{tabular}{|c|c|c|c|c|c|c|c|c|}
	\hline
	\multicolumn{2}{|c|}{\multirow{2}{*}{\diagbox{Actions}{Techniques}}} & \multicolumn{3}{c|}{Modality}                                                                    & \multicolumn{4}{c|}{Interaction type}                                                                                       \\ \cline{3-9} 
	\multicolumn{2}{|c|}{}                                    & \multicolumn{1}{c|}{6-DoF controller} & \multicolumn{1}{c|}{Hands} & \multicolumn{1}{c|}{Speech} & \multicolumn{1}{c|}{Direct} & \multicolumn{1}{c|}{Indirect} & \multicolumn{1}{c|}{Grasping} & \multicolumn{1}{c|}{Pointing} \\ \hline
	\multirow{3}{*}{Component} 
	& Add  & \textcolor{blue}{$P_1$} & \textcolor{red}{$P_2$} & \textcolor{cyan}{$P_{1}^{speech}$}, \textcolor{red}{$P_2^{*}$} & \textcolor{blue}{$P_1$}, \textcolor{red}{$P_2$} & \textcolor{cyan}{$P_{1}^{speech}$}, \textcolor{red}{$P_2^{*}$} & \textcolor{blue}{$P_1$}, \textcolor{red}{$P_2$} & \textcolor{red}{$P_2$} \\ \cline{2-9} 
	& Remove & \textcolor{blue}{$P_1$} & \textcolor{red}{$P_2$}  &  \textcolor{red}{$P_2^{*}$}   & \textcolor{blue}{$P_1$}, \textcolor{red}{$P_2$}  &   \textcolor{red}{$P_2^{*}$}   & \textcolor{blue}{$P_1$}, \textcolor{red}{$P_2$}   & \textcolor{red}{$P_2$}  \\ \cline{2-9} 
	& Move & \textcolor{blue}{$P_1$}  & \textcolor{red}{$P_2$}  &  \textcolor{red}{$P_2^{*}$}   & \textcolor{blue}{$P_1$}, \textcolor{red}{$P_2$} &   \textcolor{red}{$P_2^{*}$}   & \textcolor{blue}{$P_1$}, \textcolor{red}{$P_2$}   & \textcolor{red}{$P_2$}  \\ \hline
	\multirow{2}{*}{Link}  
	& Add & \textcolor{blue}{$P_1$}  & \textcolor{red}{$P_2$}  &  \textcolor{red}{$P_2^{*}$}   & \textcolor{blue}{$P_1$}, \textcolor{red}{$P_2$} &   \textcolor{red}{$P_2^{*}$}   & \textcolor{blue}{$P_1$}, \textcolor{red}{$P_2$}   & \textcolor{red}{$P_2$}  \\ \cline{2-9} 
	& Remove  & \textcolor{blue}{$P_1$} & \textcolor{red}{$P_2$}  &   \textcolor{red}{$P_2^{*}$}   & \textcolor{blue}{$P_1$}, \textcolor{red}{$P_2$}  &   \textcolor{red}{$P_2^{*}$}   & \textcolor{blue}{$P_1$}, \textcolor{red}{$P_2$}   & \textcolor{red}{$P_2$} \\ \hline
\end{tabular}
}
\caption{Modalities and interaction types we explored to implement \gls{vr}-based graph-editing actions in the variants of GHVRGraph, \textcolor{blue}{$P_1$}, \textcolor{cyan}{$P_{1}^{speech}$} and \textcolor{red}{$P_2$}. The asterisk sometimes used for $P_2^{*}$ denotes that the modality or interaction type is used for the corresponding action but needs to be combined with another modality or interaction type to produce a result.
}
\label{tab:enterface}
\end{table}

\subsection{Variant $P_1$: grasping metaphor with 6-DoF controllers}

\begin{figure}
	\centering
	\begin{subfigure}[t]{.33\textwidth}
		\centering
		\includegraphics[height=2.3cm]{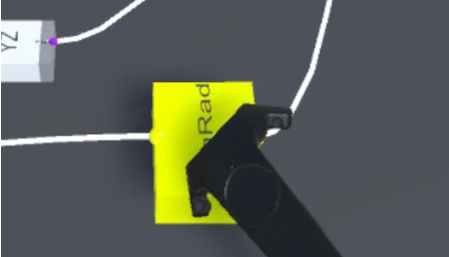}
		\centering
		\caption{\label{fig:p1-grab} Grabbing a component.}
	\end{subfigure}
	\begin{subfigure}[t]{.4\textwidth}
		\centering
		\includegraphics[height=2.3cm]{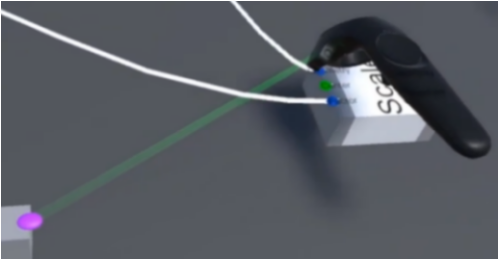}
		\caption{\label{fig:p1-addlink} Adding a link.}
	\end{subfigure}
	\begin{subfigure}[t]{.25\textwidth}
		\centering
		\includegraphics[height=2.3cm]{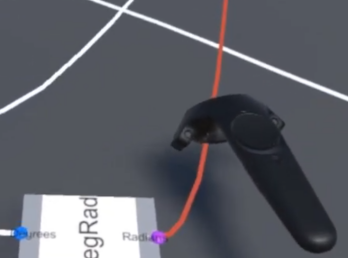}
		\caption{\label{fig:p1-removelink} Removing a link.}
	\end{subfigure}
	\caption[]{A user interacting with a \gls{6dof} controller in variant $P_1$ of GHVRGraph.}
	\label{fig:p1}
\end{figure}

Variant $P_1$ of GHVRGraph relies on the default \gls{6dof} controllers provided with the HTC Vive headset. We chose for an isomorphic interaction technique based on the grasping metaphor: the user simply touches the element he wants to interact with, and presses a button to trigger the corresponding action. Figure \ref{fig:p1-grab} shows a user that is about to grasp and start moving a component, while Figures \ref{fig:p1-addlink} and \ref{fig:p1-removelink} show a user interacting with links.

The user selects and interacts with an element by pressing a button on the controller when in reach of the said element. If the selected element is a link, we simply remove that link from the graph. If it is a component, we attach it to the controller (that type of interaction is often referred to as the grasping metaphor). The user can then choose to release it elsewhere on the table (realising the ``move vertex'' action) or throw it away (``remove vertex'' action). In order to add a link between two components, the user needs to select two ports consecutively. After selecting the first port and prior to selecting the second one, a temporary line between the selected port and the controller is rendered so as to give feedback to the user on the port that has been interacted with. Note that adding a link is prevented if that link would create a cycle in the graph structure (since Grasshopper does not allow that to happen).

A video demonstrating variant $P_1$ of GHVRGraph is available online\footnote{\url{http://informatique.umons.ac.be/staff/Coppens.Adrien/?video=eNTERFACE2019}} and our codebase is hosted as open-source software on a GitHub repository\footnote{\url{https://github.com/qdrien/Grasshopper-VR-graph}}, available under the MIT\footnote{\url{https://opensource.org/licenses/MIT}} license. We also published a paper that further describes the system and how we explored different interaction techniques \citep{coppens2019exploration}.

\subsection{Variant $P_{1}^{speech}$: Speech recognition}

In addition to direct interaction techniques, specific actions can be simplified by relying on indirect approaches such as speech recognition. Since specifying an arbitrary position or selecting an existing object is easily and rather naturally done with a direct technique (e.g., relying on a grasping or pointing metaphor), speech recognisers are often used in a multimodal context when applied to 3D selection or manipulation tasks (e.g., the ``Put-That-There'' metaphor \citep{bolt1980put}). We decided to explore this modality and integrate it as part of and extended version of $P_1$: variant $P_{1}^{speech}$.

An important distinction between speech engines is whether (and how much) they restrict potential input. Free speech recognisers can output any text, whereas directed dialogue \citep{pieraccini2005we} systems are limited to a set of predefined words or commands. Directed approaches can mostly be found in two forms: keyword-spotting solutions that extract specific words; and grammar-based tools that produce phrases defined by specific rules.

GHVRGraph, as a \gls{vr}-based graph editing application, would benefit from vocal commands such as ``Add component X'' to create new components in the graph.
Even though free speech and keyword-based approaches could be used for that purpose, they would not guarantee that a valid output is returned by the speech recogniser and would require post-processing of that output to parse it (which sequences of words or keywords are valid, and what action they correspond to). Grammar-based engines therefore seem to be the best option as only valid vocal commands, with regards to the grammar, can be recognised. 

Since GHVRGRaph is developed with the Unity game engine, we can benefit from the engine's built-in support for the Windows Speech Recognition \gls{api}\footnote{\url{https://docs.microsoft.com/en-us/windows/apps/design/input/speech-recognition}}, that includes an XML-based grammar recogniser for \gls{srgs}. \gls{srgs} is a W3C standard\footnote{\url{https://www.w3.org/TR/speech-grammar/#S1}} that describes a grammar format. Similarly to the grammars from compiler theory \citep{aho1986compilers}, a \gls{srgs} grammar describes a set of rules composed of tokens, using either an XML or an augmented BNF (Backus-Naur Format) syntax.

\gls{srgs} grammars can be augmented with Semantic Interpretation for Speech Recognition (SISR\footnote{\url{https://www.w3.org/TR/semantic-interpretation/}}) tags that contain ECMAScript (JavaScript) code to be executed when the corresponding grammar rule is matched. Those tags are typically used to assign values to variables for a matched rule. For instance, a boolean variable can see its value set to \verb|true| when the matched text is ``yes'', ``ok'' or ``yeah''. Similarly, such tags can handle numbered values, so that a user saying ``three'' could assign the value \verb|3| to a certain variable (e.g., called \verb|outValue|). It is also possible to go further and perform specific operations on such a variable when specific words are recognised. For example, saying ``thousands'' could multiply the previous \verb|outValue| variable by \verb|1000|.

While \gls{srgs} alone is enough to define commands such as ``Add component circle'' or ``Add boolean toggle'', the capabilities of SISR are interesting to allow GHVRGraph users to add valued components directly (in one go), with commands such as ``Add slider with value 7'' or ``Add boolean toggle with value true''. While alphanumeric input in grammars is non-trivial \citep{wang2004creating}, we relied on an existing set of rules provided in the Microsoft Speech Platform SDK\footnote{\url{https://www.microsoft.com/en-us/download/details.aspx?id=27226}}.

\begin{figure}
	\centering
	\begin{subfigure}[t]{.5\textwidth}
		\centering
		\includegraphics[width=.35\linewidth]{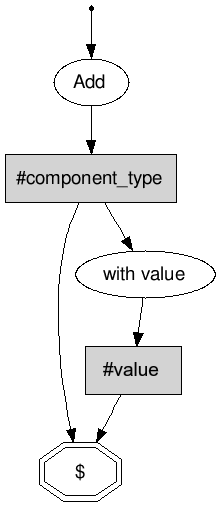}
		\caption{A visual representation subsuming the grammar rules used to add components.}
		\label{fig:enterface-speech-diagram}
	\end{subfigure}\hfill%
	\begin{subfigure}[t]{.45\textwidth}
		\centering
		\includegraphics[width=.85\linewidth]{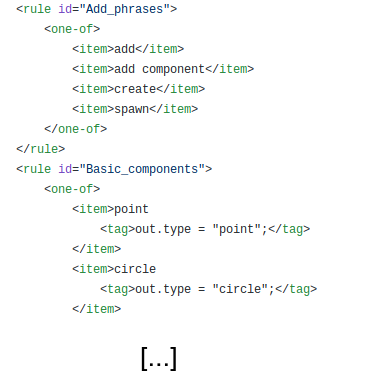}
		\caption{Part of the corresponding definition in the XML-based \gls{srgs} format.}
		\label{fig:enterface-speech-xml}
	\end{subfigure}
	\caption[]{The \gls{srgs}/SISR rules used to provide vocal commands allowing users to add components in $P_1^{speech}$.}
	\label{fig:enterface-speech}
\end{figure}

Figure \ref{fig:enterface-speech} presents part of the rules we use to provide vocal commands allowing users to add both valued and non-valued components in $P_{1}^{speech}$. Figure \ref{fig:enterface-speech-xml} correspond to a small portion of the rules in the textual (\gls{srgs}) format, while Figure \ref{fig:enterface-speech-diagram} clarifies the general structure of the ``add component'' command.

As users may want to assign an arbitrary text value to a vertex (e.g., a panel component), we also incorporated a free speech recogniser, that starts listening to user input only when specific grammar rules have been processed. In the meantime, the grammar recognition engine is paused, and it only resumes when the user stops providing free speech input.

A potential lead to further improve the integration of the speech modality in GHVRGraph would be to rely on machine learning techniques. Such techniques could be used to process free speech input and directly convert it to valid commands (with regards to the \gls{srgs} grammar). There would be no need to split the recognition of text-valued components from other components. This could also increase the flexibility of the system, in the sense that alternative wording for the same command could be learned by the machine learning algorithm, that would convert that wording to its equivalent in the grammar specification.

\subsection{Variant $P_2$: grasping and pointing metaphor with a hand-tracking system}
\label{sec:ghvrgraph-p2}

The goal of variant $P_2$ of GHVRGraph is to explore interaction techniques while relying on a hand-tracking sensor. We used the Leap Motion\footnote{\url{https://www.leapmotion.com/}} sensor, that is composed of two standard colour cameras and infrared LEDs, that give it the ability to scan the space above its surface, up to $\sim$60 cm away from the device. The development of this variant was started during the previously mentioned eNTERFACE workshop and continued as part of a Computer Science master student project, under my direct supervision \citep{robin}.

$P_2$ comes in two modes: grasping and pointing. The former replicates the grasping metaphor used by $P_1$, replacing \gls{vr} controllers by the user's hands with a grasping gesture acting as a substitute for the controller's button press to start a manipulation. Once grabbed, a component can be moved and the manipulation stops when the user releases his hand.
Interacting by grasping objects is natural to humans, but it only allows users to interact with objects that are sufficiently close to the user, at hand's reach.

The pointing-based mode mitigates that limitation by expanding the range of interaction for \gls{vr} users to objects they can point to, by using a raycasting method (see Section \ref{sec:sota-3dui}). The tip of the forefinger from the user's main hand is used as the starting point of the beam, using the closest phalanx to derive the beam's direction (from the phalanx to the tip). To trigger an action, the user speaks a simple ``action'' vocal command. Depending on the type of object that is pointed at and the state of the application, the corresponding action is performed. For instance, if nothing is selected and the user triggers the vocal command while pointing at a component, that component will be selected and become moveable, until it is released by another ``action'' command.

\begin{figure}
	\centering
	\includegraphics[height=4.5cm]{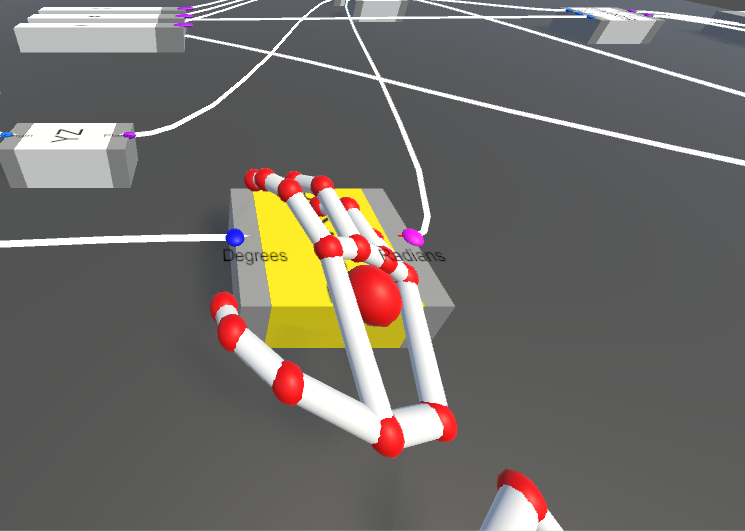}
	\centering
	\caption{\label{fig:p2-grab} A user about to grab a component in variant $P_2$ of GHVRGraph, that relies on a hand-tracking device.}
\end{figure}

\begin{figure}
	\centering
	\begin{subfigure}[t]{.49\textwidth}
		\centering
		\includegraphics[height=3.8cm]{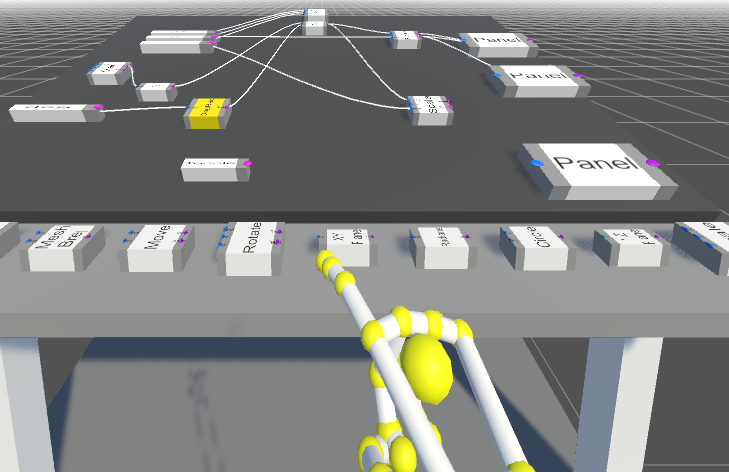}
		\caption{\label{fig:p2-point-select} Pointing to a \mbox{component}.}
	\end{subfigure}
	\begin{subfigure}[t]{.49\textwidth}
		\centering
		\includegraphics[height=3.8cm]{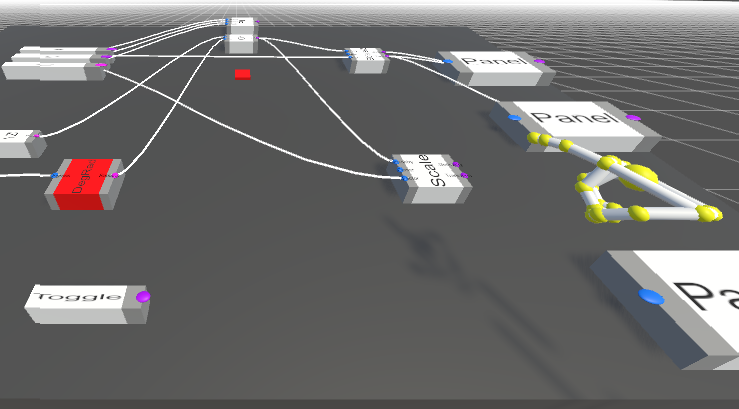}
		\caption{\label{fig:p2-point-place} Placing the selected component.}
	\end{subfigure}
	\caption[]{The two-step process to move a component in variant $P_2$ of GHVRGraph, illustrating the pointing mode of that variant.}
	\label{fig:p2-point}
\end{figure}

The grasping mode is illustrated in Figure \ref{fig:p2-grab}, where a user is about to grasp a component to manipulate it. The component is highlighted in yellow to indicate that the hand is currently touching it and it could consequently be interacted with if the user were to perform the grasping gesture.
The pointing mode is depicted in Figure \ref{fig:p2-point}, split into sub-figures \ref{fig:p2-point-select} and \ref{fig:p2-point-place}, where a user respectively selects and moves a component based on his forefinger's position.
Implementing both approaches allows to see how they compare in the context of graph-based 3D interaction. 

To do so, evaluations based on both interaction modes of variant $P_2$ were conducted in \citep{robin}. These evaluations were limited to 4 participants, partly due to the sanitary situation at the time, but showed the main advantages and drawbacks of each method. The grasping metaphor turned out to be intuitive and precise for interacting with elements at close range, but not ideal for objects further away. On the other hand, the pointing mode of $P_2$ was considered easy to use for both short and close range, but suffered from precision issues, especially for interactions on edges (since they are thin and therefore hard to point at).

Videos demonstrating this prototype are available online, for the grasping\footnote{\url{https://www.youtube.com/watch?v=Su3y1tnZk1s}} and pointing mode\footnote{\url{https://www.youtube.com/watch?v=oxgMxh4Fkx4}}.

\section{Immersive visualisation of resulting structures} 

Similarly to the two prototypes presented in Chapter \ref{chap:immersive-adjustment} (GHVRBridge and GHXR), providing \gls{vr}-based editing capabilities is much more interesting if the output result (the geometrical representation) can be visualised from within the same immersive experience.
We can unfortunately not directly borrow the mesh streaming approach from GHVRBridge (of Section \ref{sec:ecaade-vr}). That is because, unlike GHVRBridge, GHVRGraph does not simply apply user modifications through a custom Grasshopper component, but instead produces a new file in the Grasshopper format. In order to generate the new geometrical representation, we therefore need to reload the file in Grasshopper, and that would disconnect the custom component from the \gls{vr} experience.

We consequently approached the issue in a different manner and relied on Rhino.Inside\footnote{\url{http://rhino3d.com/inside}}, a project that embeds Rhino and Grasshopper into other applications. Using Rhino.Inside, we can run an instance of a Grasshopper within our Unity application and communicate with it. 

On launch, GHVRGraph starts an instance of Rhino and Grasshopper, through Rhino.Inside. When modifications made to the visual algorithm via the \gls{vr} interface are saved, a new Grasshopper file is produced and sent to the Grasshopper instance that runs within GHVRGraph. Once received, Grasshopper generates the geometrical representation and it therefore becomes available for sharing. To do so, GHVRGraph includes a custom sharing component inside the file sent to Grasshopper, that thereby automatically sends the geometry (in the same mesh-based format as in Section \ref{sec:ecaade-vr}) to the \gls{vr} experience.

\begin{figure}[h]
	\includegraphics[width=.95\textwidth]{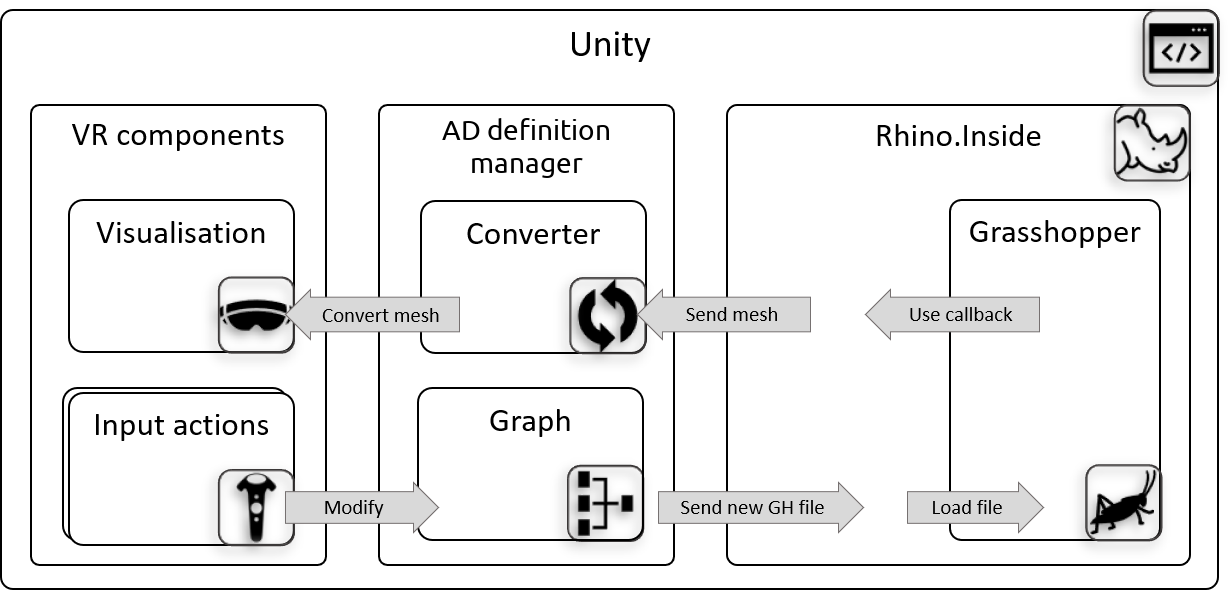}
	\centering
	\caption{\label{fig:ghvrgraph-viz} The proposed approach for \gls{vr}-based editing of \gls{ad} definitions with GHVRGraph, enabling geometry visualisation through Rhino.Inside.}
\end{figure}

This rather complex process is pictured in Figure \ref{fig:ghvrgraph-viz}.

\section{Validation}

Since most of GHVRGraph's development happened during the aforementioned eNTERFACE workshop, we were able to ask other workshop participants (with different backgrounds) to try the prototype along the way and give their feedback. As previously mentioned, this led us to use a table metaphor with a slightly inclined table for our graph visualisation.

Additionally, as part of the survey discussed in Section \ref{sec:survey} and as mentioned there, we included a video in order to show respondents the GHVRGraph prototype. All respondents that indicated being aware of \gls{ad} tools were shown the video, that demonstrated the prototype, with a \gls{vr} user editing a Grasshopper project. After watching the video, respondents were asked about the usefulness of such \gls{vr}-enabled functionality. 

\begin{figure}
	\centering
	\begin{subfigure}[t]{0.48\textwidth}
		\centering
		\includegraphics[width=\textwidth]{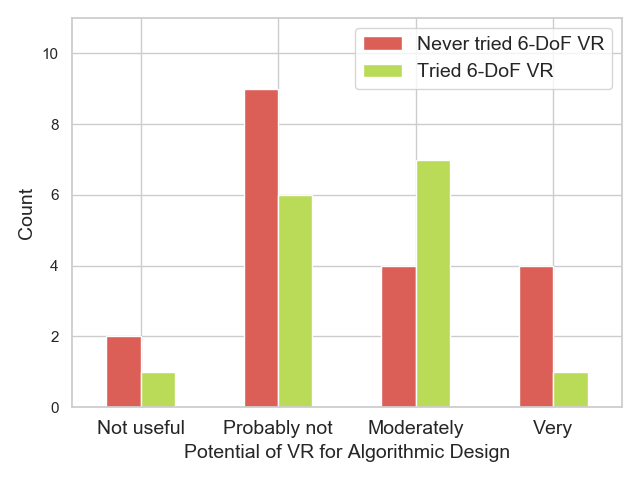}
		\caption{Usefulness of \gls{vr} for \gls{ad}, broken down by respondents' prior exposure to \mbox{\gls{6dof}} \gls{vr} devices and experiences.}
		\label{fig:survey-ad-vs-tried-vr-bis}
	\end{subfigure}
	\hfill
	\begin{subfigure}[t]{0.48\textwidth}
		\centering
		\includegraphics[width=\textwidth]{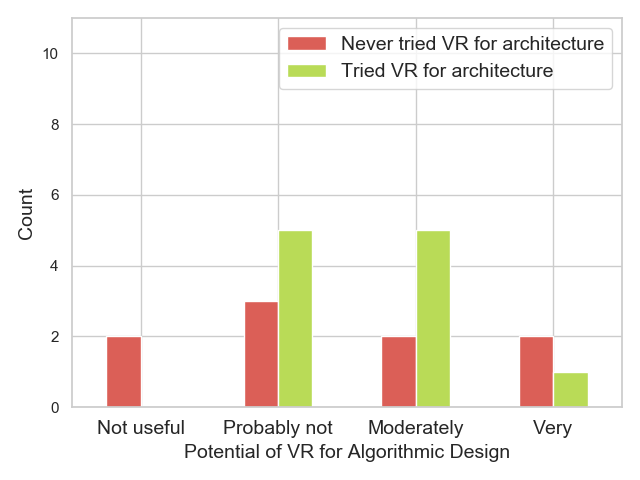}
		\caption{Usefulness of \gls{vr} for \gls{ad}, broken down by respondents' prior exposure to \gls{vr} tools for architecture (if they were familiar with \gls{vr}).}
		\label{fig:survey-ad-vs-tried-vr-archi-bis}
	\end{subfigure}
	\caption{Usefulness of \gls{vr} for Architectural Design, reproduced from Figure \ref{fig:survey-ad}.}
	\label{fig:survey-ad-bis}
\end{figure}

The answers we received for that question revealed a mixed reception, with slightly over half of the respondents (18 out of 34) considering VR functionality as ``probably not useful'' or ``not useful'', as shown on Figure \ref{fig:survey-ad-vs-tried-vr} (reproduced here on Figure \ref{fig:survey-ad-vs-tried-vr-bis}). 

We however noted a difference depending on the respondents' prior exposure to \gls{6dof} \gls{vr}, and therefore further compared the responses with whether the respondents had tried \gls{vr} tools for architecture-related activities. The corresponding results are plotted on Figure \ref{fig:survey-ad-vs-tried-vr-archi} (reproduced here on Figure \ref{fig:survey-ad-vs-tried-vr-archi-bis}), that only includes 20 respondents since the question on \gls{vr} for architecture was only asked to those who indicated they were familiar with \gls{vr} and had tried the technology in the first place.

While GHVRGraph seemed to be slightly better received amongst respondents with prior exposure to \gls{vr}, the population size did not allow us to reach statistical significance.

\section{Discussion}

GHVRGraph, the proof-of-concept research prototype presented in this chapter, goes beyond the work presented in Chapter \ref{chap:immersive-adjustment}, by enabling \gls{vr}-based \emph{Level~3} control over \gls{ad} definitions created with Grasshopper. The prototype allows control over the graph-based \gls{ad} representation itself, and not just parameter values.

This type of control allows \gls{ad} architects to make deeper modifications than what \emph{Level 2} offers, and could even lead to architectural geometries being created only through the \gls{vr} interface. While we do not believe that the latter is a realistic scenario at the moment, the ability to manipulate components and links from the immersive environment certainly opens new possibilities for architects to integrate immersive technologies in their practice.

Such tools however do not truly concern other stakeholders than architects themselves, since modifying the graph-based representation (corresponding to the visual algorithm) requires expertise with the \gls{ad} paradigm. While this limits the number of potential users for \gls{vr}-based \emph{Level 3} applications, a more polished version of GHVRGraph that would be delivered as commercial software could be affordable for architectural firms of all sizes, since the hardware requirements are the same as those of GHVRBridge, described in Section \ref{sec:ecaade}.

The following hypothetical use case help clarify a potential usage scenario for GHVRGraph, where the prototype tool is likely to improve the design process.

\begin{usecase}{for GHVRGraph}[]
	\label{usecase-3}
	A small architectural firm is in charge of the construction of a monument for a big garden (10,000 square meters) at the entrance of a small castle converted to a modern art gallery. Commissioned by the gallery's board, the monument should inspire a sense of prestige. The architects are still in an early design phase with multiple options being considered. Based on the current state of the \gls{ad} definition she created with Grasshopper, one of these architects wants to get a better idea of what these options look like from a human perspective. She therefore starts GHVRGraph with the intent of checking the options in \gls{vr}, switching between options by editing links in the graph-based \gls{vr} representation. Once immersed and after checking the initial options, the architect has a new idea: she wants to combine two of the initially proposed options. After rebranching the definition from the \gls{vr}-based representation, she visualises the result. She then moves back to Grasshopper to refine the design idea based on the version she modified from GHVRGraph.
\end{usecase}


\subsection{Towards a collaborative variant of GHVRGraph}
While GHVRGraph is a single-user application, it would be straightforward to extend it to enable multi-user co-presence using the solution presented in Section \ref{sec:copresence}. If all collaborators are placed in the same virtual environment and interact with the same \gls{ad} definition, solutions similar to the ones discussed in \ref{sec:conflicts-params} would be sufficient to enable conflict-free collaboration or at least provide a clear procedure for conflict resolution.

However, if users collaborate through separate representations of the same \gls{ad} definition, there is no guarantee that these representations are in the same state at any point in time. Changes made to the \gls{ad} definition by different collaborators must therefore be integrated to form the main joint definition. Despite the graph-based nature of the \gls{ad} definitions that GHVRGraph works with, it is clear that this problem requires solutions based on software merging \citep{mens2002state}, or more generally model-based merging techniques \citep{brunet2006manifesto}.

Since collaborators are not guaranteed to produce changes that are compatible with each other, model merging is sometimes not sufficient, and conflicts between incompatible local modifications need to be resolved. This is practically unavoidable when collaborative development occurs in parallel \citep{mckee2017software}. The tooling at developers' disposal consequently should integrate visual code comparisons to allow for conflict resolutions to become as smooth as possible.

As for \gls{ad} and visual programming, none of the popular commercial tools include merging capabilities. However, a system called MACE (Multiple  Alternatives - Comparison  and  Editing), proposed in \citep{zaman2017mace}, allows for comparison between \glspl{dag}. Using MACE, a designer can choose different versions of an \gls{ad} definition and the tool will highlight changes between the selected base version and its compared alternatives. It will colour vertices and parameters depending on the type of changes that were made to them (e.g., a green vertex means that the vertex was added). This is comparable to code differencing tools used in software developments to compare different versions of the same code base (e.g., to solve merge conflicts when they occur).

Pushing the inspiration from software engineering further with capabilities that are inspired by version control systems, \citep{cristie2019ghshot} proposed a system called GHShot, that works with snapshots of Grasshopper definitions. Using that system, a Grasshopper user can push a version of a definition to a centralised server. When doing so, the user indicates whether this new version is a simple progression of the previous one, or an alternative solution that should be included in a separate new branch. Reusing the ideas of MACE, GHShot also proposes a ``diff view'' that highlights changes between different versions of a definition. To our knowledge, GHShot is the \gls{ad} system whose features most resemble the capabilities of the version control systems used in software development.

A potential approach, that has been ruled out in GHShot but may still be a viable lead to support merging \gls{ad} definitions, is to integrate versions of a definition via the textual representation of these versions (the \texttt{.ghx} files). Since these contain a textual description of the definition, they allow for text merging techniques to be used. This should work well for collaborators that work on separate parts of a definition, since the XML entities they would create or modify are then different. However, conflicts are unavoidable when collaborators interact with the same components, and solutions such as MACE and GHShot's diff views would still be needed to allow designers to resolve these conflicts.

\subsection{Opportunities for visualisation enhancements}
\label{sec:ghvrgraph-opportunities-visualisation}

The current version of GHVRGraph displays the full graph representation to the \gls{vr} user. This means that bigger graphs will lead to smaller components, since the dimensions of the table are fixed. This effectively makes the current version of GHVRGraph unable to work with large Grasshopper definitions. To make the prototype scalable, zooming and panning (moving the view to another part of the graph) would need to be added to the set of available actions. It would therefore be interesting to explore interaction techniques to perform these actions in that \gls{vr} context.

In addition to the exploration of more modalities and interaction techniques that could lead to a better user experience, we could also explore different types of visualisations for \gls{ad} definitions in \gls{vr}. We could try to better capitalise on the three-dimensional aspect of immersive environments. A potential approach would be to ``unleash'' the graphs and allow users to move vertices anywhere in the 3D space, instead of constraining graph components to a (table) surface. As discussed in Section \ref{sec:ghvrgraph-representation}, one of the reasons we imposed that restriction was to preserve a mapping between the \gls{vr}-based representation and the desktop tool's canvas. If we would let go of that constraint, \gls{vr} users would get access to more space, and may organise the graph using the whole environment around them, placing vertices in different directions and at different heights.

In addition to the lack of mapping with the desktop software, a 3D graph representation is not necessarily beneficial, at least when the represented data is non-spatial, i.e., not inherently three-dimensional itself \citep{elmqvist-blog}, as is the case for visual algorithms (as opposed to flight data for instance). We do however note that there does not seem to be a consensus on that, since contributions such as \citep{halpin2008exploring} positively evaluate 3D and \gls{vr} general data visualisations.

As stated in \citep{drogemuller2018evaluating}, there is a trade-off between aspects supporting \gls{vr} visualisation (increased engagement, added dimension, etc) and specific issues with them (occlusion of information, navigation, etc).

Another way to capitalise on the third dimension offered by immersive experiences is to constrain the graph to a cuboid instead of a simple surface. This allows visualisations to use the upwards axis (that we will call the $Z$ axis) to relay information about the vertices but preserves the mapping with the desktop canvas, since the graph can always be projected back to the surface by ignoring the Z value in vertices' position. 

Example of information that can be conveyed by such a technique includes the depth of a vertex. This was implemented as part of the student project \citep{robin}, previously mentioned in Section \ref{sec:ghvrgraph-p2}.
The depth of a vertex is the number of edges needed to reach that vertex, starting from the root vertex in a tree (a \gls{dag} where each vertex has exactly one parent, except one vertex, the root vertex, that has no parent). We mapped \gls{ad} definitions to \glspl{dag} that are not necessarily trees, but we can still define depth in that case by choosing a convention. In the aforementioned project, we chose to define the depth of a vertex as the minimum number of edges needed to reach that vertex from any vertex that does not have incoming connections.

We can then compute the depth value for each vertex and map its $Z$ value accordingly (e.g., the deeper the vertex, the higher it is being placed). This particular visualisation if for example useful to identify input parameters easily, since they will be at a depth of 0, even if they might not be positioned on the left of the definition. In general, such a representation helps designers to visually identify the extent to which a certain vertex depends on other vertices. This is in a way similar to automated code formatting, that relies on indentation (horizontal offset) or specific characters to clarify that code blocks can be considered as an entity that is to be executed depending on the instruction generally specified above it.

While the same approach can be replicated with other metrics mapped to the $z$ axis, it is also possible to convey a different type of information. For instance, groups of related components could (automatically or based on user input) be placed at a certain height. This would allow designers to easily visualise related elements. A \gls{vr} application implementing that idea could allow its users to show or hide certain groups.

\subsection{Genericity and adaptability}

As discussed in Section \ref{sec:ghvrgraph-interoperability}, the choice to rely on a graph-based structure to represent Grasshopper definitions in \gls{vr} means that GHVRGraph could easily be adapted to other \gls{ad} modelling tools, or even to other domains that rely on models that can be represented as graphs (e.g., Unified Modelling Language diagrams \citep{uml} or Entity Relationship models \citep{chen1976entity}). Through a simple converter for the target software, it would indeed be possible to reuse the interaction mechanisms we developed. 

As mentioned in Section \ref{sec:immersive-modelling}, visual programming has been applied to other domains, such as animation design or for the development of interactive applications. 
Such domains may be able to benefit from immersive visualisations. This is typically the case when the output produced by the visual program can take advantage of the immersive context.
Adapting a solution such as GHVRGraph to these domains therefore allows ``visual programmers'' to visualise these resulting outputs alongside the generative program itself.

A potential example would be diagrams such as statecharts \citep{harel1987statecharts}, that define the behaviour of software systems and could therefore be employed to generate applications that execute in a 3D context. In that case, the immersive visualisation of what the application produces could be coupled with editing features for these statecharts, within the same immersive environment. Other potential applications for immersive visual programming are the design of transportation networks, (gaming) scenes, or robotics with three-dimensional path planning activities.

\chapter{Conclusion}
\label{chap:conclusion}
\epigraph{
	``Without the fun, none of us would go on!'' }{Ivan Sutherland}

The evolution of Computer-Aided Architectural Design was highly influenced by advances in computing technologies, as we have shown in the field's history, covered in Chapter \ref{chap:sota}. As PCs and computer networks improved and became more affordable, the discipline evolved with them, with new tools and design paradigms appearing along the way. While immersive technologies have been around for more than 50 years, they matured and became available to the masses only recently.

The current use of \gls{ar} and \gls{vr} in architecture is mostly limited to visualisation purposes and we have shown that these technologies should not be confined to such purposes. We therefore explored the integration of immersive technologies in the architectural design process, with a focus on \gls{ad}. 
While it is too early to reflect on the impact of today's developments on architectural practice, the present dissertation supports the following thesis statement:

\begin{center}
	\begin{tcolorbox}[sharp corners, colback=blue!10, colframe=gray!80!blue, coltitle=white,fonttitle=\sffamily\large, title=Thesis statement, width=.97\textwidth]
		
		Integrating \gls{ar} and \gls{vr} technologies into the Algorithmic Design toolset provides opportunities for architects to improve their workflow and to better present their creations to clients.
		
	\end{tcolorbox}
\end{center}

This chapter summarises our contributions and shows where they are located in the \gls{ad} landscape. We also present some practical usage scenarios for the proof-of-concept prototypes we developed and discuss their purpose. Finally, we discuss perspectives we envision for the field of immersive \gls{ad} and how the research presented in this dissertation could be continued.



\section{Research contributions}
\label{sec:contributions}

To support the aforementioned thesis statement, we conducted a series of experiments, developed a number of tools, and produced results to answer our research questions.

To verify the claim that immersive technologies should be integrated into the architectural design process, we conducted a survey (Section \ref{sec:survey}) on the potential of \gls{vr} in that context. The survey results support our assumption that the usage of \gls{vr} early on in the design process is desirable, motivating the need for conducting research on the topic. This section summarises the main contributions our research work brings to the \gls{caad} domain.

We followed a prototype-based process, described in Section \ref{sec:methodology}, to provide evidence for our thesis statement. Our principal realisations are three proof-of-concept software prototypes. They all provide immersive experiences based on \gls{ad} definitions. As indicated by their names (respectively GHVRBridge, GHXR and GHVRGraph), they work with Grasshopper, a popular \gls{ad} editor letting architects design models with visual algorithms. We presented these prototypes in detail in Chapters \ref{chap:immersive-adjustment} and \ref{chap:immersive-design} but we summarise hereafter their essence and their purpose.

GHVRBridge, presented in Section \ref{sec:ecaade}, enables architects to connect their \gls{ad} definitions in Grasshopper to a \gls{vr} experience. After choosing which parameters to share with the \gls{vr} application, the corresponding values can be adjusted from within the immersive environment. The same virtual environment provides a visualisation at scale of the geometry shared by the architect. Changes to the parameter values produce automatic updates on the visualised geometrical representation. GHVRBridge therefore allows architects to simultaneously visualise geometries designed with Grasshopper and adjust the parameters that are used to generate them. This all happens from a single \gls{vr} application, with no need to leave the \gls{vr} experience to interact with the model. 
We explored a collaborative extension to GHVRBridge presented in Section \ref{sec:cdve} to enable a similar experience with multiple stakeholders. It can therefore be used to show a project to a client, immersed in the same virtual environment as the architect, who still has control over parameter values used for that project. Another possible usage scenario for such an extension is the potentially remote collaboration between architects, to allow them to discuss over a definition in a \gls{vr} environment, while adjusting parameter values based on their discussion.

In Section \ref{sec:list}, we presented another prototype, GHXR. It also proposes immersive parameter adjustment and geometry visualisation features, but additionally explores the use of \gls{ar} and \acrlongpl{tui} to do so. GHXR is based on a table display on which tangible items are placed and mapped to Grasshopper parameters. Collaborators placed around the table can therefore naturally interact with these tangible parameters. The system comes with a circular screen setup that displays the geometrical representation, generated using these parameters, from different viewpoints. The visualisation also includes the surrounding buildings in the geometry's future location. At any point in time, users can wear either an \gls{ar} or a \gls{vr} headset, to visualise a three-dimensional rendering of the project.

The third developed prototype, GHVRGraph, was presented in Chapter \ref{chap:immersive-design}. GHVRGraph pushes the control over the \gls{ad} definition further, by allowing its users to interact with a graph-based representation of a Grasshopper \gls{ad} definition. We relied on an internal representation based on typed graphs to convert Grasshopper definitions to \gls{vr}-based interactable models. GHVRGraph allows designers to make deeper changes to definitions than simple parameter adjustments, through a virtual table metaphor that represents Grasshopper components as boxes placed on such a table. We explored different interaction techniques to manipulate these boxes and the links between them, via several variants that have been described in Section \ref{sec:ghvrgraph-interactions}.

\section{Discussion}
\label{sec:discussion}

This section will first discuss the immersive \gls{ad} landscape, locating the three developed prototypes in that landscape. Using it, we will clarify the relations between these prototypes as well as how they compare with other immersive \gls{ad} tools. 

On this basis, we will point out some of the limitations of the research we conducted.

\subsection{Overview of the immersive Algorithmic Design landscape} 

\begin{figure}
	\includegraphics[width=.75\textwidth]{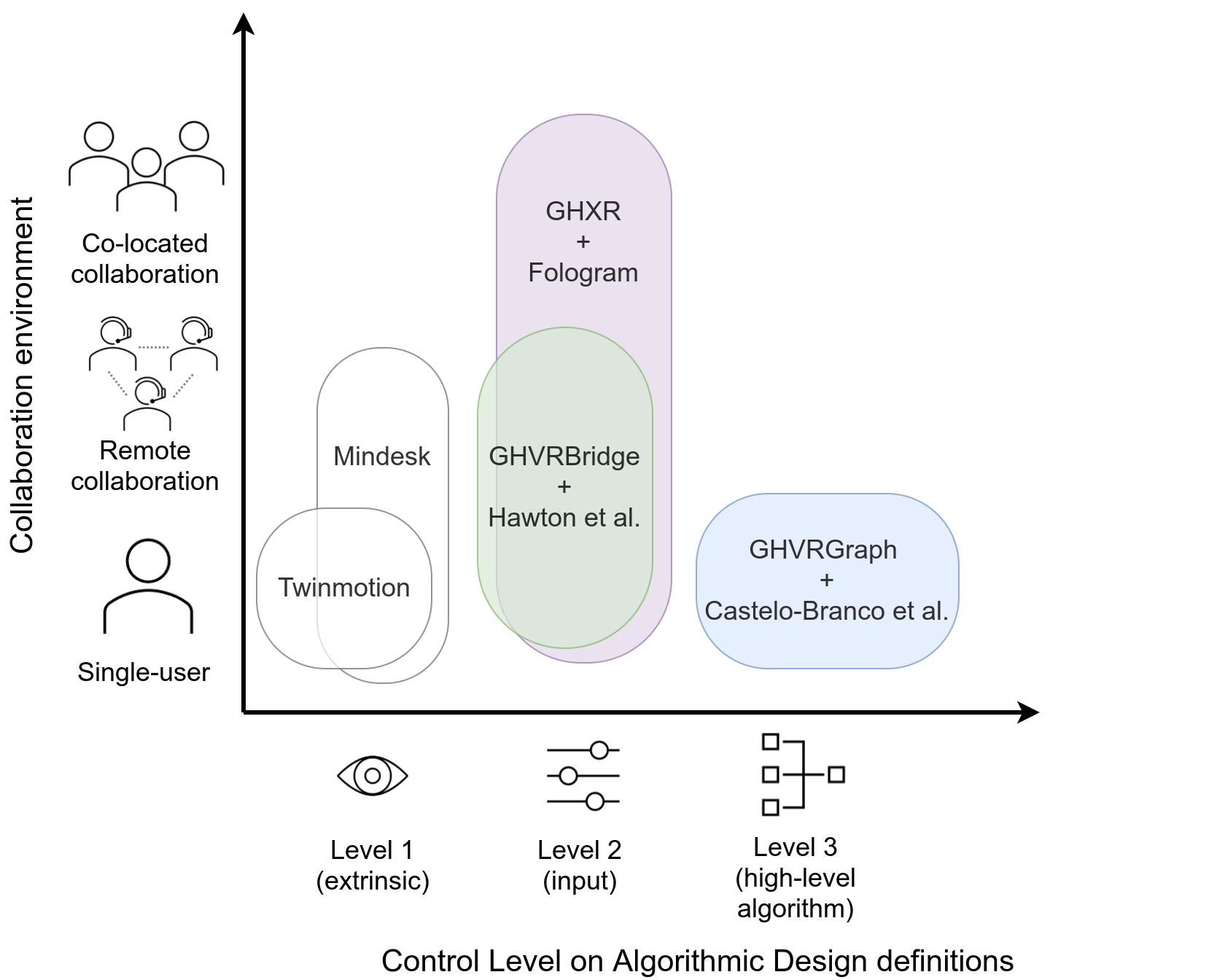}
	\centering
	\caption{\label{fig:comparison-control-collab} Relations between the developed prototypes and recent related work, based on differences in control level and collaboration environment.}
\end{figure}

A first point of comparison for the prototypes we developed is the level of control provided by these systems. This corresponds to how we structured the presentation of the prototypes in the dissertation, with Chapters \ref{chap:immersive-adjustment} and \ref{chap:immersive-design}.
The horizontal axis of the landscape in Figure \ref{fig:comparison-control-collab} corresponds to that criterion of comparison, using the control levels we previously defined (see Table \ref{tab:control-levels} on page \pageref{tab:control-levels}). On the figure's vertical axis, we put the collaboration environment, to differentiate between single-user and collaborative applications, and further subdivide the latter into remote or co-located collaboration.

The landscape shows that commercial tools such as Mindesk and Twinmotion are limited to extrinsic control and can therefore only modify the rendered version of a geometry or its environment. The two prototypes from chapter \ref{chap:immersive-adjustment}, namely GHVRBride and GHXR, allow users to adjust parameter data, and can therefore have an impact on the input of the algorithm generating the geometrical representation (intrinsic modifications). They, however, do not cover the manipulation of the visual algorithm itself; this being the purpose of GHVRGraph (chapter \ref{chap:immersive-design}).

Figure \ref{fig:comparison-control-collab} also includes other software tools, from researchers as well as companies working on similar topics. The next few paragraphs justify their presence in the landscape and discusses how the developed prototypes compare to them.

GHVRBridge's equivalent system described in \citep{hawtonshared}, that was developed at the same time and has been presented in Section \ref{sec:sota-immersive-ad}, covers the same purpose with a very similar approach. Both systems cover remote collaboration scenarios since multiple users can connect to the same virtual experience using them.

Fologram\footnote{\url{www.fologram.com}} is the \gls{ar} equivalent to GHVRBridge, and we thereby put it on the same location as GHXR in the two-dimensional landscape of Figure \ref{fig:comparison-control-collab}, since working in \gls{ar} means it is naturally adapted for co-located collaboration. We however note that GHXR has additional capabilities (multiple views, \gls{vr} support, \gls{tui}-based interaction, etc.) as compared to Fologram.

While the interaction modalities are different, GHVRGraph is comparable to the work described in \citep{castelo2020inside} (previsouly discussed in Section \ref{sec:sota-immersive-ad}) in that they both enable control over the visual algorithm for a given Grasshopper definition. We additionally note that \citeauthor{castelo2020inside}'s tool technically allows users to modify parameter values but we did not reflect that fact on Figure \ref{fig:comparison-control-collab} because it would be particularly hard to achieve a sufficient level of precision with the mirroring technique they proposed.

The coloured elements of Figure \ref{fig:comparison-control-collab} contain tools that were all published or released after the start of our research work, showing that the belief that immersive technologies need to be integrated into the \gls{ad} process is shared by many others.

\begin{figure}
	\includegraphics[width=.65\textwidth]{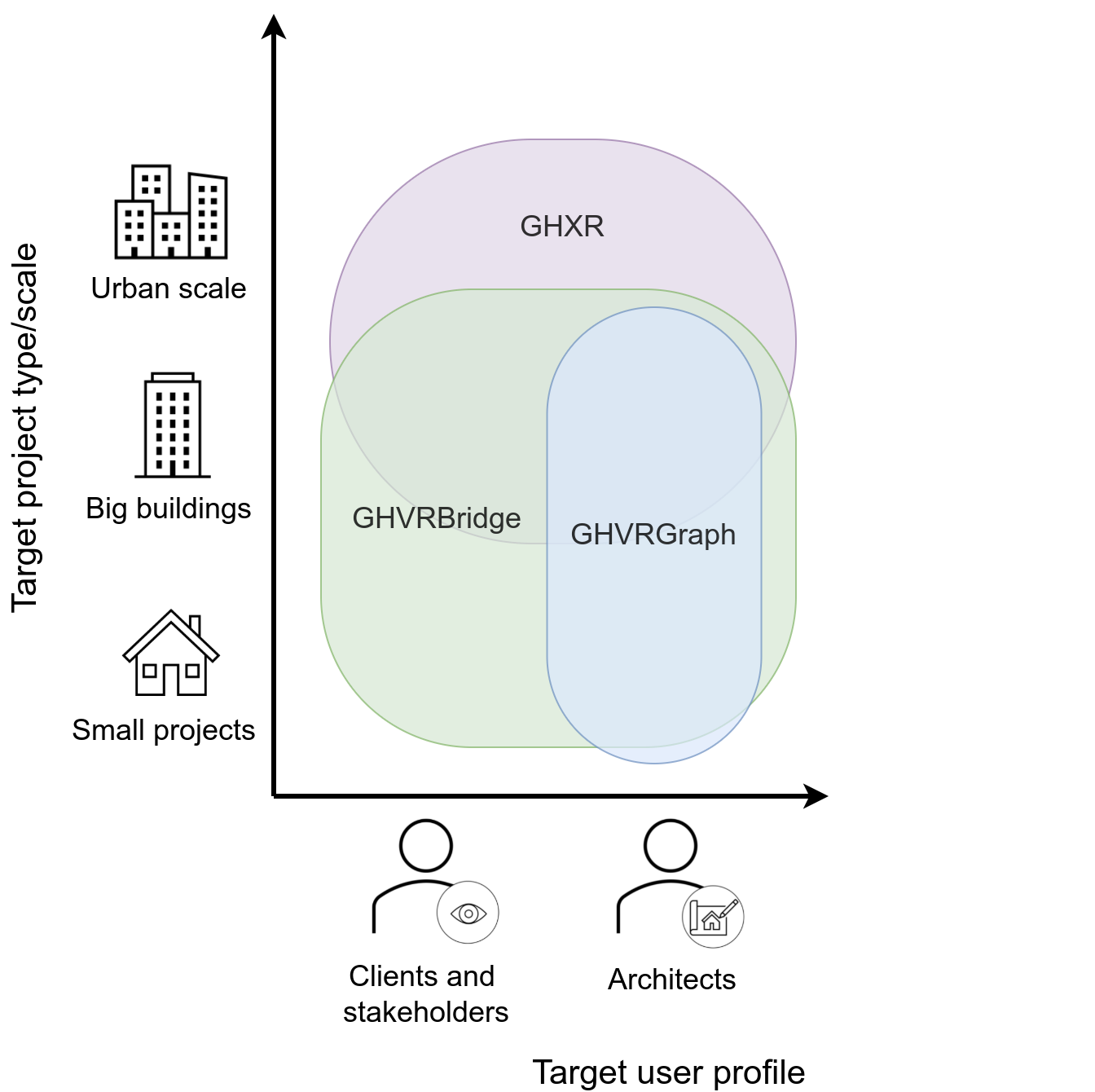}
	\centering
	\caption{\label{fig:comparison-role-project} Comparison between the three developed prototypes, with regards to target user profiles and types of architectural projects being covered.}
\end{figure}

The three prototypes we developed differ from each other in the potential usage scenarios they aim to cover, as shown in Figure \ref{fig:comparison-role-project}. While these systems all target architects, GHVRBridge and GHXR can also involve clients or other stakeholders, since these prototypes allow them to join shared experiences with the architects and visualise the same geometries. This means that both these prototypes suggest a user-centred approach for \gls{ad}, where architects preserve control over the \gls{ad} definition while immersed with these stakeholders.

While GHVRBridge and GHVRGraph can be used for all types of architectural projects, they both provide full-scale visualisations (to take advantage of the immersive aspect of \gls{vr}) and are likely not adapted for designing at the urban scale. GHXR fills that void through the various viewpoints (including non-\gls{vr} ones) that are part of the system. These viewpoints allow for such design activities because users can therefore choose the appropriate viewpoint when needed.
However, the hardware and space requirements of GHXR imply that it only makes sense to rely on such a system for projects that are big enough to justify investing in its usage. It also follows that GHXR targets bigger companies than GHVRBridge and GHVRGraph, which are accessible to any type of company or individual.

To make the previous claims more concrete, we recall here (and assign to a particular system) the 3 hypothetical use cases we introduced in Chapters \ref{chap:immersive-adjustment} and \ref{chap:immersive-design}, since they help clarify some usage scenarios where the developed prototypes are likely to make an appreciable difference.

\begin{usecase}{for GHVRGraph}[]
	\label{usecase-3bis}
A small architectural firm is in charge of the construction of a monument for a big garden (10,000 square meters) at the entrance of a small castle converted to a modern art gallery. Commissioned by the gallery's board, the monument should inspire a sense of prestige. The architects are still in an early design phase with multiple options being considered. Based on the current state of the \gls{ad} definition she created with Grasshopper, one of these architects wants to get a better idea of what these options look like from a human perspective. She therefore starts GHVRGraph with the intent of checking the options in \gls{vr}, switching between options by editing links in the graph-based \gls{vr} representation. Once immersed and after checking the initial options, the architect has a new idea: she wants to combine two of the initially proposed options. After rebranching the definition from the \gls{vr}-based representation, she visualises the result. She then moves back to Grasshopper to refine the design idea based on the version she modified from GHVRGraph.
\end{usecase}

\begin{usecase}{for GHVRBridge}[] 
\label{usecase-1bis}
An architectural firm is in charge of the construction of a modern-looking clubhouse for a golf club. Based on the club's demands, the architects have designed a first version of the building that will welcome the 300 registered members. The design process in quite advanced and the building is expected to be close to its final form. The architects would like to get input from the clients (the club's board members), so they invite them to a \gls{vr} session where both the lead architect and board members visualise the virtual building in its current state. The \gls{vr} visualisation helps the clients understand what the actual clubhouse will look like. The architect guides the clients around the virtual environment and adjusts parameter values along the way to accommodate the remarks from the clients.
\end{usecase}

\begin{usecase}{for GHXR}[]
\label{usecase-2bis}
The municipal council of a city wants to create a public space in a disused area of 1,000 square meters, nestled amid a few apartments and shopping buildings. 
They appointed an architectural firm that works with Algorithmic Design to do so and asked them to include a big sculpture to be placed somewhere in the centre of that space. The project is already well advanced in Grasshopper and the architects now want the opinion of the council to adjust parameter values. They therefore invite council members for a meeting at their office, that themselves invite representatives from commercial and residential buildings in the project's vicinity with them. Using GHXR, all these stakeholders are able to visualise the project from different angles and can take an active part in the remaining (parameter adjustment) design decisions. The \gls{ar} and \gls{vr} visualisations help non-architects to better appreciate the dimensions of the sculpture and how it integrates with the surroundings.
\end{usecase}

\subsection{Limitations}
The three proof-of-concept prototypes that we developed extend the integration of immersive technologies beyond their current established usage in the field of \gls{caad}, but it is always possible to push the boundaries further. 
This section will discuss limitations related to the developed prototypes, pointing out areas where they could improve. We will also cover aspects and challenges that were considered out of scope of the present dissertation, but that would need to be tackled for the potential commercialisation of similar software tools. Finally, we will discuss limitations that pertain to the methodology we followed.

\subsubsection{Limitations related to interfaces and proposed interactions}

The first series of limitations are related to the user interface and the interaction modalities we proposed in the three developed prototypes.
GHVRBridge is the first prototype we developed. As presented in Section \ref{sec:ecaade}, the type of interface we used is similar to those of most desktop tools, with a \gls{6dof} controller replacing the usual mouse or trackpad. Due to the three-dimensional context, it is desirable to go beyond such two-dimensional interfaces. This partly motivated the development of the next two prototypes: GHXR and GHVRGraph.

GHVRBridge could also be extended by integrating smarter interactions for certain parameters. It would for example be possible to provide control over a point, surface, or volume, directly from within the \gls{vr}-based environment, using three-dimensional widgets similar to those used in standard 3D modellers. The use of such widgets would replace the manipulation of separate sliders (e.g., a single handle to control a point would replace three number sliders).

In a similar way, going further with interactions would also have been feasible for GHXR. We in fact did not enable control over the perspective camera, as explained in Section \ref{sec:list}. It is difficult to provide control over a 3D element using widgets whose position is only tracked in two dimensions (along the tangible table's surface). However, during the last few days of my research stay at LIST, we started to develop an extended version of a tangible, that is mounted with an infrared sensor able to evaluate the distance between the sensor and an object placed above it. Using that sensor, a user would be able to move his hand up and down to control an additional dimension (e.g., the height at which the camera is placed), thereby enabling three-dimensional control. This is left as a future work that would extend the capabilities of GHXR.

As for GHVRGraph, the feature set could be expanded by supporting additional actions on the graph-based representation. This includes moving and zooming on the canvas (the table surface where the graph representation is placed). The user would then be able to choose how much of the representation needs to be visualised at any given time. More advanced interaction techniques (such as the ones presented in Section \ref{sec:sota-3dui}) could also be employed, to take advantage of different approaches and mitigate their individual shortcomings.

It would also be interesting to integrate the parameter adjustment feature from GHVRBridge and GHXR into GHVRGraph, to provide a complete experience to the architect, who would thereby gain access to multiple control levels for different design activities.

\subsubsection{Additional untackled aspects}

The scope of this dissertation focused on specific aspects, inevitably leading to other aspects that could not be covered.
As mentioned in Chapters \ref{chap:immersive-adjustment} and \ref{chap:immersive-design}, GHVRBridge and GHVRGraph are based on the assumption that the number of users, collaborating in a concurrent session with these prototypes, is limited. A commercial product based on our work would likely need to tackle the concurrent modification issue (and the conflict resolution) in a more advanced way than the one we implemented. The proposals inspired from different fields that we discussed in the corresponding chapters could be a good start to reflect on how to do so.

Other elements that would need to be perfected in order to bring the developed prototypes to a broader audience include the general look and feel of the prototypes. Better-looking \gls{vr} menus and interactable objects would need to be created. The representation of the surrounding buildings we show in GHXR could also be improved and a commercial solution may want to rely on another (paid) provider to obtain more accurate or up-to-date building data than the one we used.

Another important feature that would need to be implemented for a commercial usage of such tools is audio communication between distant collaborators (often referred to as Voice over IP \citep{davidson2000voice}). Many solutions exist to achieve such communication on game engines such as Unity, including paid out-of-the-box options that would be easy to integrate with our code.

Another limitation we mention here because it corresponds to some of the feedback we received is the potential delay between a parameter modification in GHXR or GHVRBridge, and the update of the corresponding geometrical representation rendered in the virtual environment. GHVRBridge was conceived with particular attention to network aspects, with adapted transfer protocols and optimised packet sizes. On the other hand, GHXR was designed to be run from a local setup, and should therefore typically benefit from very high speed data transfers between components of the system that do not require the same kind of optimisations as GHVRBridge. 

Regardless of the level of attention given to data transfer, the main bottleneck remains the generation of the geometry by the \gls{ad} software itself, at least for complex definitions. It is not uncommon to encounter complicated designs in which modifying a parameter value leads to multiple seconds of re-generation time. The added delay introduced by our solutions is typically negligible for such models. There is unfortunately no real solution to that issue, since it pertains to the optimisation of the \gls{ad} software itself and because some designs are simply too complex to be computed rapidly. GHVRBridge and GHXR propose a rapid \gls{vr}-based evaluation of design variations, and architects should therefore focus their updates on parameters whose adjustment does not yield a long re-generation of the geometrical representation. More complex designs that require lengthy computation times are simply not adapted to a rapid iteration process (\gls{vr}-based or not). 

\subsubsection{Methodological limitations}

Each of the developed prototypes was subject to evaluations with architectural students and researchers, through workshops and demo sessions that we organised during course sessions, seminars and other events. However, most of these evaluations were rather informal and limited in the number of participants by the nature of the events. This means that the evaluations we conducted were subject to threats related to the small population size (less diversity, more susceptible to outliers, etc\textit{}).

Furthermore, we did not validate the prototypes with professional practitioners, so the feedback we received was biased towards a student or academic point of view. As pointed out in \citep{stals2020state}, while it is common to find studies that do not involve professional architects as part of their testing sessions, doing so would help in making sure that the developed tools answer their needs.

The lack of a more systematic validation of the developed prototypes with more professional participants was initially due to a shortage of contacts. Once we had identified and solicited enough potential participants that indicated being interested in evaluating these prototypes, we had difficulty to actually perform these evaluations, due to the sanitary situation (linked to the COVID-19 crisis) that affected the last two years of our research. We nevertheless created evaluation protocols for the developed prototypes and it would be possible to conduct such evaluations rapidly, should we have access to a pool of qualified participants and the possibility to perform on-site tests. Remote validation is hardly possible because of both hardware and software requirements.

While it is difficult to transpose the evaluation of immersive systems to an online setting, since they involve \gls{vr} or \gls{ar} components, we did so for GHXR. As mentioned in Section \ref{sec:validation-adjustment}, we relied on a simulator for the display table and its tangible widgets. We in fact conducted three testing sessions where the participant had to perform a particular task using that simulator before answering a post-task questionnaire. In addition to the limited number of participants, it would be difficult to assert the irrefutable nature of such online evaluations, since the user experience is then transferred to a two-dimensional environment and the user must interact with software objects that simulate the actual tangible items of the immersive system.

\section{Perspectives}
\label{sec:perspectives}

The developed prototypes demonstrate the potential of using immersive technologies for \gls{ad} activities, and can be used as a stepping stone for discussion over the topic, so that architects can engage with these technologies. They can be seen as a precursor for the development of fully-fledged commercial products that should improve the design process for many architects. That impact should however not be limited to the architects, since the use of such technologies should lead to clients and other stakeholders being more involved during the design process, shifting it to a user-centred process. Independently from the previous discussion on necessary refinements to the developed prototypes to prepare them for broader and commercial usage, this section discusses future opportunities for \gls{ad} and its use of immersive technologies.

\subsection{Enhancing the aid for Algorithmic Design}

As stated in Section \ref{sec:ghvrgraph-opportunities-visualisation}, there are opportunities to use the third dimension offered by immersive technologies to convey additional information to the designer working on an \gls{ad} definition. Various representations should be explored. The usage of height variation to carry additional information appears to be the most promising since it preserves a natural mapping with the standard representation of \gls{ad} models in desktop-based editors such as Grasshopper. Experiments should be run with different metrics used as the conveyed information, including indicators inspired by the software engineering literature \citep{fenton2000software} (e.g., to evaluate the complexity of a definition, using metrics such as cyclomatic complexity \citep{ebert2016cyclomatic}). Such work on quality analysis of \gls{ad} definitions has in fact already begun \citep{davis2013modelled} but has yet to reach actual tools in use by practitioners.

It would also be interesting to try such alternative visualisations within desktop-based \gls{ad} tools as well. The user should in that case be able to switch from these extended visualisations to the basic representation easily, depending on that user's needs. More generally, the aids offered by most textual-based development environments (e.g., syntax colouring or code autocompletion) could be adapted to the visual \gls{ad} editors. In fact, desktop \gls{ad} editors typically include an autocompletion feature that suggests components based on what the user starts to type. This could be extended to provide suggestions based on the context (e.g., what component was placed or clicked last) and suggestions could also appear when one end of a link has been selected (only suggesting components that contain an input port with an appropriate type for the given link).

In addition to such aids, software development tools often include automated refactoring features \citep{mens2004survey}, allowing the developer to restructure the code, usually to improve on the design, while preserving its behaviour. Refactorings sometimes occur to introduce software design patterns \citep{gamma1995elements}, that are generic solutions to solve regular issues in the development of software systems. Interestingly, the concept of patterns as reusable proven solutions to common problems originated from the architectural design field \citep{alexander1977pattern}, where they serve as an aid to designers. We believe it is time for the pattern concept to return to architecture and \gls{ad} in particular, as suggested in \citep{woodbury2010elements}, a seminal book on \gls{ad} where the author defined \gls{ad} patterns inspired by the software development equivalents. Based on these patterns, \gls{ad} tools could propose refactorings to their users, so as to improve the overall quality (e.g., in terms of readability, maintainability or reusability) of the designed definitions.

Returning to the topic of immersive support for \gls{ad} activities, the way we developed the prototypes makes their interoperability with other \gls{ad} software possible, through the development of appropriate converters between such software and our representations. A combined solution blending the developed prototypes, and extending them to support other \gls{ad} software than Grasshopper, would have the potential to become a single immersive platform for \gls{ad}, providing a uniform experience to practitioners using different software tools.

\subsection{Technological improvements}

Until now, portable standalone headsets for both \gls{ar} and \gls{vr}, that do not need to be connected to a computer to function, lack computing power. This impacted the development of GHXR, since the \gls{ar} headset that we used (the Hololens) was not powerful enough to display the surrounding buildings that our \gls{vr}-based component employs (that component uses the HTC Vive, a headset linked to a computer that performs the computations).
While it appears clear that relying on a bigger computer to perform heavy computations will lead to better performances for connected headsets, an increase in computing power for portable headsets would allow for additional compute-intensive features. This would in our case allow to add holograms of surrounding buildings to the \gls{ar} component of GHXR.

Another possible direction to extend the use of \gls{ar} for architectural design would be to rely on other kinds of augmentations than visual ones. Olfactory \gls{ar} could be employed to further immerse users through smells and audio \gls{ar} would allow for sound simulations to be run, so that users would get an idea of expected noise disturbances linked to a project.

A potential future (r)evolution for \gls{ar} and \gls{vr} displays is the advent of Retinal Projection Displays (RPDs) \citep{pryor1998virtual}, that project content directly onto the wearer's retina. Once such technology matures, it has the potential to lead to immersive ``glasses'' in a small form factor, that would become wearable for significantly longer periods than the current headsets. These glasses could also potentially turn into devices capable of switching between \gls{ar} and \gls{vr} modes, and would therefore be an ideal multi-purpose immersive display. Architects would easily get access to the types of visualisations we developed, with nothing but such glasses.

We also think and hope that immersive technologies will keep on becoming more affordable, so that they can reach architects from firms of all sizes, and thereby become integrated into architectural practice.

\begin{appendices}
	\chapter{GHXR evaluation protocol}	
	\label{apx:ghxr-protocol}
	Removed because arxiv does not like pdfs
	\chapter{GHXR System Usability Scale questionnaire}	
	\label{apx:ghxr-sus}
	Removed because arxiv does not like pdfs
	
\end{appendices}

\backmatter

\bibliography{sources}

\clearpage

\thispagestyle{empty}
\newpage
\textcolor{white}{.}

\thispagestyle{empty}
\newpage
\textcolor{white}{.}

\thispagestyle{empty}
\newpage
\textcolor{white}{.}

\thispagestyle{empty}
\newpage

\thispagestyle{empty}

\vspace{1em}

\vspace{0.4cm}
\noindent {\bf \large Integrating Immersive Technologies for Algorithmic Design in Architecture} \\ \vspace{0.4cm} 

\vspace*{1mm}

Architectural design practice has radically evolved over the course of its history, due to technological improvements that gave rise to advanced automated tools for many design tasks. Traditional paper drawings and scale models are now accompanied by 2D and 3D Computer-Aided Architectural Design (CAAD) software.

While such tools improved in many ways, including performance and accuracy improvements, the modalities of user interaction have mostly remained the same, with 2D interfaces displayed on 2D screens. The maturation of Augmented Reality (AR) and Virtual Reality (VR) technology has led to some level of integration of these immersive technologies into architectural practice, but mostly limited to visualisation purposes, e.g. to show a finished project to a potential client.

We posit that there is potential to employ such technologies earlier in the architectural design process and therefore explore that possibility with a focus on Algorithmic Design (AD), a CAAD paradigm that relies on (often visual) algorithms to generate geometries. The main goal of this dissertation is to demonstrate that AR and VR can be adopted for AD activities. 

To verify that claim, we follow an iterative prototype-based methodology to develop research prototype software tools and evaluate them. The three developed prototypes provide evidence that integrating immersive technologies into the AD toolset provides opportunities for architects to improve their workflow and to better present their creations to clients. Based on our contributions and the feedback we gathered from architectural students and other researchers that evaluated the developed prototypes, we additionally provide insights as to future perspectives in the field.

\vspace*{20mm}

\begin{tabular}{lcccr}
\includegraphics[width=8em]{Figures/UMONS.jpg}  & \hspace{3em} &
\end{tabular}

\end{document}